\documentclass[fleqn,10pt]{wlscirep}
\usepackage[utf8]{inputenc}
\usepackage[T1]{fontenc}
\usepackage{physics}
\usepackage{amsmath}
\usepackage{cancel}
\usepackage{graphicx}
\usepackage{xcolor}
\usepackage{soul}

\newcommand{\pati}[1]{}

\title{Electron Scattering at a Potential \\ Temporal Step Discontinuity}

\author[1]{Furkan Ok}
\author[1]{Amir Bahrami}
\author[1,*]{Christophe Caloz}
\affil[1]{Department of Electrical Engineering, KU Leuven, Leuven, 3000,  Belgium}

\affil[*]{christophe.caloz@kuleuven.be}

\begin{abstract}
We solve the problem of electron scattering at a potential temporal step discontinuity. For this purpose, instead of the Schr\"{o}dinger equation, we use the Dirac equation, for access to back-scattering and relativistic solutions. We show that back-scattering, which is associated with gauge symmetry breaking, requires a vector potential, whereas a scalar potential induces only Aharonov–Bohm type energy transitions. We derive the scattering probabilities, which are found to be of later-forward and later-backward nature, with the later-backward wave being a relativistic effect, and compare the results with those for the spatial step and classical electromagnetic counterparts of the problem. Given the unrealizability of an infinitely sharp temporal discontinuity -- which is of the same nature as its spatial counterpart! -- we also provide solutions for a smooth potential step and demonstrate that the same physics as for the infinitely sharp case is obtained when the duration of the potential transition is sufficiently smaller than the de Broglie period of the electron (or deeply sub-period).
\end{abstract}

\begin{document}

\flushbottom
\maketitle
%
%
\thispagestyle{empty}

\noindent


\section*{Introduction}
\label{sec:intro}

\pati{Spatial Step}
Electron scattering at a potential \emph{spatial} step is a canonical problem that is treated in the introductory section of most textbooks on quantum mechanics~\cite{Griffiths_2018_Quantum,zettili2009quantum,shankar2012principles,miller2008quantum,landau_2013_quantum,sakurai_napolitano_2020} and that underpins uncountable phenomena (e.g., quantum reflection, transmission and interference, quantum tunneling, quantum wells and scattering resonances, quantum coherent transport) 
and applications (e.g., p-n junction diodes, transistors, semiconductor lasers, and detectors, scanning tunneling microscopy, quantum computing, particle accelerators). The problem is typically addressed by resolving the  Schr\"{o}dinger equation~\cite{Sch_1926} for non-relativistic particles, but requires promotion to the Klein-Gordon equation~\cite{klein1926quantentheorie,gordon1926comptoneffekt} or to the Dirac equation~\cite{Dirac_1928_paper} for relativistic particles, of spin~0 or $1/2$, respectively.

\pati{Temporal Step Discontinuity}
The problem of electron scattering at a potential \emph{temporal} step is arguably, from space-time duality, as fundamental as that of the spatial step. A number of works on quantum phenomena occurring at step discontinuities have been reported in the literature~\cite{Mendonca_2000_quan_time_refrac,Mendonca_2003_temporalsplitter,Goldman_2014_periodically,Reck_2017_dirac_time_mirror,xu_2018_spacetime,Junk_2020_floquet_oscillations_dirac,Gao_2021_floquet_zener,schultheiss_2021_time,peng_2022_topological,Gao_2022_semiclassical,vazquez_2022_shaping,Lu_2022_floquet_dirac_bands,Kim_2023_temporal_Dirac}, but the specific problem of electron scattering at a potential temporal step has surprisingly not been explicitly resolved yet. Such a gap needs to be filled. This is all the more obvious when considering the promising opportunities of transposing to the quantum realm recent concepts developed in the booming field of \emph{classical} electromagnetic modulation-based time-varying~\cite{Morgenthaler_1958_puretime,Plansinis_2015_TemporalAnalog_Refl_Refr,Mazor_2021_unitary,Tretyakov_2023_Temporal_discon} and space-time varying~\cite{Chamanara_2017_optical,Engheta_2021_Metamat,Huidobro_2021_homogenization,Li_2023_gener_total_ref,Bahrami_2023_Accelerated} metamaterials~\cite{Caloz_2020_spacetimeI,Caloz_2020_spacetimeII,Engheta_2021_Metamat,Caloz_2022_GSTEM}, which have already led to a wealth of novel effects and applications, including the inverse prism~\cite{Akbarzadeh_2018_inverse_prism}, linear-time invariance bound breaking~\cite{Shlivinski_2018_Beyond}, temporal aiming~\cite{Pacheco_2020_temporalaming}, extreme energy transformation~\cite{Li_2021_temporal_parity}, temporal antireflection coating~\cite{Pena_2020_tempcoating}, temporal polarization conversion~\cite{Xu_2021_complete_pol}, temporal analog computing~\cite{Rizza_2022_short_pulsed, Castaldi_2023_multiple_short_pulsed}, static-to-dynamic field conversion~\cite{Mencagli_2022_static}, temporal Faraday rotation~\cite{Li_2022_nonreciprocity,Li_2023_Faradaycrystal,Li_2023_stationary_charge}, arbitrary transfer function emulation~\cite{Ptitcyn_2023_timecircuit}, optimization-free filter and matched-filter~\cite{Silbiger_2023_filter}, broadband parametric amplification~\cite{Tien_JAP_1958,Galiffi_2019_broadband}, wave deflection and shifted refocusing~\cite{Deck_2018_wavedeflection} and nonreciprocity and optical isolation~\cite{Yu_2009_opticalisolation,Correas_2016_NonreciGraphene,chamanara2017optical,Taravati_2017_nonreciprocal,guo2019nonreciprocal}.

\pati{Contribution}
We present here an exact and comprehensive resolution of the problem of electron scattering at a potential temporal step discontinuity. We first show that the Schr\"{o}dinger equation cannot account for electron back-scattering for that problem and therefore decide to resort to the (more general) Dirac equation. We then demonstrate that a \emph{scalar} potential temporal step does not produce back-scattering, whereas a \emph{vector} potential temporal step does (see Supplementary Sec.~1), and explain this fact in terms of related gauge symmetry and symmetry breaking. We next derive formulas for the scattering coefficients, probabilities, and energy transitions of the electronic wave. Finally, we demonstrate that the corresponding scattering is a relativistic effect. Throughout the report, we systematically compare the problem with its spatial counterpart, and also point out some similarities and differences with corresponding electromagnetic problems~\cite{Caloz_2020_spacetimeI,Caloz_2020_spacetimeII,Caloz_2022_GSTEM}. Finally, we also provide solutions for a smooth step potential and investigate the related physics versus the transition duration of the step with respect to the de Broglie period of the electron.

\section*{Spatial and Temporal Sharp Potential Step Discontinuities}
\label{sec:STPotentials}

\begin{figure}[ht!]
    \centering
    \includegraphics[width=0.8\textwidth]{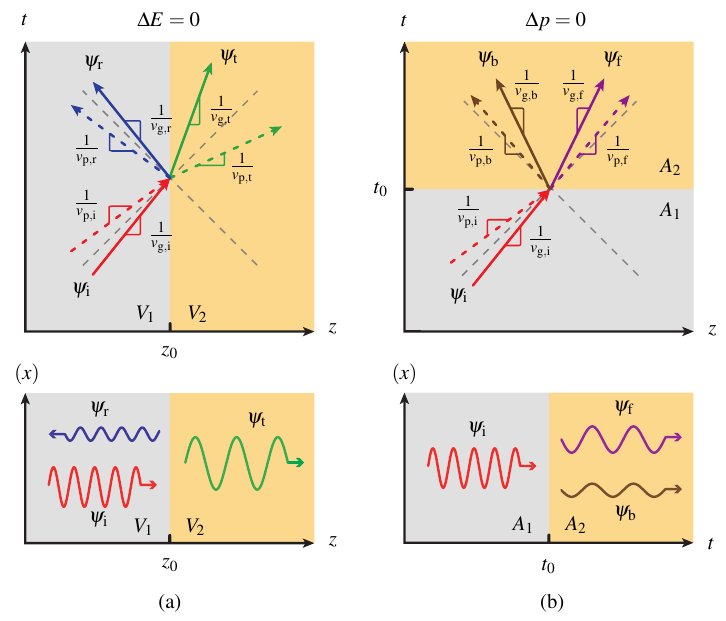}
    \caption{Electron scattering at a potential (a)~spatial and (b)~temporal step discontinuity, in spacetime (top panels) and space/time-transverse coordinates (bottom panels). The subscripts i, r, t, b, and f stand for incident, reflected, transmitted, later-backward, and later-forward, while the subscripts p and g stand for phase and group (velocity), respectively.}\label{fig:space_time_discon}
\end{figure}

\pati{Structures}
Figure~\ref{fig:space_time_discon} represents the problem of electron scattering at a potential step discontinuity, with the discontinuity being spatial in Fig.~\ref{fig:space_time_discon}(a) and temporal in Fig.~\ref{fig:space_time_discon}(b). The latter is the problem at hand in the report while the former is considered as its dual reference. In both cases, the changing parameter is a component of the four-vector potential $A^\mu=(V,\vb{A})$ and we shall later see why, as indicated in the figure, $V$ and $\vb{A}$ are the most relevant components for the spatial and temporal cases, respectively. 

\pati{Announcement of Main Results}
As known from textbooks, the scattered electronic waves in the spatial problem [Fig.~\ref{fig:space_time_discon}(a)] are reflected and transmitted waves, with conserved energy ($\Delta E=0$) and transformed momentum ($\Delta{p}\neq{0}$), as in classical electromagnetics. We shall show that scattering is quite different in the counterpart temporal problem [Fig.~\ref{fig:space_time_discon}(b)]. First, the scattered electronic waves are generally \emph{later-backward} and \emph{later-forward} waves, as opposed to reflected and transmitted waves, where the term \emph{later} means ``after the temporal discontinuity'', contrasting with the term \emph{earlier}, which refers to the wave launched before the temporal discontinuity, while the terms \emph{forward} and \emph{backward} denote corresponding propagation directions in space~\cite{Caloz_2020_spacetimeI}. Second, it is now the momentum that is conserved ($\Delta p=0$), while the energy is transformed ($\Delta E\neq{0}$). These two aspects parallel the situation that prevails in classical electromagnetics~\cite{Morgenthaler_1958_puretime,Caloz_2020_spacetimeII}, but with a number of differences, such as the fact the phase and group velocities are generally distinct, as represented in the figure, and the fact that the later-backward wave exists only in the relativistic regime, whereas it is always present in classical electromagnetics. Throughout the paper, we shall restrict our attention to the 1+1-dimensional case, with one dimension of space ($z$) and the dimension of time ($t$). Moreover, we shall use natural units ($\hbar=c=1$) and the Minkowski metric $\eta^{\mu \nu}=\mathrm{diag}(1,-1,-1,-1)$ throughout the report.

\section*{Choice of an Appropriate Equation}
\label{sec:analysis}

\subsection*{Limitations of the Schr{\"o}dinger Equation}

\pati{Derivative Characterization of the Equation}
One may first be tempted to address the problem of the sharp temporal potential step [Fig.~\ref{fig:space_time_discon}(b)] with the Schr{\"o}dinger equation, as typically done for the sharp spatial potential step [Fig.~\ref{fig:space_time_discon}(a)] in the non-relativistic regime. The Schr{\"o}dinger equation reads $i\partial_t\psi=-{\nabla}^2\psi/(2m)$~\cite{Griffiths_2018_Quantum,zettili2009quantum,shankar2012principles,miller2008quantum,landau_2013_quantum,sakurai_napolitano_2020}, where $m$ is the mass of the particle, which we shall consider from now on as being the electron. This equation has unpaired spatial and temporal derivatives, with the former (${\nabla}^2$) being of the second order and the latter ($\partial_t$) of the first order. 

\pati{Spatial Problem Double Condition}
In the case of the sharp spatial step, where one typically assumes the monochromatic ansatz $\psi\propto\textrm{e}^{-iEt}$, the Schr{\"o}dinger equation reduces to \mbox{$\psi=-{\nabla}^2\psi/(2mE)$}. Therefore, ${\nabla}^2\psi$ must be finite to ensure finite $\psi$, and hence $\grad\psi$ must be continuous. In addition, $\psi$ must also be continuous, for otherwise, $\grad\psi$ would be singular, and so would then also be $\psi \propto {\nabla}^2\psi$, and hence $\psi$. We have then the double boundary condition that both $\psi$ and ${\grad}\psi$ must be continuous at the spatial discontinuity. Thus, the second-order derivative operator ${\nabla}^2$ provides two boundary conditions, viz., the continuity of both $\psi$ and of ${\grad}\psi$ at the spatial discontinuity, which leads to a fully determined problem whose resolution provides the usual reflected and transmitted scattered electronic waves [Fig.~\ref{fig:space_time_discon}(a)]. 

\pati{Temporal Problem Single Condition and Possible Uncompleteness}
In contrast, the \emph{first-order derivative} $\partial_t$ provides only one boundary condition in the sharp temporal potential step problem [Fig.~\ref{fig:space_time_discon}(b)]. Assuming the plane-wave ansatz $\psi\propto\textrm{e}^{ipz}$, the Schr{\"o}dinger equation reduces to $\psi=i(2m/p^2)\partial_t\psi$ and $\partial_t\psi$ must therefore be finite to ensure finite $\psi$, which entails that $\psi$ must be continuous at the temporal discontinuity. However, this is indeed the \emph{the only} boundary condition, due to the absence of a higher-order temporal derivative. Consequently, the Schr{\"o}dinger equation does not readily include sufficient information to account for more than one scattered wave, which would involve more unknowns, specifically two unknowns in the possible case of later-forward and later-backward waves. Moreover, the Schr{\"o}dinger equation is not relativistic, and would hence miss related solutions.

\subsection*{Selection of the Dirac Equation}

\pati{Promotion to DE}
The Dirac equation, which reads for the free-electron case $i\partial_t\psi=-(i\alpha^{i}\partial_{i}-\gamma^{0}m)\psi$, where $\psi$ is a ($4\times{}1$) spinor and where $\alpha^{i}$ and $\gamma^{0}$ are ($4\times{}4$) matrices~\cite{Dirac_1928_paper,shankar2012principles,Greiner_2000_rel_quant,Peskin_1995_QFT} (see Supplementary Sec.~2), seems to represent a safer approach for obtaining a complete solution to our temporal step problem. It also has, as the Schr{\"o}dinger equation, a first-order temporal derivative order ($\partial_t$), but it involves multiple sub-equations that might together support sufficient information to account for more than one scattered wave, including relativistic ones.

\pati{Minimal-Coupling Dirac Equation \& Solution}
Let us then try to address the problem with the Dirac equation. In order to account for the potentials in Fig.~\ref{fig:space_time_discon}, we extend the free-electron Dirac equation to its minimal-coupling form~\cite{Dirac_1928_paper,shankar2012principles,Greiner_2000_rel_quant,Peskin_1995_QFT} (see Supplementary Sec.~2)
\begin{subequations}\label{eq:Dirac_equation}
\begin{equation}
    [\gamma^{\mu}\left(i \partial_{\mu} - q A_{\mu}\right)-m] \psi=0,
\end{equation}
where $\gamma^\mu$ are the matrices (Dirac-Pauli representation)
\begin{equation}\label{eq:Gamma_matrices}
\gamma^{0}=\begin{pmatrix}
        I & 0 \\
        0 & -I
    \end{pmatrix}
    \quad \text { and } \quad
\gamma^{i}=\begin{pmatrix}
    0 & \sigma^{i} \\ 
    -\sigma^{i} & 0
    \end{pmatrix},
\end{equation} 
with $I$ and $\sigma_i$ being the ($2\times{2}$) unit and Pauli matrices, respectively. 
\end{subequations}
Inserting the positive-energy monochromatic plane traveling wave ansatz $\psi=
        \begin{pmatrix}
            \varphi \\
            \vartheta
        \end{pmatrix}
        \mathrm{e}^{-i(E t - p z)}$ into Eqs.~\eqref{eq:Dirac_equation} yields the general solution form (see Supplementary Sec.~3.1)
\begin{equation}\label{eq:plane_wave_sol}
    \psi=
        \begin{pmatrix}
        1\\
        0\\
        \frac{E-qV-m}{p-qA}\\
        0
        \end{pmatrix}
    \mathrm{e}^{-i(E t - p z)},
\end{equation}
where $q=-e$ ($e>0$) is the charge of the electron. We make here the choice of a non-localized, continuous-wave ansatz because it is both the simplest and most appropriate regime to reveal the fundamental physics of the problem. The localized, wave-packet regime would be the next interesting regime to consider, with expected interesting novel time-delay physics, such as the quantum analog of the temporal Goos-H\"{a}nchen shift~\cite{ponomarenko2022goos}.

\section*{Scalar Potential Discontinuity} \label{sec:ScalarPot}

\pati{Absence of Back Scattering for $V(t)$}
One may first attempt to apply the general solution~\eqref{eq:plane_wave_sol} to the case of a pure-scalar potential, i.e., $A^\mu=(V,0)$, as typically done for the spatial step [Fig.~\ref{fig:space_time_discon}(a)], which corresponds in the problem at hand to a temporal scalar potential step $V(t)$, with $V(t<t_0)=V_1$ and $V(t>t_0)=V_2=V_1+\Delta{V}$, with $\Delta{V}=V_2-V_1$, where $t_0$ is the switching time. However, it may be easily verified (see Supplementary Sec.~3.2.1) that, although providing the expected energy shift (from $E_\mathrm{i}=\sqrt{p^2+m^2}+qV_1$ to $E_\mathrm{f}=\sqrt{p^2+m^2}+qV_2$), of potential interest for amplification applications~\cite{sakurai_napolitano_2020}, such a potential does not produce any later-backward wave scattering! This result, which might a priori appear surprising, may be explained in terms of gauge invariance symmetry.

\pati{Gauge Invariance and Symmetry Explanation}
The (external) electric and magnetic fields, $\vb{E}$ and $\vb{B}$, associated with the potential modulation, are generally related to the potentials as $\vb{E}=-{\grad}V-\partial_t\vb{A}$ and $\vb{B}=\curl\vb{A}$, which are invariant under the gauge transformation~\cite{Jackson_2000_electrodyn,Jackson_2001_Historical_Gauge}
\begin{equation}\label{eq:gauge_transf}
    V\rightarrow V^\prime=V-\pdv{\Lambda}{t}
    \quad\textrm{and}\quad
    \vb{A}\rightarrow\vb{A}^\prime=\vb{A}+{\grad}\Lambda,
\end{equation}
where $\Lambda$ is an arbitrary scalar function. The sharp temporal potential  step $V(t)$ considered in the previous paragraph is equivalent to the transformation $V'=V_1+\Delta{V}\theta(t-t_0)$, where $\theta(t-t_0)$ is the Heaviside step function, and $\vb{A}^\prime=0$, which is a particular case of the gauge transformation~$\eqref{eq:gauge_transf}$ with $V=V_1$, $-\partial_t\Lambda=\Delta{}V\theta(t-t_0)$, $\vb{A}=0$ and ${\grad}\Lambda=0$, corresponding to $\Lambda=-\Delta{}Vt\theta(t-t_0)$. Therefore, this sharp potential step does not involve any change in the external fields, which explains why we found that it produces no later-backward scattering.~[In fact, a similar result -- unchanged external fields and the consequent absence of back-scattering -- is found in the case of the sharp spatial step for the potential $A(z)$ (see Supplementary Secs.~4.2 and~3.2.3)]. The external fields are actually zero, since $\vb{A}=0$ and ${\grad}V={\grad}V(t)=0$; the related energy transition due to potential without field is therefore somewhat akin to the Aharonov–Bohm effect~\cite{Aharonov_1959_AB_effect}. This absence of back-scattering contrasts with the situation of the pure-scalar sharp spatial potential step $V(z)$ [Fig.~\ref{fig:space_time_discon}(a)], whose (reflected wave) back-scattering results from the breaking of the gauge condition~\eqref{eq:gauge_transf} (see Supplementary Sec.~4.1).

\section*{Vector Potential Discontinuity} \label{sec:VectorPot}

\pati{Gauge Symmetry Breaking for $\vb{A}(t)$}
We may suspect at this point that, since $V(t)$ fails to break the gauge symmetry~\eqref{eq:gauge_transf}, its pure-vector potential counterpart $\vb{A}(t)$ should break it, and hence bring about back-scattering, as the familiar (reflection) sharp spatial step $V(z)$. That this is indeed the case is shown as follows. Assuming $\vb{A}(t)=A(t)\vb{\hat{z}}$, the step function reads now $A(t<t_0)=A_1$ and $A(t>t_0)=A_2=A_1+\Delta{A}$, with $\Delta{A}=A_2-A_1$. The corresponding transformation is $A^\prime=A_1+\Delta{}A$$\theta(t-t_0)$ and $V'=V_1+\Delta{V}$ with $V_1=\Delta{V}=0$. Consistency with the gauge~\eqref{eq:gauge_transf}, given the mapping $A=A_1$, $\nabla\Lambda=\Delta\vb{A}(t)$ or $\partial_z\Lambda=\Delta{A}\theta(t-t_0)$, $V=V_1=0$ and $-\partial_t\Lambda=0$, would now demand that $\Lambda=\Delta{}A\theta(t-t_0)z=\Lambda(z,t)$ along with $\Lambda\neq\Lambda(t)$. The incompatibility between the last two conditions on $\Lambda$ indicates that the transformation indeed breaks the symmetry of the gauge~\eqref{eq:gauge_transf}, which entails transformed external fields and which may hence lead to electron back-scattering.

\pati{Dispersion and Transition Solution}
We can now solve the problem of interest, for the potential, $A(t)$, using the later-forward and later-backward ans\"{a}tze corresponding to the related temporal-step classical electromagnetic solutions~\cite{Morgenthaler_1958_puretime,Caloz_2020_spacetimeII} [Fig.~\ref{fig:space_time_discon}(b)]. According to Noether's theorem~\cite{Noether_1918}, for such a potential, momentum is conserved ($\Delta p=0$) due to spatial translational symmetry, viz., $p_\mathrm{i}=p_\mathrm{f}=p_\mathrm{b}=p$, while broken temporal translational symmetry leads to energy transitions, which are given by the dispersion relation (see Supplementary Secs.~3.1 and~3.2.4)
\begin{equation}\label{eq:disp_rel_temporal}
    E_{1,2}^{2}=\left(p-qA_{1,2}\right)^{2}+m^{2},
\end{equation}
where the subscript labels~1 and~2 refer to the earlier and later potential regions, respectively. Equation~\eqref{eq:disp_rel_temporal} leads to the energy relations
\begin{subequations}\label{eq:trans_temporal}
    \begin{equation}\label{eq:Ei_pA1}    
        E_\mathrm{i}=\sqrt{(p-qA_1)^2+m^2},
    \end{equation}
    \begin{equation}\label{eq:Efb_pA2}     
        E_\mathrm{f}=\sqrt{(p-qA_2)^2+m^2}
        \quad\textrm{and}\quad
        E_\mathrm{b}=-E_\mathrm{f},
    \end{equation}
\end{subequations}
where, assuming $E_\mathrm{f}>0$, the apparent negative energy $E_\mathrm{b}<0$ in the last relation simply represents propagation in the negative $z$ direction ($v_\mathrm{g,b}<0$), with positive energy ($|E_\mathrm{b}|>0$) (see Supplementary Sec.~5).

\section*{Dispersion and Transition Diagrams} \label{sec:DispDiagrams}

\begin{figure}[ht!]
    \centering
    \includegraphics[width=0.7\textwidth]{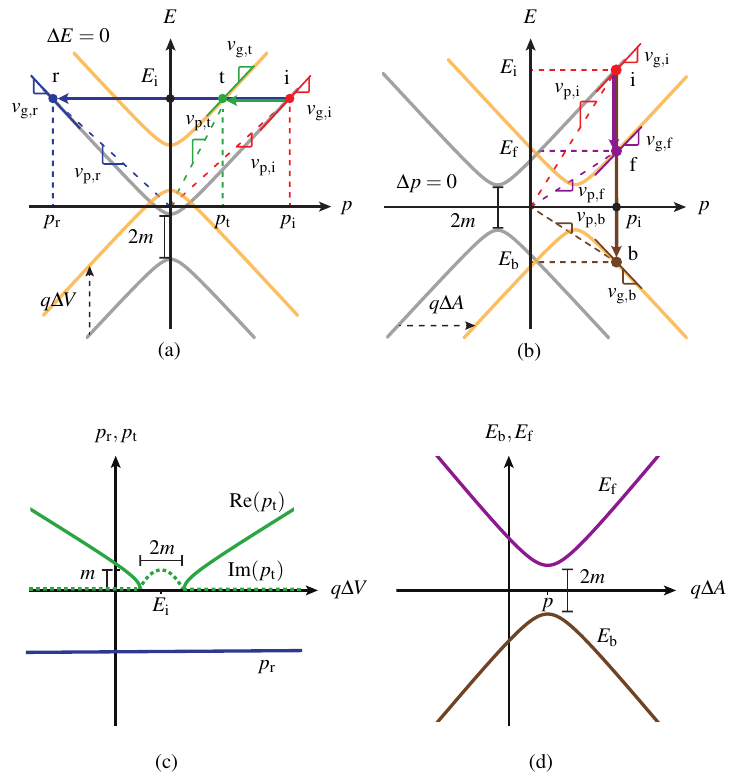}
    \caption{Dispersion relations and electronic transitions corresponding to Fig.~\ref{fig:space_time_discon} for (a)~the spatial step $V(z)$ [Fig.~\ref{fig:space_time_discon}(a)], with (horizontal) momentum transitions, and (b)~the temporal step $A(t)$ [Fig.~\ref{fig:space_time_discon}(b)], with (vertical) energy transitions, as well as corresponding (c)~momenta and (d)~energies versus potential steps.}\label{fig:dispersions}
\end{figure}

\pati{Dispersion and Transition Diagrams}
Figure~\ref{fig:dispersions} plots the dispersion relations and electronic transitions for the two problems in Fig.~\ref{fig:space_time_discon}, with Figs.~\ref{fig:dispersions}(a) and~\ref{fig:dispersions}(b) corresponding to the (reference) spatial step and temporal step problems in Figs.~\ref{fig:space_time_discon}(a) and~(b), respectively, and with indications of the phase and group velocities~(see Supplementary Sec.~5), corresponding to those in Fig.~\ref{fig:space_time_discon}.

The spatial step [Fig.~\ref{fig:dispersions}(a)] features the well-known vertical dispersion shifting with horizontal (momentum) transitions from the incident to the reflected and transmitted states (see Supplementary Sec.~3.2.2). The temporal step [Figs.~\ref{fig:dispersions}(b)] [Eqs.~\eqref{eq:disp_rel_temporal} and~\eqref{eq:trans_temporal}] exhibits \emph{perfectly dual} characteristics, with horizontal dispersion shifting and vertical (energy) transitions from the earlier to the later-backward and later-forward states, whose energy levels versus $E_\mathrm{i}$ are obtained by solving Eq.~\eqref{eq:Ei_pA1} for $p$ and inserting the result into Eqs.~\eqref{eq:Efb_pA2}, which yields
\begin{equation}\label{eq:EfbvsEi}
    E_\mathrm{f}=-E_\mathrm{b}=\sqrt{\left(\sqrt{E_\mathrm{i}^2-m^2}-q\Delta{A}\right)^2+m^2}.
\end{equation}

Note that the orthogonal dispersion shifting in electronic scattering in Fig.~\ref{fig:dispersions} is a feature that does not exist in the classical electromagnetic counterparts of these problems, which rather involve (refractive index) dispersion curves that are rotated with respect to each other and that do not differ between the space and time cases~\cite{Caloz_2020_spacetimeII} (see Supplementary Sec.~6).

\section*{Scattering Coefficients} \label{sec:ScatteringCoeff}

\pati{Potential Temporal Step Solution}
Upon the basis of the energy relations~\eqref{eq:trans_temporal}, the scattering amplitudes and probabilities may be easily found by inserting the expression for the vector potential step function $A(t)$ into the general solution form~\eqref{eq:plane_wave_sol} and enforcing the continuity condition $\left.\psi_{1}\right|_{t=t_0}=\left.\psi_{2}\right|_{t=t_0}$. The resulting later-backward and later-forward amplitude coefficients are (see Supplementary Sec.~3.2.4)
\begin{subequations}\label{eq:coef_prob_Gamma_temporal}
\begin{equation}\label{eq:coef_temporal}
    b = \frac{\Gamma_\mathrm{t}-1}{2 \Gamma_\mathrm{t}}  
        \quad \text { and } \quad f = \frac{1+\Gamma_\mathrm{t}}{2 \Gamma_\mathrm{t}},
\end{equation}
    \text{corresponding to the probabilities}
\begin{equation}\label{eq:prob_temporal}    
    B = \left|b\right|^2 \frac{2\Gamma_\mathrm{t}^2}{1+\Gamma_\mathrm{t}^2}
    \quad \text { and } \quad
    F = \left|f\right|^2 \frac{2\Gamma_\mathrm{t}^2}{1+\Gamma_\mathrm{t}^2},
\end{equation}
    \text{where}
\begin{equation}\label{Gamma_temporal}
    \Gamma_\mathrm{t}  = \frac{\sqrt{\left(\sqrt{E_\mathrm{i}^2-m^2}-(qA_2-qA_1)\right)^2+m^2}}{\left(\sqrt{E_\mathrm{i}^2-m^2}-(qA_2-qA_1)\right)\left(\dfrac{E_\mathrm{i}-m}{\sqrt{E_\mathrm{i}^2-m^2}}\right)+m}.
\end{equation}
\end{subequations}
Interestingly, the amplitude coefficients in Eqs.~\eqref{eq:coef_temporal} are formally identical to those for classical electromagnetic scattering at a refractive index temporal step discontinuity~\cite{Morgenthaler_1958_puretime,Caloz_2020_spacetimeII}, with the parameter $\Gamma_\mathrm{t}$ in Eq.~\eqref{Gamma_temporal} replacing the refractive index contrast $N=n_2/n_1$ (see Supplementary Sec.~6).

\begin{figure}[ht!]
    \centering
    \includegraphics[width=1\textwidth]{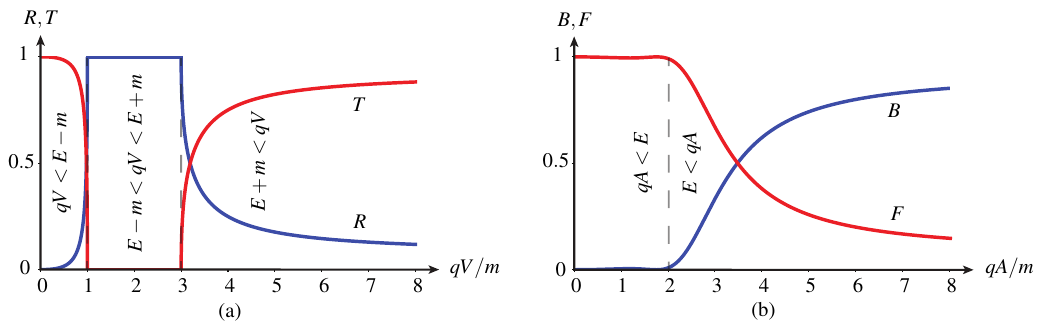}
    \caption{Electron scattering probabilities versus potential strength corresponding to Fig.~\ref{fig:space_time_discon} for (a)~reflection and transmission at the spatial step $V(z)$ [Fig.~\ref{fig:space_time_discon}(a)] with $(V_1,V_2)=(0,V)$ and (b)~later-backward and later-forward propagation at the temporal step $A(t)$ [Fig.~\ref{fig:space_time_discon}(b)] with $(A_1,A_2)=(0,A)$, for the (incident) energy to rest mass ratio $E/m=2$.}\label{fig:probs_Dirac}
\end{figure}

\pati{Probability Graphs}
Figure~\ref{fig:probs_Dirac} plots the electron scattering probabilities versus potential strength for the two problems in Fig.~\ref{fig:space_time_discon}, with Figs.~\ref{fig:probs_Dirac}(a) and~\ref{fig:probs_Dirac}(b) corresponding to the (reference) spatial step and temporal step problems in Figs.~\ref{fig:space_time_discon}(a) and~(b), respectively. The probabilities for the spatial step [Fig.~\ref{fig:probs_Dirac}(a)], also computed here from the Dirac equation (see Supplementary Sec.~3.2.2), show the well-known Klein paradox~\cite{Klein_1929_paper,Greiner_2000_rel_quant}, corresponding to the transmission gap in the range $qV=[E-m,E+m]$ and increasing transmission with increasing potential beyond the gap. In contrast, the probabilities for the temporal step [Fig.~\ref{fig:probs_Dirac}(b)] do not exhibit such a gap; they follow a monotonic trend of exchange from forward propagation at low potentials to backward propagation at high potentials.~These observations interestingly suggest that a shifted Klein gap may be expected in the case of a space-time (traveling) step. The asymptotic response at high potentials ($qV/m,qA/m\gtrsim{5}$) is another fundamental difference: while the temporal step is mostly ``reflective'' (backward-wave) there, the spatial discontinuity is mostly transmissive, as a result of the double reflection-transmission crossing due to the Klein effect. Otherwise, the temporal step supports quasi-total forward transmission up to energies ($qA/m\approx{2}$) more than twice the cutoff of the quasi-total transmission in the spatial case ($qV/m<1$) and a forward-backward crossing point ($qA/m\approx{3.4}$) almost identical to the transmission-reflection crossing point in the spatial case ($qV/m\approx{3.2}$); these two observations correspond to trends that are generally valid when the (incident) energy is sufficient to produce a transition to the backward state, as understandable from the dispersion diagram in Fig.~\ref{fig:dispersions}(b).

\section*{Smooth Temporal Potential Step}\label{sec:smooth_pot}
\begin{figure}[ht!]
    \centering
    \includegraphics[width=1\textwidth]{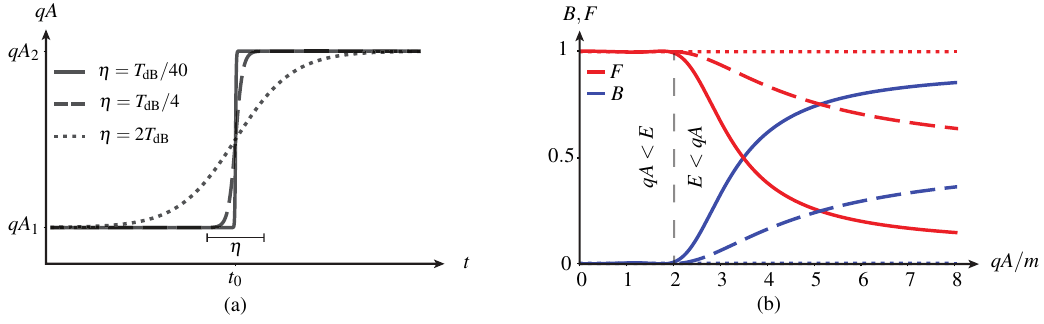}
    \caption{Alternative problem of a smooth temporal potential step for three different transition durations in terms of the de Broglie period, $T_\mathrm{dB}$,  for an electron with energy to rest mass ratio of $E/m=2$, as in Fig.~\ref{fig:probs_Dirac}. (a)~Hyperbolic tangent function [Eq.~\eqref{eq:smooth_trans}] of the related transition, between the potentials $A_1$ and $A_2$, over the time $\eta$. (b)~Corresponding later-backward and later-forward probabilities, $B$ and $F$.}\label{fig:smootht_pot_prob}
\end{figure}

\pati{Unphysicality of the Sharp Potential Step}
The infinitely sharp temporal potential step discussed so far is a canonical structure, but it is not practically realizable, given its instantaneous transition from $A_1$ to $A_2$ and corresponding electric field singularity [$\vb{E}=-\partial\vb{A}/\partial{t}=-\delta(t-t_0)$]. Note that the same unrealizability issue occurs in the infinitely sharp spatial (scalar) potential discontinuity, given the dual contiguous transition from $V_1$ to $V_2$ and corresponding electric field singularity [$\vb{E}=-\nabla{V}=-\delta(z-z_0)$]. What really matters then is to determine whether the interesting physics predicted for the infinitely sharp (unphysical) discontinuity survives as its transition is replaced by a smooth one.\\
\pati{Demonstration of the Smooth Structure}
For this purpose, we choose a smooth-transition potential corresponding to the hyperbolic tangent function
\begin{equation}\label{eq:smooth_trans}
    \vb{A}(t) = \left[A_1 + \frac{A_2-A_1}{2} \left(1 + \tanh{\frac{t-t_0}{\eta}}\right)\right]\vb{\hat{z}}, 
\end{equation}
where the $\eta$ parameter is proportional to the transition time and whose exact Dirac solution is derived in Supplementary Sec.~7. The corresponding results are provided in Fig.~\ref{fig:smootht_pot_prob}, with Fig.~\ref{fig:smootht_pot_prob}(a) plotting the hyperbolic-tangent potential and  Fig.~\ref{fig:smootht_pot_prob}(b) plotting the scattering probabilities for three representative transition times in terms of the ``de Broglie period'' of the electron, $T_\mathrm{dB}$. In the sharpest case, $\eta=T_\mathrm{dB}/40$, the scattering probabilities are indistinguishable from those for the infinitely sharp discontinuity in Fig.~\ref{fig:probs_Dirac}(b), because the transition is \emph{deeply sub-period}, the temporal dual regime of deep sub-wavelength. At $\eta=T_\mathrm{dB}/4$, which may be considered as the temporal dual of the (spatial) \emph{Fabry-P\'{e}rot condition}, the back-scattering level is less than half of that in the former case. Finally, in the smoothest case, $\eta=2T_\mathrm{dB}$, the transition has become so slow with respect to the period, that the electron does not ``see'' it anymore, which results in zero back-scattering. 

The deeply sub-period regime ($\eta<T_\mathrm{dB}/10$) can unfortunately not readily be attained in current technologies, because of the extremely short de Broglie period, $T_\mathrm{dB}=h/E=h/(2mc^2)\sim{4}\times{10}^{-21}$~s, but it might be reachable soon, given recent spectacular progress in attosecond lasers. Moreover, the present investigation can be easily extended to Dirac-type materials, such as graphene, whose de Broglie period is much smaller (e.g., $T_\mathrm{dB}^\mathrm{graphene}=(3a/2)/v_\textrm{F}\approx{3.7}\times{10}^{-16}$~s, where $a=0.246$~nm is the lattice constant and $v_\textrm{F}\approx{10}^{6}$~m/s is the Fermi velocity).

\section*{Discussion} \label{DiscussionConclusion}
\pati{Relativisticality of The Backward Wave}
\subsection*{Relativistic Nature of Back-Scattering} 

The Dirac solutions in Eqs.~\eqref{eq:coef_prob_Gamma_temporal} and Fig.~\ref{fig:probs_Dirac}(b) confirm the validity of the later-forward wave and later-backward wave ansatz for the scattering potential $\vb{A}=A(t)\vb{\hat{z}}$. In the non-relativistic regime, where $\Gamma_\textrm{t}=1$ (see Supplementary Sec.~8), these solutions reduce to $b=0$ and $f=1$, actually corresponding to the purely later-forward solution of the Schr\"{o}dinger equation. This fact reveals that later-backward scattering is a \emph{relativistic effect}. Note that the temporal step problem, which may be seen as the infinite-velocity limit of a superluminal space-time modulation medium~\cite{Caloz_2020_spacetimeII}, does not seem to be relativistic per se. Indeed, the corresponding Lorentz factor is $\gamma=1/\sqrt{1-(v_\mathrm{f}/c)^2}$ with $v_\mathrm{f}=c^2/v_\mathrm{m}$, where $v_\mathrm{f}$ is the velocity of the (instantaneous) frame and $v_\mathrm{m}$ is the velocity of the modulation~\cite{Zoe_2018_superlu}, so that $\gamma\stackrel{v_\mathrm{m}\rightarrow\infty}=1$ (no boost); it is really the speed of the electron ($v$) (not that of the modulation ($v_\mathrm{m}$) in which it propagates) that may be relativistic in our problem. The conditional (relativistic) nature of the later-backward wave may a priori seem contradictory, given that the electromagnetic-counterpart problem unconditionally supports back-scattering~\cite{Morgenthaler_1958_puretime,Caloz_2020_spacetimeII}. However, considering that the particle (photon) in the latter case is inherently relativistic ($v_\mathrm{photon}=c$), whereas it is not necessarily in the former case ($v_\mathrm{electron}<c$), makes the finding a posteriori much less surprising.
\pati{Experiment}
\subsection*{Experimental Perspectives}
\subsubsection*{Potential Generation}
It is well-known that producing a magnetic vector potential ($\vb{A}$) to our liking might be a difficult task. This is at least the case when $\vb{A}$ is produced from a $\vb{B}$-field source, according to the relation $\vb{B}=\nabla\times\vb{A}$, as in the experiment originally proposed by Aharonov and Bohm~\cite{Aharonov_1959_AB_effect} and later realized by Tonomura et al.~\cite{tonomura1982observation,tonomura1986evidence}, where $\vb{A}$ is a distant effect of an enclosed $\vb{B}$ field, theoretically requiring an infinite solenoid and producing an inconvenient curved potential~\cite{Aharonov_1959_AB_effect} or requiring a toroid with cumbersome superconductor shielding~\cite{tonomura1982observation,tonomura1986evidence}. Fortunately, our interest here is not to produce a potential in a field-free region, as in the Aharonov-Bohm effect, but just a temporal step potential with a short transition, without any further specific restriction. This might be realized from an $\vb{E}$-field source, according to the relation $\vb{E}=-\partial\vb{A}/\partial{t}$. Indeed, inserting Eq.~\eqref{eq:smooth_trans} into this relation leads to the pulse function
\begin{equation}\label{eq:Ezpulse}
    \vb{E}(t) = -\frac{(A_2-A_1)}{2\eta}\sech^2\left(\frac{t-t_0}{\eta}\right)\vb{\hat{z}},
\end{equation}
which may be produced by an ultrashort-pulse laser~\cite{krausz2009attosecond} to provide the desired potential step function [Eq.~\eqref{eq:smooth_trans}] -- collocated and aligned with the electric field. Achieving maximal back-scattering as in Fig.~\ref{fig:probs_Dirac}(b) requires, as shown in the previous section, a sub-period ($\eta<T_\mathrm{dB}/10$) pulse, but a broader (larger $\eta$) pulse, simply generating a smoother step, may still produce some back-scattering, as shown in Fig.~\ref{fig:smootht_pot_prob}.
\subsubsection*{Measurement}
Once the potential has been generated, as just described, electrons should be shot by an electron gun parallel to the electric field (perpendicular to the laser beam axis) and an appropriate detection procedure should be used to measure the scattering probabilities predicted by Eqs.~\eqref{eq:prob_temporal} [Fig.~\ref{fig:probs_Dirac}(b)]. That detection procedure might be delicate because the physical interpretation of Dirac spinors [Eq.~\eqref{eq:plane_wave_sol}] and related quantities is not trivial: while the operators (position, momentum, energy, spin, etc.) associated with the Schr\"{o}dinger equation directly correspond to observables, those associated with the Dirac equation do not! However, the \emph{scattering probabilities} to measure here are not problematic. Their magnitudes can be obtained by placing an electron counter in the forward region of the setup, measuring the related electron count, $N_\mathrm{F}$, and deducing the corresponding electron count in the backward direction as $N_\mathrm{B}=N_\mathrm{gun}-N_\mathrm{F}$, to obtain $F_\mathrm{meas}=N_\mathrm{F}/N_\mathrm{gun}$ and $B_\mathrm{meas}=N_\mathrm{B}/N_\mathrm{gun}$. Moreover, their energy levels, predicted by Eqs.~\eqref{eq:EfbvsEi} [Fig.~\ref{fig:dispersions}(d)], can be obtained via the phase differences $\Delta\phi_\mathrm{f,b}=-\left(E_\mathrm{i}-E_\mathrm{f,b}\right)t/\hbar$ measured by an electron interferometer.
\pati{Summary and Outlook}
\subsection*{Summary and Outlook}

In this report, we have resolved the fundamental problem of electron scattering at a potential temporal step discontinuity, with a systematic comparison to the spatial counterpart of the problem and mention of similarities and differences with the classical electromagnetic counterparts of the two problems.
The related effects described in this report might lead to a wide range of new concepts and applications in semiconductor electronics, quantum computing and information processing, and attosecond physics. A simple application would be a versatile (aligned-outputs) beam splitter, with tunable splitting ratio and splitting angle, consisting of a rotatable laser with varying intensity, where the latter controls the splitting ratio, according to Fig.~\ref{fig:probs_Dirac}(b), and the former controls the output splitting direction.

\section*{Data availability}

All data generated or analyzed during this study are included in this published article and its supplementary information files.

\newpage
\bibliography{Sci_Rep_e_scat_temporal}

\providecommand{\noopsort}[1]{}\providecommand{\singleletter}[1]{#1}%
\begin{thebibliography}{10}
\urlstyle{rm}
\expandafter\ifx\csname url\endcsname\relax
  \def\url#1{\texttt{#1}}\fi
\expandafter\ifx\csname urlprefix\endcsname\relax\def\urlprefix{URL }\fi
\expandafter\ifx\csname doiprefix\endcsname\relax\def\doiprefix{DOI: }\fi
\providecommand{\bibinfo}[2]{#2}
\providecommand{\eprint}[2][]{\url{#2}}

\bibitem{Griffiths_2018_Quantum}
\bibinfo{author}{Griffiths, D.~J.} \& \bibinfo{author}{Schroeter, D.~F.}
\newblock \emph{\bibinfo{title}{Introduction to Quantum Mechanics}} (\bibinfo{publisher}{Cambridge University Press}, \bibinfo{year}{2018}), \bibinfo{edition}{3} edn.

\bibitem{zettili2009quantum}
\bibinfo{author}{Zettili, N.}
\newblock \emph{\bibinfo{title}{Quantum {M}echanics: {C}oncepts and {A}pplications}} (\bibinfo{publisher}{Wiley}, \bibinfo{year}{2009}).

\bibitem{shankar2012principles}
\bibinfo{author}{Shankar, R.}
\newblock \emph{\bibinfo{title}{Principles of {Q}uantum {M}echanics}} (\bibinfo{publisher}{Springer US}, \bibinfo{year}{2012}).

\bibitem{miller2008quantum}
\bibinfo{author}{Miller, D.~A.}
\newblock \emph{\bibinfo{title}{Quantum {M}echanics for {S}cientists and {E}ngineers}} (\bibinfo{publisher}{Cambridge University Press}, \bibinfo{year}{2008}).

\bibitem{landau_2013_quantum}
\bibinfo{author}{Landau, L.} \& \bibinfo{author}{Lifshitz, E.}
\newblock \emph{\bibinfo{title}{Quantum {M}echanics: {A} {S}horter {C}ourse of {T}heoretical {P}hysics}} (\bibinfo{publisher}{Elsevier Science}, \bibinfo{year}{2013}).

\bibitem{sakurai_napolitano_2020}
\bibinfo{author}{Sakurai, J.~J.} \& \bibinfo{author}{Napolitano, J.}
\newblock \emph{\bibinfo{title}{Modern Quantum Mechanics}} (\bibinfo{publisher}{Cambridge University Press}, \bibinfo{year}{2020}), \bibinfo{edition}{3} edn.

\bibitem{Sch_1926}
\bibinfo{author}{Schr\"odinger, E.}
\newblock \bibinfo{journal}{\bibinfo{title}{An undulatory theory of the mechanics of atoms and molecules}}.
\newblock {\emph{\JournalTitle{Phys. Rev.}}} \textbf{\bibinfo{volume}{28}}, \bibinfo{pages}{1049--1070}, \doiprefix\url{10.1103/PhysRev.28.1049} (\bibinfo{year}{1926}).

\bibitem{klein1926quantentheorie}
\bibinfo{author}{Klein, O.}
\newblock \bibinfo{journal}{\bibinfo{title}{Quantentheorie und f{\"u}nfdimensionale {R}elativit{\"a}tstheorie}}.
\newblock {\emph{\JournalTitle{Z. Phys.}}} \textbf{\bibinfo{volume}{37}}, \bibinfo{pages}{895--906} (\bibinfo{year}{1926}).

\bibitem{gordon1926comptoneffekt}
\bibinfo{author}{Gordon, W.}
\newblock \bibinfo{journal}{\bibinfo{title}{Der {C}omptoneffekt nach der {S}chr{\"o}dingerschen {T}heorie}}.
\newblock {\emph{\JournalTitle{Z. Phys.}}} \textbf{\bibinfo{volume}{40}}, \bibinfo{pages}{117--133} (\bibinfo{year}{1926}).

\bibitem{Dirac_1928_paper}
\bibinfo{author}{Dirac, P. A.~M.}
\newblock \bibinfo{journal}{\bibinfo{title}{The quantum theory of the electron}}.
\newblock {\emph{\JournalTitle{Proc. R. Soc. Lond. A}}} \textbf{\bibinfo{volume}{117}}, \bibinfo{pages}{610--624} (\bibinfo{year}{1928}).

\bibitem{Mendonca_2000_quan_time_refrac}
\bibinfo{author}{Mendon\ifmmode~\mbox{\c{c}}\else \c{c}\fi{}a, J.~T.}, \bibinfo{author}{Guerreiro, A.} \& \bibinfo{author}{Martins, A.~M.}
\newblock \bibinfo{journal}{\bibinfo{title}{Quantum theory of time refraction}}.
\newblock {\emph{\JournalTitle{Phys. Rev. A}}} \textbf{\bibinfo{volume}{62}}, \bibinfo{pages}{033805}, \doiprefix\url{10.1103/PhysRevA.62.033805} (\bibinfo{year}{2000}).

\bibitem{Mendonca_2003_temporalsplitter}
\bibinfo{author}{Mendon\ifmmode~\mbox{\c{c}}\else \c{c}\fi{}a, J.~T.}, \bibinfo{author}{Martins, A.~M.} \& \bibinfo{author}{Guerreiro, A.}
\newblock \bibinfo{journal}{\bibinfo{title}{Temporal beam splitter and temporal interference}}.
\newblock {\emph{\JournalTitle{Phys. Rev. A}}} \textbf{\bibinfo{volume}{68}}, \bibinfo{pages}{043801}, \doiprefix\url{10.1103/PhysRevA.68.043801} (\bibinfo{year}{2003}).

\bibitem{Goldman_2014_periodically}
\bibinfo{author}{Goldman, N.} \& \bibinfo{author}{Dalibard, J.}
\newblock \bibinfo{journal}{\bibinfo{title}{Periodically driven quantum systems: Effective {H}amiltonians and engineered gauge fields}}.
\newblock {\emph{\JournalTitle{Phys. Rev. X}}} \textbf{\bibinfo{volume}{4}}, \bibinfo{pages}{031027}, \doiprefix\url{10.1103/PhysRevX.4.031027} (\bibinfo{year}{2014}).

\bibitem{Reck_2017_dirac_time_mirror}
\bibinfo{author}{Reck, P.} \emph{et~al.}
\newblock \bibinfo{journal}{\bibinfo{title}{{D}irac quantum time mirror}}.
\newblock {\emph{\JournalTitle{Phys. Rev. B}}} \textbf{\bibinfo{volume}{95}}, \bibinfo{pages}{165421}, \doiprefix\url{10.1103/PhysRevB.95.165421} (\bibinfo{year}{2017}).

\bibitem{xu_2018_spacetime}
\bibinfo{author}{Xu, S.} \& \bibinfo{author}{Wu, C.}
\newblock \bibinfo{journal}{\bibinfo{title}{Space-time crystal and space-time group}}.
\newblock {\emph{\JournalTitle{Phys. Rev. Lett.}}} \textbf{\bibinfo{volume}{120}}, \bibinfo{pages}{096401}, \doiprefix\url{10.1103/PhysRevLett.120.096401} (\bibinfo{year}{2018}).

\bibitem{Junk_2020_floquet_oscillations_dirac}
\bibinfo{author}{Junk, V.}, \bibinfo{author}{Reck, P.}, \bibinfo{author}{Gorini, C.} \& \bibinfo{author}{Richter, K.}
\newblock \bibinfo{journal}{\bibinfo{title}{Floquet oscillations in periodically driven {D}irac systems}}.
\newblock {\emph{\JournalTitle{Phys. Rev. B}}} \textbf{\bibinfo{volume}{101}}, \bibinfo{pages}{134302}, \doiprefix\url{10.1103/PhysRevB.101.134302} (\bibinfo{year}{2020}).

\bibitem{Gao_2021_floquet_zener}
\bibinfo{author}{Gao, Q.} \& \bibinfo{author}{Niu, Q.}
\newblock \bibinfo{journal}{\bibinfo{title}{Floquet-{B}loch oscillations and intraband {Z}ener tunneling in an oblique spacetime crystal}}.
\newblock {\emph{\JournalTitle{Phys. Rev. Lett.}}} \textbf{\bibinfo{volume}{127}}, \bibinfo{pages}{036401}, \doiprefix\url{10.1103/PhysRevLett.127.036401} (\bibinfo{year}{2021}).

\bibitem{schultheiss_2021_time}
\bibinfo{author}{Schultheiss, K.} \emph{et~al.}
\newblock \bibinfo{journal}{\bibinfo{title}{Time refraction of spin waves}}.
\newblock {\emph{\JournalTitle{Phys. Rev. Lett.}}} \textbf{\bibinfo{volume}{126}}, \bibinfo{pages}{137201}, \doiprefix\url{10.1103/PhysRevLett.126.137201} (\bibinfo{year}{2021}).

\bibitem{peng_2022_topological}
\bibinfo{author}{Peng, Y.}
\newblock \bibinfo{journal}{\bibinfo{title}{Topological space-time crystal}}.
\newblock {\emph{\JournalTitle{Phys. Rev. Lett.}}} \textbf{\bibinfo{volume}{128}}, \bibinfo{pages}{186802}, \doiprefix\url{10.1103/PhysRevLett.128.186802} (\bibinfo{year}{2022}).

\bibitem{Gao_2022_semiclassical}
\bibinfo{author}{Gao, Q.} \& \bibinfo{author}{Niu, Q.}
\newblock \bibinfo{journal}{\bibinfo{title}{Semiclassical dynamics of electrons in a space-time crystal: Magnetization, polarization, and current response}}.
\newblock {\emph{\JournalTitle{Phys. Rev. B}}} \textbf{\bibinfo{volume}{106}}, \bibinfo{pages}{224311}, \doiprefix\url{10.1103/PhysRevB.106.224311} (\bibinfo{year}{2022}).

\bibitem{vazquez_2022_shaping}
\bibinfo{author}{Vázquez-Lozano, J.~E.} \& \bibinfo{author}{Liberal, I.}
\newblock \bibinfo{journal}{\bibinfo{title}{Shaping the quantum vacuum with anisotropic temporal boundaries}}.
\newblock {\emph{\JournalTitle{Nanophotonics}}} \textbf{\bibinfo{volume}{12}}, \bibinfo{pages}{539--548}, \doiprefix\url{doi:10.1515/nanoph-2022-0491} (\bibinfo{year}{2023}).

\bibitem{Lu_2022_floquet_dirac_bands}
\bibinfo{author}{Lu, M.}, \bibinfo{author}{Reid, G.~H.}, \bibinfo{author}{Fritsch, A.~R.}, \bibinfo{author}{Pi\~neiro, A.~M.} \& \bibinfo{author}{Spielman, I.~B.}
\newblock \bibinfo{journal}{\bibinfo{title}{Floquet engineering topological {D}irac bands}}.
\newblock {\emph{\JournalTitle{Phys. Rev. Lett.}}} \textbf{\bibinfo{volume}{129}}, \bibinfo{pages}{040402}, \doiprefix\url{10.1103/PhysRevLett.129.040402} (\bibinfo{year}{2022}).

\bibitem{Kim_2023_temporal_Dirac}
\bibinfo{author}{Kim, S.} \& \bibinfo{author}{Kim, K.}
\newblock \bibinfo{journal}{\bibinfo{title}{Propagation of {D}irac waves through various temporal interfaces, slabs, and crystals}}.
\newblock {\emph{\JournalTitle{Phys. Rev. Res.}}} \textbf{\bibinfo{volume}{5}}, \bibinfo{pages}{023162}, \doiprefix\url{10.1103/PhysRevResearch.5.023162} (\bibinfo{year}{2023}).

\bibitem{Morgenthaler_1958_puretime}
\bibinfo{author}{{Morgenthaler}, F.~R.}
\newblock \bibinfo{journal}{\bibinfo{title}{{Velocity Modulation of Electromagnetic Waves}}}.
\newblock {\emph{\JournalTitle{IEEE Trans. Microw. Theory Tech.}}} \textbf{\bibinfo{volume}{6}}, \bibinfo{pages}{167--172}, \doiprefix\url{10.1109/TMTT.1958.1124533} (\bibinfo{year}{1958}).

\bibitem{Plansinis_2015_TemporalAnalog_Refl_Refr}
\bibinfo{author}{Plansinis, B.~W.}, \bibinfo{author}{Donaldson, W.~R.} \& \bibinfo{author}{Agrawal, G.~P.}
\newblock \bibinfo{journal}{\bibinfo{title}{What is the temporal analog of reflection and refraction of optical beams?}}
\newblock {\emph{\JournalTitle{Phys. Rev. Lett.}}} \textbf{\bibinfo{volume}{115}}, \bibinfo{pages}{183901}, \doiprefix\url{10.1103/PhysRevLett.115.183901} (\bibinfo{year}{2015}).

\bibitem{Mazor_2021_unitary}
\bibinfo{author}{Mazor, Y.}, \bibinfo{author}{Cotrufo, M.} \& \bibinfo{author}{Al\`u, A.}
\newblock \bibinfo{journal}{\bibinfo{title}{Unitary excitation transfer between coupled cavities using temporal switching}}.
\newblock {\emph{\JournalTitle{Phys. Rev. Lett.}}} \textbf{\bibinfo{volume}{127}}, \bibinfo{pages}{013902}, \doiprefix\url{10.1103/PhysRevLett.127.013902} (\bibinfo{year}{2021}).

\bibitem{Tretyakov_2023_Temporal_discon}
\bibinfo{author}{Wang, X.}, \bibinfo{author}{Mirmoosa, M.~S.} \& \bibinfo{author}{Tretyakov, S.~A.}
\newblock \bibinfo{journal}{\bibinfo{title}{Controlling surface waves with temporal discontinuities of metasurfaces}}.
\newblock {\emph{\JournalTitle{Nanophotonics}}} \textbf{\bibinfo{volume}{12}}, \bibinfo{pages}{2813--2822}, \doiprefix\url{doi:10.1515/nanoph-2022-0685} (\bibinfo{year}{2023}).

\bibitem{Chamanara_2017_optical}
\bibinfo{author}{Chamanara, N.}, \bibinfo{author}{Taravati, S.}, \bibinfo{author}{Deck-L\'eger, Z.-L.} \& \bibinfo{author}{Caloz, C.}
\newblock \bibinfo{journal}{\bibinfo{title}{Optical isolation based on space-time engineered asymmetric photonic band gaps}}.
\newblock {\emph{\JournalTitle{Phys. Rev. B}}} \textbf{\bibinfo{volume}{96}}, \bibinfo{pages}{155409}, \doiprefix\url{10.1103/PhysRevB.96.155409} (\bibinfo{year}{2017}).

\bibitem{Engheta_2021_Metamat}
\bibinfo{author}{Engheta, N.}
\newblock \bibinfo{journal}{\bibinfo{title}{Metamaterials with high degrees of freedom: space, time, and more}}.
\newblock {\emph{\JournalTitle{Nanophotonics}}} \textbf{\bibinfo{volume}{10}}, \bibinfo{pages}{639--642}, \doiprefix\url{doi:10.1515/nanoph-2020-0414} (\bibinfo{year}{2021}).

\bibitem{Huidobro_2021_homogenization}
\bibinfo{author}{Huidobro, P.~A.}, \bibinfo{author}{Silveirinha, M.~G.}, \bibinfo{author}{Galiffi, E.} \& \bibinfo{author}{Pendry, J.~B.}
\newblock \bibinfo{journal}{\bibinfo{title}{Homogenization theory of space-time metamaterials}}.
\newblock {\emph{\JournalTitle{Phys. Rev. Appl.}}} \textbf{\bibinfo{volume}{16}}, \bibinfo{pages}{014044}, \doiprefix\url{10.1103/PhysRevApplied.16.014044} (\bibinfo{year}{2021}).

\bibitem{Li_2023_gener_total_ref}
\bibinfo{author}{Li, Z.}, \bibinfo{author}{Ma, X.}, \bibinfo{author}{Bahrami, A.}, \bibinfo{author}{Deck-L\'eger, Z.-L.} \& \bibinfo{author}{Caloz, C.}
\newblock \bibinfo{journal}{\bibinfo{title}{Generalized total internal reflection at dynamic interfaces}}.
\newblock {\emph{\JournalTitle{Phys. Rev. B}}} \textbf{\bibinfo{volume}{107}}, \bibinfo{pages}{115129}, \doiprefix\url{10.1103/PhysRevB.107.115129} (\bibinfo{year}{2023}).

\bibitem{Bahrami_2023_Accelerated}
\bibinfo{author}{Bahrami, A.}, \bibinfo{author}{Deck-L\'eger, Z.-L.} \& \bibinfo{author}{Caloz, C.}
\newblock \bibinfo{journal}{\bibinfo{title}{Electrodynamics of accelerated-modulation space-time metamaterials}}.
\newblock {\emph{\JournalTitle{Phys. Rev. Appl.}}} \textbf{\bibinfo{volume}{19}}, \bibinfo{pages}{054044}, \doiprefix\url{10.1103/PhysRevApplied.19.054044} (\bibinfo{year}{2023}).

\bibitem{Caloz_2020_spacetimeI}
\bibinfo{author}{Caloz, C.} \& \bibinfo{author}{Deck-L\'{e}ger, Z.-L.}
\newblock \bibinfo{journal}{\bibinfo{title}{Spacetime metamaterials, part {I}: {G}eneral concepts}}.
\newblock {\emph{\JournalTitle{IEEE Trans. Antennas Propag.}}} \textbf{\bibinfo{volume}{68}}, \bibinfo{pages}{1569--1582}, \doiprefix\url{10.1109/TAP.2019.2944225} (\bibinfo{year}{2020}).

\bibitem{Caloz_2020_spacetimeII}
\bibinfo{author}{Caloz, C.} \& \bibinfo{author}{Deck-L\'{e}ger, Z.-L.}
\newblock \bibinfo{journal}{\bibinfo{title}{Spacetime metamaterials, part {II}: {T}heory and applications}}.
\newblock {\emph{\JournalTitle{IEEE Trans. Antennas Propag.}}} \textbf{\bibinfo{volume}{68}}, \bibinfo{pages}{1583--1598}, \doiprefix\url{10.1109/TAP.2019.2944216} (\bibinfo{year}{2020}).

\bibitem{Caloz_2022_GSTEM}
\bibinfo{author}{Caloz, C.}, \bibinfo{author}{Deck-Léger, Z.-L.}, \bibinfo{author}{Bahrami, A.}, \bibinfo{author}{Vicente, O.~C.} \& \bibinfo{author}{Li, Z.}
\newblock \bibinfo{journal}{\bibinfo{title}{Generalized space-time engineered modulation ({GSTEM}) metamaterials: {A} global and extended perspective.}}
\newblock {\emph{\JournalTitle{IEEE Antennas and Propagation Magazine}}} \bibinfo{pages}{2--12}, \doiprefix\url{10.1109/MAP.2022.3216773} (\bibinfo{year}{2022}).

\bibitem{Akbarzadeh_2018_inverse_prism}
\bibinfo{author}{Akbarzadeh, A.}, \bibinfo{author}{Chamanara, N.} \& \bibinfo{author}{Caloz, C.}
\newblock \bibinfo{journal}{\bibinfo{title}{Inverse prism based on temporal discontinuity and spatial dispersion}}.
\newblock {\emph{\JournalTitle{Opt. Lett.}}} \textbf{\bibinfo{volume}{43}}, \bibinfo{pages}{3297--3300}, \doiprefix\url{10.1364/OL.43.003297} (\bibinfo{year}{2018}).

\bibitem{Shlivinski_2018_Beyond}
\bibinfo{author}{Shlivinski, A.} \& \bibinfo{author}{Hadad, Y.}
\newblock \bibinfo{journal}{\bibinfo{title}{Beyond the {B}ode-{F}ano bound: Wideband impedance matching for short pulses using temporal switching of transmission-line parameters}}.
\newblock {\emph{\JournalTitle{Phys. Rev. Lett.}}} \textbf{\bibinfo{volume}{121}}, \bibinfo{pages}{204301}, \doiprefix\url{10.1103/PhysRevLett.121.204301} (\bibinfo{year}{2018}).

\bibitem{Pacheco_2020_temporalaming}
\bibinfo{author}{Pacheco-Pe{\~n}a, V.} \& \bibinfo{author}{Engheta, N.}
\newblock \bibinfo{journal}{\bibinfo{title}{Temporal aiming}}.
\newblock {\emph{\JournalTitle{Light Sci. Appl.}}} \textbf{\bibinfo{volume}{9}}, \bibinfo{pages}{129}, \doiprefix\url{10.1038/s41377-020-00360-1} (\bibinfo{year}{2020}).

\bibitem{Li_2021_temporal_parity}
\bibinfo{author}{Li, H.}, \bibinfo{author}{Yin, S.}, \bibinfo{author}{Galiffi, E.} \& \bibinfo{author}{Al\`u, A.}
\newblock \bibinfo{journal}{\bibinfo{title}{Temporal parity-time symmetry for extreme energy transformations}}.
\newblock {\emph{\JournalTitle{Phys. Rev. Lett.}}} \textbf{\bibinfo{volume}{127}}, \bibinfo{pages}{153903}, \doiprefix\url{10.1103/PhysRevLett.127.153903} (\bibinfo{year}{2021}).

\bibitem{Pena_2020_tempcoating}
\bibinfo{author}{Pacheco-Pe{\~n}a, V.} \& \bibinfo{author}{Engheta, N.}
\newblock \bibinfo{journal}{\bibinfo{title}{Antireflection temporal coatings}}.
\newblock {\emph{\JournalTitle{Optica}}} \textbf{\bibinfo{volume}{7}}, \bibinfo{pages}{323--331}, \doiprefix\url{10.1364/OPTICA.381175} (\bibinfo{year}{2020}).

\bibitem{Xu_2021_complete_pol}
\bibinfo{author}{Xu, J.}, \bibinfo{author}{Mai, W.} \& \bibinfo{author}{Werner, D.~H.}
\newblock \bibinfo{journal}{\bibinfo{title}{Complete polarization conversion using anisotropic temporal slabs}}.
\newblock {\emph{\JournalTitle{Opt. Lett.}}} \textbf{\bibinfo{volume}{46}}, \bibinfo{pages}{1373--1376}, \doiprefix\url{10.1364/OL.415757} (\bibinfo{year}{2021}).

\bibitem{Rizza_2022_short_pulsed}
\bibinfo{author}{Rizza, C.}, \bibinfo{author}{Castaldi, G.} \& \bibinfo{author}{Galdi, V.}
\newblock \bibinfo{journal}{\bibinfo{title}{Short-pulsed metamaterials}}.
\newblock {\emph{\JournalTitle{Phys. Rev. Lett.}}} \textbf{\bibinfo{volume}{128}}, \bibinfo{pages}{257402}, \doiprefix\url{10.1103/PhysRevLett.128.257402} (\bibinfo{year}{2022}).

\bibitem{Castaldi_2023_multiple_short_pulsed}
\bibinfo{author}{Castaldi, G.}, \bibinfo{author}{Rizza, C.}, \bibinfo{author}{Engheta, N.} \& \bibinfo{author}{Galdi, V.}
\newblock \bibinfo{journal}{\bibinfo{title}{Multiple actions of time-resolved short-pulsed metamaterials}}.
\newblock {\emph{\JournalTitle{Appl. Phys. Lett.}}} \textbf{\bibinfo{volume}{122}}, \bibinfo{pages}{021701}, \doiprefix\url{10.1063/5.0132554} (\bibinfo{year}{2023}).

\bibitem{Mencagli_2022_static}
\bibinfo{author}{Mencagli, M.~J.}, \bibinfo{author}{Sounas, D.~L.}, \bibinfo{author}{Fink, M.} \& \bibinfo{author}{Engheta, N.}
\newblock \bibinfo{journal}{\bibinfo{title}{Static-to-dynamic field conversion with time-varying media}}.
\newblock {\emph{\JournalTitle{Phys. Rev. B}}} \textbf{\bibinfo{volume}{105}}, \bibinfo{pages}{144301}, \doiprefix\url{10.1103/PhysRevB.105.144301} (\bibinfo{year}{2022}).

\bibitem{Li_2022_nonreciprocity}
\bibinfo{author}{Li, H.}, \bibinfo{author}{Yin, S.} \& \bibinfo{author}{Al\`u, A.}
\newblock \bibinfo{journal}{\bibinfo{title}{Nonreciprocity and faraday rotation at time interfaces}}.
\newblock {\emph{\JournalTitle{Phys. Rev. Lett.}}} \textbf{\bibinfo{volume}{128}}, \bibinfo{pages}{173901}, \doiprefix\url{10.1103/PhysRevLett.128.173901} (\bibinfo{year}{2022}).

\bibitem{Li_2023_Faradaycrystal}
\bibinfo{author}{He, H.}, \bibinfo{author}{Zhang, S.}, \bibinfo{author}{Qi, J.}, \bibinfo{author}{Bo, F.} \& \bibinfo{author}{Li, H.}
\newblock \bibinfo{journal}{\bibinfo{title}{Faraday rotation in nonreciprocal photonic time-crystals}}.
\newblock {\emph{\JournalTitle{Appl. Phys. Lett.}}} \textbf{\bibinfo{volume}{122}}, \bibinfo{pages}{051703}, \doiprefix\url{10.1063/5.0131818} (\bibinfo{year}{2023}).

\bibitem{Li_2023_stationary_charge}
\bibinfo{author}{Li, H.} \emph{et~al.}
\newblock \bibinfo{journal}{\bibinfo{title}{Stationary charge radiation in anisotropic photonic time crystals}}.
\newblock {\emph{\JournalTitle{Phys. Rev. Lett.}}} \textbf{\bibinfo{volume}{130}}, \bibinfo{pages}{093803}, \doiprefix\url{10.1103/PhysRevLett.130.093803} (\bibinfo{year}{2023}).

\bibitem{Ptitcyn_2023_timecircuit}
\bibinfo{author}{Ptitcyn, G.}, \bibinfo{author}{Mirmoosa, M.~S.}, \bibinfo{author}{Hrabar, S.} \& \bibinfo{author}{Tretyakov, S.}
\newblock \bibinfo{journal}{\bibinfo{title}{Time-modulated circuits and metasurfaces for emulating arbitrary transfer functions}}.
\newblock {\emph{\JournalTitle{arXiv preprint arXiv:2302.14657}}}  (\bibinfo{year}{2023}).

\bibitem{Silbiger_2023_filter}
\bibinfo{author}{Silbiger, O.} \& \bibinfo{author}{Hadad, Y.}
\newblock \bibinfo{journal}{\bibinfo{title}{Optimization-free filter and matched-filter design through spatial and temporal soft switching of the dielectric constant}}.
\newblock {\emph{\JournalTitle{Phys. Rev. Appl.}}} \textbf{\bibinfo{volume}{19}}, \bibinfo{pages}{014047}, \doiprefix\url{10.1103/PhysRevApplied.19.014047} (\bibinfo{year}{2023}).

\bibitem{Tien_JAP_1958}
\bibinfo{author}{Tien, P.~K.}
\newblock \bibinfo{journal}{\bibinfo{title}{Parametric amplification and frequency mixing in propagating circuits}}.
\newblock {\emph{\JournalTitle{J. Appl. Phys.}}} \textbf{\bibinfo{volume}{29}}, \bibinfo{pages}{1347--1357}, \doiprefix\url{10.1063/1.1723440} (\bibinfo{year}{1958}).

\bibitem{Galiffi_2019_broadband}
\bibinfo{author}{Galiffi, E.}, \bibinfo{author}{Huidobro, P.~A.} \& \bibinfo{author}{Pendry, J.~B.}
\newblock \bibinfo{journal}{\bibinfo{title}{Broadband nonreciprocal amplification in luminal metamaterials}}.
\newblock {\emph{\JournalTitle{Phys. Rev. Lett.}}} \textbf{\bibinfo{volume}{123}}, \bibinfo{pages}{206101}, \doiprefix\url{10.1103/PhysRevLett.123.206101} (\bibinfo{year}{2019}).

\bibitem{Deck_2018_wavedeflection}
\bibinfo{author}{Deck-L{\'e}ger, Z.-L.}, \bibinfo{author}{Akbarzadeh, A.} \& \bibinfo{author}{Caloz, C.}
\newblock \bibinfo{journal}{\bibinfo{title}{Wave deflection and shifted refocusing in a medium modulated by a superluminal rectangular pulse}}.
\newblock {\emph{\JournalTitle{Phys. Rev. B}}} \textbf{\bibinfo{volume}{97}}, \bibinfo{pages}{104305}, \doiprefix\url{10.1103/PhysRevB.97.104305} (\bibinfo{year}{2018}).

\bibitem{Yu_2009_opticalisolation}
\bibinfo{author}{Yu, Z.} \& \bibinfo{author}{Fan, S.}
\newblock \bibinfo{journal}{\bibinfo{title}{Complete optical isolation created by indirect interband photonic transitions}}.
\newblock {\emph{\JournalTitle{Nat. Photonics}}} \textbf{\bibinfo{volume}{3}}, \bibinfo{pages}{91--94}, \doiprefix\url{10.1038/nphoton.2008.273} (\bibinfo{year}{2009}).

\bibitem{Correas_2016_NonreciGraphene}
\bibinfo{author}{Correas-Serrano, D.} \emph{et~al.}
\newblock \bibinfo{journal}{\bibinfo{title}{Nonreciprocal graphene devices and antennas based on spatiotemporal modulation}}.
\newblock {\emph{\JournalTitle{IEEE Antennas Wirel. Propag. Lett.}}} \textbf{\bibinfo{volume}{15}}, \bibinfo{pages}{1529--1532}, \doiprefix\url{10.1109/LAWP.2015.2510818} (\bibinfo{year}{2016}).

\bibitem{chamanara2017optical}
\bibinfo{author}{Chamanara, N.}, \bibinfo{author}{Taravati, S.}, \bibinfo{author}{Deck-L\'eger, Z.-L.} \& \bibinfo{author}{Caloz, C.}
\newblock \bibinfo{journal}{\bibinfo{title}{Optical isolation based on space-time engineered asymmetric photonic band gaps}}.
\newblock {\emph{\JournalTitle{Phys. Rev. B}}} \textbf{\bibinfo{volume}{96}}, \bibinfo{pages}{155409}, \doiprefix\url{10.1103/PhysRevB.96.155409} (\bibinfo{year}{2017}).

\bibitem{Taravati_2017_nonreciprocal}
\bibinfo{author}{Taravati, S.}, \bibinfo{author}{Chamanara, N.} \& \bibinfo{author}{Caloz, C.}
\newblock \bibinfo{journal}{\bibinfo{title}{Nonreciprocal electromagnetic scattering from a periodically space-time modulated slab and application to a quasisonic isolator}}.
\newblock {\emph{\JournalTitle{Phys. Rev. B}}} \textbf{\bibinfo{volume}{96}}, \bibinfo{pages}{165144}, \doiprefix\url{10.1103/PhysRevB.96.165144} (\bibinfo{year}{2017}).

\bibitem{guo2019nonreciprocal}
\bibinfo{author}{Guo, X.}, \bibinfo{author}{Ding, Y.}, \bibinfo{author}{Duan, Y.} \& \bibinfo{author}{Ni, X.}
\newblock \bibinfo{journal}{\bibinfo{title}{Nonreciprocal metasurface with space--time phase modulation}}.
\newblock {\emph{\JournalTitle{Light Sci. Appl.}}} \textbf{\bibinfo{volume}{8}}, \bibinfo{pages}{123}, \doiprefix\url{10.1038/s41377-019-0225-z} (\bibinfo{year}{2019}).

\bibitem{Greiner_2000_rel_quant}
\bibinfo{author}{Greiner, W.}
\newblock \emph{\bibinfo{title}{Relativistic Quantum Mechanics. Wave Equations}} (\bibinfo{publisher}{Springer, Berlin}, \bibinfo{year}{2000}), \bibinfo{edition}{3rd} edn.

\bibitem{Peskin_1995_QFT}
\bibinfo{author}{Peskin, M.} \& \bibinfo{author}{Schroeder, D.}
\newblock \emph{\bibinfo{title}{An {I}ntroduction to {Q}uantum {F}ield {T}heory}} (\bibinfo{publisher}{CRC press}, \bibinfo{year}{2018}).

\bibitem{ponomarenko2022goos}
\bibinfo{author}{Ponomarenko, S.~A.}, \bibinfo{author}{Zhang, J.} \& \bibinfo{author}{Agrawal, G.~P.}
\newblock \bibinfo{journal}{\bibinfo{title}{Goos-h{\"a}nchen shift at a temporal boundary}}.
\newblock {\emph{\JournalTitle{Physical Review A}}} \textbf{\bibinfo{volume}{106}}, \bibinfo{pages}{L061501} (\bibinfo{year}{2022}).

\bibitem{Jackson_2000_electrodyn}
\bibinfo{author}{Jackson, J.~D.}
\newblock \emph{\bibinfo{title}{Classical Electrodynamics}} (\bibinfo{publisher}{Wiley}, \bibinfo{year}{1998}), \bibinfo{edition}{3rd} edn.

\bibitem{Jackson_2001_Historical_Gauge}
\bibinfo{author}{Jackson, J.~D.} \& \bibinfo{author}{Okun, L.~B.}
\newblock \bibinfo{journal}{\bibinfo{title}{Historical roots of gauge invariance}}.
\newblock {\emph{\JournalTitle{Rev. Mod. Phys.}}} \textbf{\bibinfo{volume}{73}}, \bibinfo{pages}{663--680}, \doiprefix\url{10.1103/RevModPhys.73.663} (\bibinfo{year}{2001}).

\bibitem{Aharonov_1959_AB_effect}
\bibinfo{author}{Aharonov, Y.} \& \bibinfo{author}{Bohm, D.}
\newblock \bibinfo{journal}{\bibinfo{title}{Significance of electromagnetic potentials in the quantum theory}}.
\newblock {\emph{\JournalTitle{Phys. Rev.}}} \textbf{\bibinfo{volume}{115}}, \bibinfo{pages}{485--491}, \doiprefix\url{10.1103/PhysRev.115.485} (\bibinfo{year}{1959}).

\bibitem{Noether_1918}
\bibinfo{author}{Noether, E.}
\newblock \bibinfo{journal}{\bibinfo{title}{Invariante {V}ariationsprobleme}}.
\newblock {\emph{\JournalTitle{Nach. Ges. Wiss. G{\"o}tt.}}} \textbf{\bibinfo{volume}{1918}}, \bibinfo{pages}{235--257} (\bibinfo{year}{1918}).

\bibitem{Klein_1929_paper}
\bibinfo{author}{Klein, O.}
\newblock \bibinfo{journal}{\bibinfo{title}{Die {R}eflexion von {E}lektronen an einem {P}otentialsprung nach der relativistischen {D}ynamik von {D}irac}}.
\newblock {\emph{\JournalTitle{Z. Phys.}}} \textbf{\bibinfo{volume}{53}}, \bibinfo{pages}{157–165} (\bibinfo{year}{1929}).

\bibitem{Zoe_2018_superlu}
\bibinfo{author}{Deck-L\'eger, Z.-L.}, \bibinfo{author}{Akbarzadeh, A.} \& \bibinfo{author}{Caloz, C.}
\newblock \bibinfo{journal}{\bibinfo{title}{Wave deflection and shifted refocusing in a medium modulated by a superluminal rectangular pulse}}.
\newblock {\emph{\JournalTitle{Phys. Rev. B}}} \textbf{\bibinfo{volume}{97}}, \bibinfo{pages}{104305}, \doiprefix\url{10.1103/PhysRevB.97.104305} (\bibinfo{year}{2018}).

\bibitem{tonomura1982observation}
\bibinfo{author}{Tonomura, A.} \emph{et~al.}
\newblock \bibinfo{journal}{\bibinfo{title}{Observation of aharonov-bohm effect by electron holography}}.
\newblock {\emph{\JournalTitle{Phys. Rev. Lett.}}} \textbf{\bibinfo{volume}{48}}, \bibinfo{pages}{1443--1446}, \doiprefix\url{10.1103/PhysRevLett.48.1443} (\bibinfo{year}{1982}).

\bibitem{tonomura1986evidence}
\bibinfo{author}{Tonomura, A.} \emph{et~al.}
\newblock \bibinfo{journal}{\bibinfo{title}{Evidence for aharonov-bohm effect with magnetic field completely shielded from electron wave}}.
\newblock {\emph{\JournalTitle{Phys. Rev. Lett.}}} \textbf{\bibinfo{volume}{56}}, \bibinfo{pages}{792--795}, \doiprefix\url{10.1103/PhysRevLett.56.792} (\bibinfo{year}{1986}).

\bibitem{krausz2009attosecond}
\bibinfo{author}{Krausz, F.} \& \bibinfo{author}{Ivanov, M.}
\newblock \bibinfo{journal}{\bibinfo{title}{Attosecond physics}}.
\newblock {\emph{\JournalTitle{Rev. Mod. Phys.}}} \textbf{\bibinfo{volume}{81}}, \bibinfo{pages}{163--234}, \doiprefix\url{10.1103/RevModPhys.81.163} (\bibinfo{year}{2009}).

\bibitem{gradshteyn2014table}
\bibinfo{author}{Gradshteyn, I.~S.} \& \bibinfo{author}{Ryzhik, I.~M.}
\newblock \emph{\bibinfo{title}{Table of Integrals, Series, and Products}} (\bibinfo{publisher}{Academic press}, \bibinfo{year}{2014}).

\bibitem{das2020lectures}
\bibinfo{author}{Das, A.}
\newblock \emph{\bibinfo{title}{Lectures On Quantum Field Theory (Second Edition)}} (\bibinfo{publisher}{World Scientific Publishing Company}, \bibinfo{year}{2020}).

\bibitem{ryder1996quantum}
\bibinfo{author}{Ryder, L.~H.}
\newblock \emph{\bibinfo{title}{Quantum Field Theory}} (\bibinfo{publisher}{Cambridge University Press}, \bibinfo{year}{1996}).

\end{thebibliography}

\noindent

\section*{Acknowledgements}

This work is supported by FWO under grant G0B0623N.

\section*{Author contributions}

F.O. conducted the calculations. F.O. and A.B. prepared the figures. C.C. wrote the manuscript, integrating scientific insights from F.O., A.B., and C.C. All authors reviewed the manuscript. C.C. supervised the overall process.

\section*{Additional information}

The authors declare no competing interests.

\section*{Supplementary Material for ``Electron Scattering at a Temporal Step Discontinuity"}




\noindent

\tableofcontents
%
%

\noindent

\clearpage
\section{Types of Scalar and Vector Potential Spatial and Temporal Steps}\label{sec:types_pot}
\begin{figure}[h!]
    \centering 
        \includegraphics[width=0.7\textwidth]{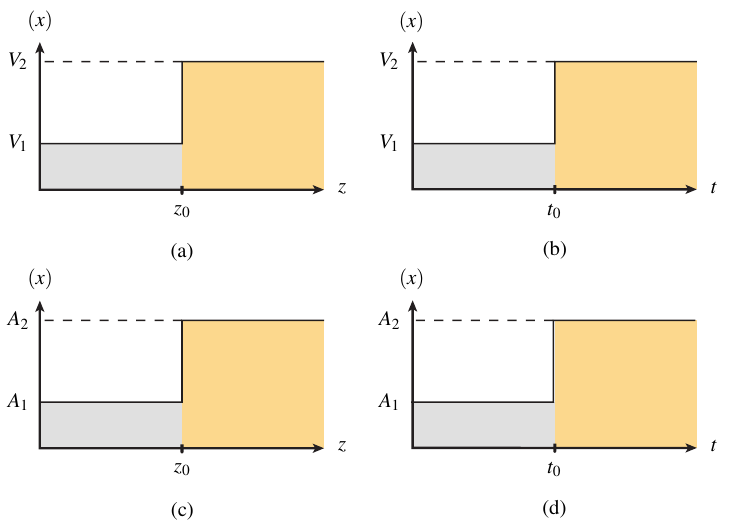}
    \caption{Types of potential steps considered. (a)~Scalar potential spatial step, $V(z)$. (b)~Scalar potential temporal step, $V(t)$. (c)~Vector potential spatial step, $A(z)$. (d)~Vector potential temporal step, $A(t)$.}
    \label{fig:four_step_pot}
\end{figure}

Figure~\ref{fig:four_step_pot} depicts the different scalar and vector potential spatial and temporal steps considered in the paper.

\clearpage
\section{The Dirac Equation}\label{sec:Dirac_equation}
The Dirac equation, in natural units ($\hbar=c=1$), reads~\cite{Dirac_1928_paper,Greiner_2000_rel_quant}
\begin{subequations}
    \begin{equation}\label{eq:s_dirac_eq}
        i\pdv{t}\psi=\mathcal{\hat{H}}\psi,
    \end{equation} 
where $\psi$ is the ($4\times{}1$) Dirac spinor wavefunction and $\mathcal{\hat{H}}$ is the ($4\times{}4$) Hamiltonian, which is in the free (fermion) particle case
    \begin{equation}\label{eq:Dirac_Hamiltonian1}
        \mathcal{\hat{H}}=-i\alpha^{i}\partial_{i}+\gamma^{0} m.
    \end{equation}
This Hamiltonian is composed of the following elements:
\begin{itemize}
    \item the ($4\times{}4$) $\alpha^i$ matrices
        \begin{equation}\label{eq:alpha_i}
            \alpha^{i}=
            \begin{pmatrix}
                0 & \sigma^i \\
                \sigma^i & 0
            \end{pmatrix},
            \quad\text{with}\quad i=1,2,3,
        \end{equation} 
whose components are the Pauli matrices
\begin{equation}\label{eq:Pauli_matrix}
        \sigma^1=
        \begin{pmatrix}
            0 & 1 \\
            1 & 0 
        \end{pmatrix},
        \quad
        \sigma^2=
        \begin{pmatrix}
            0 & -i \\
            i & 0 
        \end{pmatrix}
        \quad\text{and}\quad
        \sigma^3=
        \begin{pmatrix}
            1 & 0 \\
            0 & -1 
        \end{pmatrix},
    \end{equation} 
    \item the four-gradient
        \begin{equation}
            \partial_{\mu} \equiv\left(\partial_{0},\partial_{i}\right),
            \quad\text{with}\quad i=1,2,3,
        \end{equation}
    \item the ($4\times{}4$) $\gamma^0$ matrix (Dirac-Pauli representation)
        \begin{equation}\label{eq:gamma_0}
	   \gamma^{0}=
            \begin{pmatrix}
                I & 0 \\
                0 & -I
            \end{pmatrix},
            \quad\text{where}\quad
            I=
            \begin{pmatrix}
                1 & 0 \\
                0 & 1
            \end{pmatrix},
        \end{equation}
        \item the particle mass, $m$. 
    \end{itemize}
\end{subequations}

Inserting Eq.~\eqref{eq:Dirac_Hamiltonian1} into Eq.~\eqref{eq:s_dirac_eq} multiplied by $\gamma^0$ yields
\begin{equation}\label{eq:Dirac_interm}
    i\gamma^0\partial_0\psi
    =\left(-i\gamma^0\alpha^i\partial_i+\gamma^0\gamma^0m\right)\psi.
\end{equation}
Then combining Eqs.~\eqref{eq:gamma_0} and~\eqref{eq:alpha_i}, which gives
\begin{equation}\label{eq:gamma_i}
    \gamma^0\alpha^i=
    \begin{pmatrix}
        0 &\sigma^i \\
        -\sigma^i & 0
    \end{pmatrix}
    =\gamma^i,
\end{equation}
and noting that $\gamma^0\gamma^0=1$, simplifies Eq.~\eqref{eq:Dirac_interm} to
\begin{equation}
    i\gamma^0\partial_0\psi
    =\left(-i\gamma^i\partial_i+m\right)\psi.
\end{equation}
Finally, moving the right-hand side terms of this equation to its left-hand side provides the explicit form of the free-particle Dirac equation, 
\begin{equation}\label{eq:Dirac_explicit}
    \left(i\gamma^\mu\partial_\mu-m\right)\psi
    =0,\quad\text{with}\quad\mu=0,1,2,3,
\end{equation}
which assumes the Einstein summation convention, whereby the repetition of an index in a given term implies summation over that index. Note that Eq.~\eqref{eq:Dirac_explicit} represents four scalar equations via the $\gamma^\mu$ matrices.

When the particle is subjected to an electromagnetic field, expressed in terms of the four-potential~\cite{Jackson_2000_electrodyn}
\begin{equation}
A_{\mu}=\left(A_{0},-\vb{A}\right),
\end{equation}
the Dirac equation~\eqref{eq:Dirac_explicit} transforms according to the minimal-coupling prescription~\cite{Peskin_1995_QFT}
\begin{equation}
\partial_{\mu}\rightarrow \partial_{\mu}+iqA_{\mu},
\end{equation}
viz., 
\begin{equation}\label{eq:Dirac_potential}
    \left[\gamma^\mu(i\partial_\mu-qA_\mu)-m\right]\psi
    =0,
\end{equation} 
where $q$ is the charge of the particle.

The Dirac equation~\eqref{eq:Dirac_potential} may then be solved with the general ansatz
\begin{equation}\label{eq:pw_ansatz}
    \psi= \begin{pmatrix} 
        \varphi \\
        \vartheta
    \end{pmatrix}\mathrm{e}^{\mp{}ip\cdot x},
\end{equation}
where $\varphi$ and $\vartheta$ are two-component ($2\times{}1$) spinors and 
\begin{equation}\label{eq:s_product_of_p_and_x}
    p\cdot x=p^{\mu}x_\mu=\eta^{\mu\sigma}p_{\sigma}x_\mu,
\end{equation}
with
\begin{equation}
    \eta^{\mu\sigma}=
    \begin{pmatrix}
        1 & 0 & 0 & 0 \\
        0 & -1 & 0 & 0 \\
        0 & 0 & -1 & 0 \\
        0 & 0 & 0 & -1
    \end{pmatrix}
\end{equation}
being the Minkowski metric (West Coast convention). That the ansatz~\eqref{eq:pw_ansatz} is correct will be verified in Sec.~\ref{sec:solution_Dirac}. In Eq.~\eqref{eq:s_product_of_p_and_x}, 
\begin{equation}
p_\mu=(E, -\vb{p})
\end{equation}
is the momentum four-vector and 
\begin{equation}
x_{\mu}=(t, -\vb{x})
\end{equation}
is the position four-vector. The expression $p\cdot{x}=Et-\vb{p}\cdot\vb{x}$ is then identified as the phase of a plane wave, which reveals that the solution form in Eq.~\eqref{eq:pw_ansatz} corresponds to a plane-wave solution. We see then that two signs in that equation [Eq.~\eqref{eq:pw_ansatz}] correspond to positive and negative energies, which Dirac identified as corresponding to the particle (e.g., electron) and its antiparticle (e.g., positron). Assuming that we are exclusively dealing with the former type, we shall ignore negative energies and hence drop the positive sign in the spinor waveform~\eqref{eq:pw_ansatz}. Thus, further restricting our attention to the $1+1$ dimensional case, with the spatial dimension being $z$, we find that the ansatz~\eqref{eq:pw_ansatz} reduces to the specific form
\begin{equation}\label{eq:s_plane_wave_1D}
    \psi=
        \begin{pmatrix}
            \varphi \\
            \vartheta
        \end{pmatrix}
        \mathrm{e}^{-i(E t - p z)}.
\end{equation}

\clearpage 
\section{Solution to the Dirac Equations}\label{sec:solution_Dirac}

\subsection{General Solution in the Presence of Both Scalar and Vector Potentials}\label{sec:gen_sol_VA}

Substituting the ansatz~\eqref{eq:s_plane_wave_1D} into the Dirac equation~\eqref{eq:Dirac_potential} gives
\begin{subequations}
    \begin{equation}
        \left[\gamma^\mu(i\partial_\mu-qA_\mu)-m\right]\begin{pmatrix}
        \varphi \\
        \vartheta
        \end{pmatrix}\mathrm{e}^{-i(E t - p z)}
        =0.
    \end{equation}
Expanding this relation according to the Einstein summation convention yields
    \begin{equation}
        \left[\gamma^0(i\partial_0-qA_0)+\gamma^3(i\partial_3+qA_3)-m\right]\begin{pmatrix}
        \varphi \\
        \vartheta
        \end{pmatrix}\mathrm{e}^{-i(E t - p z)}
        =0,
    \end{equation}
which becomes after applying the derivative operators to the exponential term
    \begin{equation}
        \left[\gamma^0(E-qA_0)-\gamma^3(p-qA_3)-m\right]\begin{pmatrix}
        \varphi \\
        \vartheta
        \end{pmatrix}\mathrm{e}^{-i(E t - p z)}
        =0.
    \end{equation}
Inserting now the gamma matrices $\gamma^0$ and $\gamma^3$ from Eqs.~\eqref{eq:gamma_0} and~\eqref{eq:gamma_i}, respectively, into this relation and dropping the exponential function, leads to 
    \begin{equation}
        \left[\begin{pmatrix}
                I & 0 \\
                0 & -I
            \end{pmatrix}(E-qA_0)-\begin{pmatrix}
        0 &\sigma^3 \\
        -\sigma^3 & 0
    \end{pmatrix}(p-qA_3)-m\right]\begin{pmatrix}
        \varphi \\
        \vartheta
        \end{pmatrix}
        =0,
    \end{equation}
\end{subequations}
which splits into the two equations
\begin{subequations}
    \begin{equation}\label{eq:first_spinor_eq}
        (E-qA_0)\varphi-(p-qA_3)\sigma^3\vartheta-m\varphi=0
    \end{equation}
    \text{and}
    \begin{equation}\label{eq:second_spinor_eq}
        -(E-qA_0)\vartheta+(p-qA_3)\sigma^3\varphi-m\vartheta=0.
    \end{equation}
\end{subequations}
Separately solving these equations for $\varphi$ yields then
\begin{subequations}\label{eq:spinor_eq}
    \begin{equation}\label{eq:phi_first}
        \varphi=\frac{(p-qA_3)\sigma^3}{E-qA_0-m}\vartheta
    \end{equation}
and
    \begin{equation}\label{eq:phi_second}
        \varphi=\frac{E-qA_0+m}{(p-qA_3)\sigma^3}\vartheta.
    \end{equation}
\end{subequations}
Equating these two relations and using the identity $(\sigma^3)^2=I$ that follows from Eq.~\eqref{eq:Pauli_matrix}, entails the Dirac energy-momentum or dispersion relation
\begin{equation}\label{eq:disp_rel}
    (E-qA_0)^2=(p-qA_3)^2+m^2,
\end{equation}
which reduces for $A_0=A_3=0$ to Einstein's equation $E^2=p^2+m^2=(\gamma m)^2$, where $\gamma=(1-v^2)^{-1/2}$ is the Lorentz factor, with $v$ begin the velocity of the particle.

At this point, we may select either the spin-up solution or the spin-down solution, which correspond to the spinors

\begin{subequations}
    \begin{equation}\label{eq:phi_up}
        \varphi_\uparrow
        =\begin{pmatrix}
            1 \\
            0
        \end{pmatrix}
    \end{equation}
    and
    \begin{equation}\label{eq:phi_down}
        \varphi_\downarrow
        =\begin{pmatrix}
            0 \\
            1
            \end{pmatrix},
    \end{equation}
\end{subequations}
respectively. Choosing the spin-up solution and correspondingly inserting Eq.~\eqref{eq:phi_up} into Eq.~\eqref{eq:phi_first}, yields 
\begin{subequations}
    \begin{equation}
        \begin{pmatrix}
        1 \\
        0
        \end{pmatrix}=\frac{p-qA_3}{E-qA_0-m}\begin{pmatrix}
            1 & 0 \\
            0 & -1 
        \end{pmatrix}\begin{pmatrix}
        \vartheta_1 \\
        \vartheta_2
        \end{pmatrix}
    \end{equation}
or
    \begin{equation}
        \begin{pmatrix}
        1 \\
        0
        \end{pmatrix}=\frac{p-qA_3}{E-qA_0-m}\begin{pmatrix}
        \vartheta_1 \\
        -\vartheta_2
        \end{pmatrix},
    \end{equation}
\end{subequations}
which determines the spinor as $\vartheta$ as
\begin{equation}\label{eq:theta_up}
    \vartheta=\begin{pmatrix}
        \vartheta_1 \\
        \vartheta_2
        \end{pmatrix}=\begin{pmatrix}
        \frac{E-qA_0-m}{p-qA_3} \\
        0
        \end{pmatrix}.
    \end{equation}
Finally, substituting Eq.~\eqref{eq:phi_up} and Eq.~\eqref{eq:theta_up} into~\eqref{eq:s_plane_wave_1D} yields the solution form
\begin{equation}\label{eq:s_plane wave_sol}
    \psi=
        \begin{pmatrix}
        1\\
        0\\
        \frac{E-qA_0-m}{p-qA_3}\\
        0
        \end{pmatrix}
    \mathrm{e}^{-i(E t - p z)}.
\end{equation}
Note that the alternative choice of Eq.~\eqref{eq:phi_second} would have led to the third element of this spinor being replaced by $\frac{p-qA_3}{E-qA_0+m}$, which can easily be shown to be equal to $\frac{E-qA_0-m}{p-qA_3}$. Further note that the alternative choice of the spin-down solution [Eq.~\eqref{eq:phi_down}] would have led to the entries $1$ and $\frac{E-qA_0-m}{p-qA_3}$ being replaced by zero and shifted to the second and fourth slots of the spinor, respectively, with the sign of the latter changed.

\subsection{Particular Solutions in the Presence of Specific Potentials}

We shall now apply the general solution in Eq.~\eqref{eq:s_plane wave_sol} to the different step potential introduced in Sec.~\ref{sec:types_pot}.

\subsubsection{Scalar Potential Temporal Step, \texorpdfstring{$V(t)$}{V(t)}} \label{subsec:solution_Dirac_temporal_scalar}

The scalar potential temporal step, shown in Fig.~\ref{fig:four_step_pot}(b), may be written
\begin{equation}\label{eq:s_scalar_pot_temporal}
    A_0=V(t)=
    \begin{cases}
        V_1 &\;\textrm{for}\;t<t_{0}, \\
        V_2 &\;\textrm{for}\;t>t_{0}, 
    \end{cases}
    \quad\textrm{and}\quad A_3=0.
\end{equation}

Substituting this potential into the solution form~\eqref{eq:s_plane wave_sol} and using ans\"{a}tze corresponding to the related temporal-step electromagnetic solutions~\cite{Morgenthaler_1958_puretime,Caloz_2020_spacetimeII}, we assume the incident, or earlier ($t<t_0$), spinor wavefunction
\begin{subequations}\label{eq:s_scalar_temporal_plane_wave_sol}
\begin{equation}\label{eq:s_scalar_temporal_plane_wave_sol_before}
    \psi_{1} = \begin{pmatrix}
        1\\
        0\\
        \frac{E_\mathrm{i}-qV_1-m}{p_\mathrm{i}}\\
        0
    \end{pmatrix} \mathrm{e}^{-i E_\mathrm{i} t} \mathrm{e}^{i p_\mathrm{i} z} 
\end{equation}
and the later ($t>t_0$) spinor wavefunction
\begin{equation}\label{eq:s_scalar_temporal_plane_wave_sol_after}
    \psi_{2} =
    f \begin{pmatrix}
        1\\
        0\\
        \frac{E_\mathrm{f}-qV_2-m}{p_\mathrm{f}}\\
        0
    \end{pmatrix} \mathrm{e}^{-i E_\mathrm{f} t} \mathrm{e}^{i p_\mathrm{f} z}
    + b \begin{pmatrix}
        1\\
        0\\
        \frac{E_\mathrm{b}-qV_2-m}{p_\mathrm{b}}\\
        0
    \end{pmatrix} \mathrm{e}^{-i E_\mathrm{b} t} \mathrm{e}^{i p_\mathrm{b} z},
\end{equation}
whose first and second terms correspond to later-forward and later-backward waves, respectively, with corresponding amplitude coefficients $f$ and $b$.
\end{subequations}

\begin{subequations}\label{eq:s_scalar_temporal_energy_momentum}
According to Noether's theorem~\cite{Noether_1918}, momentum is conserved ($\Delta{p}=0$) due to spatial translational symmetry, viz.,
\begin{equation}\label{eq:s_scalar_temporal_momenta}
    p_\mathrm{i}=p_\mathrm{f}=p_\mathrm{b}=p,
\end{equation}
whereas the breaking of temporal symmetry entails energy transformations, which are found from Eq.~\eqref{eq:disp_rel} with $A_0=V(t)$ and $A_3=0$ to be
\begin{equation}    
    E_\mathrm{i} = \sqrt{p^2 + m^2}+qV_1,
\end{equation}
\begin{equation}    
    E_\mathrm{f} = \sqrt{p^2 + m^2}+qV_2
\end{equation}
and
\begin{equation}    
    E_\mathrm{b} = -E_\mathrm{f} + 2qV_2,
\end{equation}
where the apparent negative energy in the last relation simply represents propagation in the negative $z$ direction, with positive energy.
\end{subequations}

According to Eq.~\eqref{eq:Dirac_potential}, the spinor wavefunction must be continuous at the temporal discontinuity, viz.,
\begin{equation}\label{eq:s_scalar_temporal_boundary}  
    \left.\psi_{1}\right|_{t=t_0}=\left.\psi_{2}\right|_{t=t_0}.
\end{equation}
Inserting Eqs.~\eqref{eq:s_scalar_temporal_plane_wave_sol} into this relation yields
\begin{subequations}\label{eq:s_scalar_temporal_boundary_eq}
    \begin{equation}\label{eq:s_scalar_temporal_boundary_first_eq}
        1=f+b 
    \end{equation}
and
    \begin{equation}\label{eq:s_scalar_temporal_boundary_second_eq}
        \frac{E_\mathrm{i}-qV_1-m}{p_\mathrm{i}}
        =f\frac{E_\mathrm{f}-qV_2-m}{p_\mathrm{f}}+b\frac{E_\mathrm{b}-qV_2-m}{p_\mathrm{b}}.
    \end{equation}
\end{subequations}
Substituting then Eqs.~\eqref{eq:s_scalar_temporal_energy_momentum} into Eq.~\eqref{eq:s_scalar_temporal_boundary_second_eq}, and solving the system formed by the resulting equation and Eq.~\eqref{eq:s_scalar_temporal_boundary_first_eq}, finally leads, after some algebraic manipulations, to
\begin{equation}\label{eq:s_scalar_temporal_coef}
    f = 1  
    \quad \text { and } \quad 
    b = 0,
\end{equation}
which reveals that the scalar potential temporal step, $V(t)$, does not produce any back-scattering.

\subsubsection{Scalar Potential Spatial Step, \texorpdfstring{$V(z)$}{V(z)}}\label{subsec:solution_Dirac_spatial_scalar}

The scalar potential spatial step, shown in Fig.~\ref{fig:four_step_pot}(a), may be written
\begin{equation}\label{eq:s_scalar_pot_spatial}
        A_0=V(z)=
        \begin{cases}
        V_1 &\;\textrm{for}\;z<z_{0}, \\
        V_2 &\;\textrm{for}\;z>z_{0},
    \end{cases}
    \quad\textrm{and}\quad A_3=0.
\end{equation}

Substituting this potential into the solution form~\eqref{eq:s_plane wave_sol} and using ans\"{a}tze corresponding to the related spatial-step electromagnetic solutions~\cite{Jackson_2000_electrodyn,Caloz_2020_spacetimeII}, we assume the left ($z<z_0$) spinor wavefunction
\begin{subequations}\label{eq:s_scalar_spatial_plane_wave_sol}
\begin{equation}\label{eq:s_scalar_spatial_plane_wave_sol_before}
    \psi_{1} = \begin{pmatrix}
        1\\
        0\\
        \frac{E_\mathrm{i}-qV_1-m}{p_\mathrm{i}}\\
        0
    \end{pmatrix} \mathrm{e}^{-i E_\mathrm{i} t} \mathrm{e}^{i p_\mathrm{i} z}  
    +r \begin{pmatrix}
        1\\
        0\\
        \frac{E_\mathrm{r}-qV_1-m}{p_\mathrm{r}}\\
        0
    \end{pmatrix} \mathrm{e}^{-i E_\mathrm{r} t} \mathrm{e}^{i p_\mathrm{r} z}, 
\end{equation}
whose first and second terms correspond to incident and reflected waves, respectively, with reflection amplitude coefficient $r$, and the transmitted, or right ($z>z_0$), spinor wavefunction
\begin{equation}\label{eq:s_scalar_spatial_plane_wave_sol_after}
    \psi_{2} = t \begin{pmatrix}
        1\\
        0\\
        \frac{E_\mathrm{t}-qV_2-m}{p_\mathrm{t}}\\ 
        0
    \end{pmatrix} \mathrm{e}^{-i E_\mathrm{t} t} \mathrm{e}^{i p_\mathrm{t} z},
\end{equation}
\end{subequations}
with transmission amplitude coefficient $t$.

\begin{subequations}\label{eq:s_scalar_spatial_energy_momentum}
According to Noether's theorem~\cite{Noether_1918}, energy is conserved ($\Delta{E}=0$) due to temporal translational symmetry, viz.,
\begin{equation}\label{eq:s_scalar_spatial_energies}
    E_\mathrm{i}=E_\mathrm{r}=E_\mathrm{t}=E,
\end{equation}
whereas the breaking of spatial symmetry implies momentum transformations, which are found from Eq.~\eqref{eq:disp_rel} with $A_0=V(z)$ and $A_3=0$ to be
\begin{equation}\label{eq:s_scalar_spatial_momenta_before}
    p_\mathrm{i}=\sqrt{(E-qV_1)^2 - m^2},
    \end{equation}
    \begin{equation}\label{eq:negative_momentum_Vz}
    p_\mathrm{r}=-p_\mathrm{i}   
    \end{equation} 
and
    \begin{equation}\label{eq:s_scalar_spatial_momenta_after}
    p_\mathrm{t} = \sqrt{(E-qV_2)^2 - m^2},
    \end{equation}
where the apparent negative momentum in Eq.~\eqref{eq:negative_momentum_Vz} simply represents propagation in the negative $z$ direction, with positive momentum, whose momentum versus $p_\mathrm{i}$ are obtained by solving Eq.~\eqref{eq:s_scalar_spatial_momenta_before} for $E$ and inserting the result into Eqs.~\eqref{eq:negative_momentum_Vz} and~\eqref{eq:s_scalar_spatial_momenta_after}, which yields
\begin{equation}
    p_\mathrm{t}=\sqrt{\left(\sqrt{p_\mathrm{i}^2+m^2}-q\Delta{V}\right)^2-m^2}.
\end{equation}
\end{subequations}

According to Eq.~\eqref{eq:Dirac_potential}, the spinor wavefunction must be continuous at the spatial discontinuity, viz.,
\begin{equation} \label{eq:s_scalar_spatial_boundary}  
\left.\psi_{1}\right|_{z=z_{0}}=\left.\psi_{2}\right|_{z=z_{0}}.
\end{equation}
Inserting Eqs.~\eqref{eq:s_scalar_spatial_plane_wave_sol} in this relation yields 
\begin{subequations}\label{eq:s_scalar_spatial_boundary_eq}
    \begin{equation}\label{eq:s_scalar_spatial_boundary_first_eq}
        1+r=t
    \end{equation}
and
    \begin{equation}\label{eq:s_scalar_spatial_boundary_second_eq}
        \frac{E_\mathrm{i}-qV_1-m}{p_\mathrm{i}}+r\frac{E_\mathrm{r}-qV_1-m}{p_\mathrm{r}}=t\frac{E_\mathrm{t}-qV_2-m}{p_\mathrm{t}}.
    \end{equation}
\end{subequations}
Substituting then Eqs.~\eqref{eq:s_scalar_spatial_energy_momentum} into Eq.~\eqref{eq:s_scalar_spatial_boundary_second_eq}, and solving the system formed by the resulting equation and Eq.~\eqref{eq:s_scalar_spatial_boundary_first_eq}, finally leads, after some algebraic manipulations, to
\begin{subequations}\label{eq:s_scalar_spatial_coef_gamma}
\begin{equation}\label{eq:s_scalar_spatial_coef}
    r = \frac{1-\Gamma_\mathrm{s}}{1+\Gamma_\mathrm{s}}
    \quad \text { and } \quad
    t = \frac{2}{1+\Gamma_\mathrm{s}},
\end{equation}
    \text{where}
\begin{equation}\label{eq:s_scalar_spatial_gamma}
    \Gamma_\mathrm{s} = \frac{{(E-qV_2-m)}{p_\mathrm{i}}}{{{(E-qV_1-m)}{p_\mathrm{t}}}}.
\end{equation}
\end{subequations}

The probabilities associated with the reflected and transmitted waves may then be found from the ratios of the corresponding components of the conserved Dirac current. The Dirac conserved current is
\begin{equation} \label{eq:s_cuurent_formula}
    j^{\mu}=\bar{\psi} \gamma^{\mu} \psi,
\end{equation}
where $\bar{\psi}$ is the Dirac adjoint,
\begin{equation}\label{eq:s_Dirac_adjoint}
    \bar{\psi} = \psi^{\dagger} \gamma^{0}, 
\end{equation}
with $\psi^{\dagger}$ being the Hermitian conjugate of $\psi$. In our problem [Fig.~\ref{fig:four_step_pot}(a)], $j^{\mu}$ has only a $z$ spatial component, and takes then the forms
\begin{subequations}
    \begin{equation}
        j^z_{1}=\psi^{\dagger}_{1}\gamma^{0}\gamma^{3}\psi_{1}
    \end{equation}
    and
    \begin{equation}
        j^z_{2}=\psi^{\dagger}_{2}\gamma^{0}\gamma^{3}\psi_{2},
    \end{equation}
\end{subequations}
in the left and right regions, respectively. Substituting Eqs.~\eqref{eq:s_scalar_spatial_plane_wave_sol}, \eqref{eq:gamma_0} and~\eqref{eq:gamma_i} into these relations yields
\begin{subequations}\label{eq:s_currents}
    \begin{equation}\label{eq:s_current_before}
        j^z_{1} = j^z_\mathrm{i} - j^z_\mathrm{r},
    \end{equation}
    \text{with}
    \begin{equation}\label{eq:s_current_incident_reflected}
        j^z_\mathrm{i} = 2\frac{E-qV_1 - m}{p_\mathrm{i}}
        \quad \text{and} \quad
        j^z_\mathrm{r} = 2|r|^2 \frac{E-qV_1-m}{p_\mathrm{i}},
    \end{equation}  
    \text{and}
    \begin{equation}\label{eq:s_current_after}
        j^z_{2} = j^z_\mathrm{t},
    \end{equation}
    \text{with}
    \begin{equation}\label{eq:s_current_transmitted}
        j^z_\mathrm{t} = 2|t|^2\frac{E - qV_2 - m}{p_\mathrm{t}}.
    \end{equation}
\end{subequations}
The reflection and the transmission probabilities are then obtained as
\begin{subequations}\label{eq:s_scalar_spatial_probs}
    \begin{equation}\label{eq:s_scalar_spatial_prob_ref}
    R = \left|\frac{j^z_\mathrm{r}}{j^z_\mathrm{i}}\right| = |r|^2 = \left|\frac{1-\Gamma_\mathrm{s}}{1+\Gamma_\mathrm{s}}\right|^2
    \end{equation}
    and
    \begin{equation}\label{eq:s_scalar_spatial_prob_trans}
    T = \left|\frac{j^z_\mathrm{t}}{j^z_\mathrm{i}}\right| = |t|^2\Gamma_\mathrm{s} = \left|\frac{2}{1+\Gamma_\mathrm{s}}\right|^2\Gamma_\mathrm{s},
    \end{equation}
\end{subequations}
where $\Gamma_\mathrm{s}$ was defined in Eqs.~\eqref{eq:s_scalar_spatial_gamma}. Note that these probabilities verify the probability conservation formula
\begin{equation}\label{prob_total}
    R + T = 1.
\end{equation}

An alternative, pragmatic way to determine the probabilities~\eqref{eq:s_scalar_spatial_probs} is to write 
\begin{equation}
    R=|r|^2\mathcal{C}_R
    \quad\text{and}\quad
    T=|t|^2\mathcal{C}_T,
\end{equation}
where the parameters $\mathcal{C}_R$ and $\mathcal{C}_T$ are ``momentum-transition'' coefficients, associated with change of region (from $V_1$ to $V_2$). Therefore, we must have $\mathcal{C}_R=1$ and $\mathcal{C}_T\neq 1$. The coefficient $\mathcal{C}_T$ may then be determined from the probability conservation, specifically by inserting $R=|r|^2$ and $T=|t|^2\mathcal{C}_T$ with Eq.~\eqref{eq:s_scalar_spatial_coef_gamma} into Eq.~\eqref{prob_total}, which leads to $\mathcal{C}_T=\Gamma_\mathrm{s}$ and hence retrieves the results in Eqs.~\eqref{eq:s_scalar_spatial_probs}.

One may distinguish three potential regions in plotting $R$ and $T$, assuming $(E-qV_1)>m$ so that $p_\mathrm{i}$ [Eq.~\eqref{eq:s_scalar_spatial_momenta_before}] is real, depending on $p_\mathrm{t}$ [Eq.~\eqref{eq:s_scalar_spatial_momenta_after}]~\cite{Greiner_2000_rel_quant}:
\begin{enumerate}
    \item $qV_2<E-m$: $p_\mathrm{t}=$ real and hence $\Gamma_\mathrm{s}=\frac{{(E-qV_2-m)}{\sqrt{(E-qV_1)^{2}-m^{2}}}}{{{(E-qV_1-m)}{\sqrt {(E-qV_2)^{2}-m^{2}}}}}$;
    \item $E-m<qV_2<E+m$: $p_\mathrm{t}=$ is imaginary and hence $\Gamma_\mathrm{s}= \frac{{(E-qV_2-m)}{\sqrt{(E-qV_1)^{2}-m^{2}}}}{{{(E-qV_1-m)}{ (i) \sqrt {-(E-qV_2)^{2}+m^{2}}}}}$;
    \item $E+m<qV_2$: $p_\mathrm{t}=$ real. However, if we choose $p_\mathrm{t}>0$, then, according to Eq.~\eqref{eq:s_scalar_spatial_momenta_after}, $v_\mathrm{g}=dE/dp_\mathrm{t}=\frac{p_\mathrm{t}}{E-qV_2}<0$ since $E-qV_2<0$, which is contradictory to the assumption of positive-$z$ propagation for the transmitted wave. Therefore, in order to maintain a positive group velocity, we must assign a negative sign to $p_\mathrm{t}$, viz., use $\Gamma_\mathrm{s}=\frac{{(E-qV_2-m)}{\sqrt{(E-qV_1)^{2}-m^{2}}}}{{{(E-qV_1-m)}{(-)\sqrt {(E-qV_2)^{2}-m^{2}}}}}$.
\end{enumerate}

\subsubsection{Vector Potential Spatial Step, \texorpdfstring{$A(z)$}{A(z)}}\label{subsec:solution_Dirac_spatial_vector}

The vector potential spatial step, shown in Fig.~\ref{fig:four_step_pot}(c), may be written
\begin{equation}\label{eq:s_vector_pot_spatial}
    A_3=A(z)=
    \begin{cases}
        A_1 &\;\textrm{for}\;z<z_{0}, \\
        A_2 &\;\textrm{for}\;z>z_{0},
    \end{cases}
    \quad\textrm{and}\quad A_0=0.
\end{equation}

Substituting this potential into the solution form~\eqref{eq:s_plane wave_sol} and using ans\"{a}tze corresponding to the related spatial-step electromagnetic solutions~\cite{Jackson_2000_electrodyn,Caloz_2020_spacetimeII}, we assume the left ($z<z_0$) spinor wavefunction
\begin{subequations}\label{eq:s_vector_spatial_plane_wave_sol}
\begin{equation}\label{eq:s_vector_spatial_plane_wave_sol_before}
    \psi_{1} = \begin{pmatrix}
        1\\
        0\\
        \frac{E_\mathrm{i}-m}{p_\mathrm{i}-qA_1}\\
        0
    \end{pmatrix} \mathrm{e}^{-i E_\mathrm{i} t} \mathrm{e}^{i p_\mathrm{i} z}  
    +r \begin{pmatrix}
        1\\
        0\\
        \frac{E_\mathrm{r}-m}{p_\mathrm{r}-qA_1}\\
        0
    \end{pmatrix} \mathrm{e}^{-i E_\mathrm{r} t} \mathrm{e}^{i p_\mathrm{r} z},
\end{equation}
whose first and second terms correspond to incident and reflected waves, respectively, with reflection amplitude coefficient $r$, and the transmitted, or right ($z>z_0$), spinor wavefunction
\begin{equation}\label{eq:s_vector_spatial_plane_wave_sol_after}
    \psi_{2} = t \begin{pmatrix}
        1\\
        0\\
        \frac{E_\mathrm{t}-m}{p_\mathrm{t}-qA_2}\\
        0
    \end{pmatrix} \mathrm{e}^{-i E_\mathrm{t} t} \mathrm{e}^{i p_\mathrm{t} z},
\end{equation}
\end{subequations}
with transmission amplitude coefficient $t$.

\begin{subequations}\label{eq:s_vector_spatial_energy_momentum}
According to Noether's theorem~\cite{Noether_1918}, energy is conserved ($\Delta{E}=0$) due to temporal translational symmetry, viz.,
    \begin{equation}\label{eq:s_vector_spatial_energies}
    E_\mathrm{i}=E_\mathrm{r}=E_\mathrm{t}=E,
    \end{equation}
whereas the breaking of spatial symmetry implies momentum transformations, which are found from Eq.~\eqref{eq:disp_rel} with $A_3=A(z)$ and $A_0=V=0$ to be
    \begin{equation}\label{eq:s_vector_spatial_momenta_before}
    p_\mathrm{i}=\sqrt{E^2 - m^2} + qA_1,   
    \end{equation}
    \begin{equation}\label{eq:negative_momentum_Az}
    p_\mathrm{r}=-p_\mathrm{i} + 2qA_1   
    \end{equation} 
and
    \begin{equation}\label{eq:s_vector_spatial_momenta_after}
    p_\mathrm{t} = \sqrt{E^2 - m^2} + qA_2,
    \end{equation}
where the apparent negative momentum in Eq.~\eqref{eq:negative_momentum_Az} simply represents propagation in the negative $z$ direction, with positive momentum.
\end{subequations}

According to Eq.~\eqref{eq:Dirac_potential}, the spinor wavefunction must be continuous at the spatial discontinuity, viz.,

\begin{equation} \label{eq:s_vector_spatial_boundary}
    \left.\psi_{1}\right|_{z=z_{0}}=\left.\psi_{2}\right|_{z=z_{0}}.
\end{equation}
Inserting Eqs.~\eqref{eq:s_vector_spatial_plane_wave_sol} in this relation yields
\begin{subequations}\label{eq:s_vector_spatial_boundary_eq}
    \begin{equation}\label{eq:s_vector_spatial_boundary_first_eq}
        1+r=t
    \end{equation}
and
    \begin{equation}\label{eq:s_vector_spatial_boundary_second_eq}
        \frac{E_\mathrm{i}-m}{p_\mathrm{i}-qA_1}+r\frac{E_\mathrm{r}-m}{p_\mathrm{r}-qA_1}=t\frac{E_\mathrm{t}-m}{p_\mathrm{t}-qA_2}.
    \end{equation}
\end{subequations}
Substituting then Eqs.~\eqref{eq:s_vector_spatial_energy_momentum} into Eq.~\eqref{eq:s_vector_spatial_boundary_second_eq}, and solving the system formed by the resulting equation and Eq.~\eqref{eq:s_vector_spatial_boundary_first_eq}, finally leads, after some algebraic manipulations, to
\begin{equation}\label{eq:s_vector_spatial_coef}
    r = 0
    \quad \text { and } \quad
    t = 1,
\end{equation}
which reveals that the vector potential spatial step, $A(z)$, does not produce any back-scattering.

\subsubsection{Vector Potential Temporal Step, \texorpdfstring{$A(t)$}{A(t)}}\label{subsec:solution_Dirac_temporal_vector}

The vector potential temporal step, shown in Fig.~\ref{fig:four_step_pot}(d), may be written
\begin{equation}\label{eq:s_vector_pot_temporal}
    A_3=A(t)=
    \begin{cases}
        A_1 &\;\textrm{for}\;t<t_{0}, \\
        A_2 &\;\textrm{for}\;t>t_{0},
    \end{cases}
    \quad\textrm{and}\quad A_0=V=0.
\end{equation}

Substituting this potential into the solution form~\eqref{eq:s_plane wave_sol} and using ans\"{a}tze corresponding to the related temporal-step electromagnetic solutions~\cite{Morgenthaler_1958_puretime,Caloz_2020_spacetimeII}, we assume the incident, or earlier ($t<t_0$), spinor wavefunction
\begin{subequations}\label{eq:s_vector_temporal_plane_wave_sol}
\begin{equation}\label{eq:s_vector_temporal_plane_wave_sol_before}
    \psi_{1} = \begin{pmatrix}
        1\\
        0\\
        \frac{E_\mathrm{i}-m}{p_\mathrm{i}-qA_1}\\
        0
    \end{pmatrix} \mathrm{e}^{-i E_\mathrm{i} t} \mathrm{e}^{i p_\mathrm{i} z} 
\end{equation}
and the later ($t>t_0$) spinor wavefunction
\begin{equation}\label{eq:s_vector_temporal_plane_wave_sol_after}
    \psi_{2} = f \begin{pmatrix}
        1\\
        0\\
        \frac{E_\mathrm{f}-m}{p_\mathrm{f}-qA_2}\\
        0
    \end{pmatrix} \mathrm{e}^{-i E_\mathrm{f} t} \mathrm{e}^{i p_\mathrm{f} z}+ b  \begin{pmatrix}
        1\\
        0\\
        \frac{E_\mathrm{b}-m}{p_\mathrm{b}-qA_2}\\
        0
    \end{pmatrix} \mathrm{e}^{-i E_\mathrm{b} t} \mathrm{e}^{i p_\mathrm{b} z},
\end{equation}
whose first and second terms correspond to later-forward and later-backward waves, respectively, with corresponding amplitude coefficients $f$ and $b$.
\end{subequations}

\begin{subequations}\label{eq:s_vector_temporal_energy_momentum}
According to Noether's theorem~\cite{Noether_1918}, momentum is conserved ($\Delta{p}=0$) due to spatial translational symmetry, viz.,
    \begin{equation}\label{eq:s_vector_temporal_momenta}
    p_\mathrm{i}=p_\mathrm{f}=p_\mathrm{b}=p,
    \end{equation}
whereas the breaking of spatial symmetry implies energy transformations, which are found from Eq.~\eqref{eq:disp_rel} with $A_3=A(t)$ and $A_0=V=0$ to be
    \begin{equation}\label{eq:s_vector_temporal_energies_before}    
    E_\mathrm{i} = \sqrt{(p-qA_1)^2 + m^2},
    \end{equation}
    \begin{equation}\label{eq:s_vector_temporal_energies_after}
    E_\mathrm{f} = \sqrt{(p-qA_2)^2 + m^2}
    \end{equation}
and
    \begin{equation}\label{eq:s_vector_temporal_energies_after_b}
        E_\mathrm{b}=-E_\mathrm{f},
    \end{equation}
where the apparent negative energy in the last relation simply represents propagation in the negative $z$ direction, with positive energy.
\end{subequations}

According to Eq.~\eqref{eq:Dirac_potential}, the spinor wavefunction must be continuous at the temporal discontinuity, viz.,
\begin{equation}\label{eq:s_vector_temporal_boundary}  
\left.\psi_{1}\right|_{t=t_0}=\left.\psi_{2}\right|_{t=t_0}.
\end{equation}
Inserting Eqs.~\eqref{eq:s_scalar_temporal_plane_wave_sol} into this relation yields
\begin{subequations}\label{eq:s_vector_temporal_boundary_eq}
    \begin{equation}\label{eq:s_vector_temporal_boundary_first_eq}
        1=f+b
    \end{equation}
and
    \begin{equation}\label{eq:s_vector_temporal_boundary_second_eq}
        \frac{E_\mathrm{i}-m}{p_\mathrm{i}-qA_1}=f\frac{E_\mathrm{f}-m}{p_\mathrm{f}-qA_2}+b\frac{E_\mathrm{b}-m}{p_\mathrm{b}-qA_2}.
    \end{equation}
\end{subequations}
Substituting then Eqs.~\eqref{eq:s_vector_temporal_energy_momentum} into Eq.~\eqref{eq:s_vector_temporal_boundary_second_eq}, and solving the system formed by the resulting equation and Eq.~\eqref{eq:s_vector_temporal_boundary_first_eq}, finally leads, after some algebraic manipulations, to
\begin{subequations}\label{eq:s_vector_temporal_coef_gamma}
    \begin{equation}\label{eq:s_vector_temporal_coef}
        f = \frac{1+\Gamma_\mathrm{t}}{2 \Gamma_\mathrm{t}}
        \quad \text { and } \quad
        b = \frac{\Gamma_\mathrm{t}-1}{2 \Gamma_\mathrm{t}}  
    \end{equation}
where
    \begin{equation}\label{eq:GtofpEf}
        \Gamma_\mathrm{t}=
        \dfrac{\dfrac{E_\mathrm{f}}{p-qA_2}}{\dfrac{E_\mathrm{i}-m}{p-qA_1}+\dfrac{m}{p-qA_2}}.
    \end{equation}
This expression may be alternatively written in terms of $E_\mathrm{i}$ and $A_{1,2}$ only upon first using Eq.~\eqref{eq:s_vector_temporal_energies_after} to eliminate $E_\mathrm{f}$ and then substituting in the resulting expression 
\begin{equation}\label{eq:pofEi}
    p=\sqrt{E_\mathrm{i}^2-m^2}+qA_1, 
\end{equation}
which was obtained from Eq.~\eqref{eq:s_vector_temporal_energies_after}, which yields
    \begin{equation}\label{eq:s_vector_temporal_gamma}
        \Gamma_\mathrm{t}  = \dfrac{\sqrt{\left(\sqrt{E_\mathrm{i}^2-m^2}-(q A_2-qA_1)\right)^2+m^2}}{\left(\sqrt{E_\mathrm{i}^2-m^2}-(q A_2-qA_1)\right)\left(\dfrac{E_\mathrm{i}-m}{\sqrt{E_\mathrm{i}^2-m^2}}\right)+m}.
    \end{equation}
\end{subequations}

The probabilities associated with the later-forward and later-backward waves cannot be found from the ratios of the corresponding Dirac currents, contrary to the case of the $V(z)$ problem in Sec.~\ref{subsec:solution_Dirac_temporal_scalar}, because the Dirac current is \emph{not} conserved here, due to the non-conservation of energy. However, we may resort to an alternative approach similar to that also used in Sec.~\ref{subsec:solution_Dirac_temporal_scalar}, writing
\begin{equation}
    F=|f|^2\mathcal{C}_F
    \quad\text{and}\quad
    B=|b|^2\mathcal{C}_B,
\end{equation}
where the parameters $\mathcal{C}_F$ and $\mathcal{C}_B$ are now ``energy-transition'' coefficients, associated with change of region (from $A_1$ to $A_2$). Therefore, we must have $\mathcal{C}_F=\mathcal{C}_B=\mathcal{C}$, since the two probabilities correspond to the same change of region, from the earlier region ($A_1$) to the later region ($A_2$), so that
\begin{equation}\label{eq:FB_C}
    F=|f|^2\mathcal{C}
    \quad\text{and}\quad
    B=|b|^2\mathcal{C}.
\end{equation}
At the same, probability must be conserved, since the particle can only either keep moving forward or move backward, viz.,
\begin{equation}\label{eq:FB1}
    F+B=1.
\end{equation}
Substituting then Eqs.~\eqref{eq:FB_C} with Eqs.~\eqref{eq:s_vector_temporal_coef} into Eq.~\eqref{eq:FB1} yields then
\begin{equation}
    \mathcal{C} = \frac{2\Gamma_\mathrm{t}^2}{1+\Gamma_\mathrm{t}^2},
\end{equation} 
so that the later forward and backward probabilities are finally obtained from Eq.~\eqref{eq:FB_C} as
\begin{subequations}\label{eq:s_vector_temporal_probs}
\begin{equation}\label{eq:s_vector_temporal_prob_later_forward}
    F = |f|^2 \frac{2\Gamma_\mathrm{t}^2}{1+\Gamma_\mathrm{t}^2} = \left|\frac{1+\Gamma_\mathrm{t}}{2 \Gamma_\mathrm{t}}\right|^2 \frac{2\Gamma_\mathrm{t}^2}{1+\Gamma_\mathrm{t}^2}
\end{equation}
and
\begin{equation}\label{eq:s_vector_temporal_prob_later_backward}
    B = |b|^2 \frac{2\Gamma_\mathrm{t}^2}{1+\Gamma_\mathrm{t}^2} = \left|\frac{\Gamma_\mathrm{t}-1}{2 \Gamma_\mathrm{t}}\right|^2 \frac{2\Gamma_\mathrm{t}^2}{1+\Gamma_\mathrm{t}^2},
\end{equation}    
\end{subequations}
where Eq.~\eqref{eq:s_vector_temporal_coef} was used in the second equalities and where $\Gamma_\mathrm{t}$ was defined in Eq.~\eqref{eq:s_vector_temporal_gamma}.

\clearpage
\section{Gauge Transformations and Symmetries}\label{sec:gauge_sm}

\subsection{Scalar Potential Spatial Step \texorpdfstring{$V(z)$}{V(z)}}\label{sec:gauge_sm_Vz}
The electric and magnetic fields, $\vb{E}$ and $\vb{B}$, are generally related to the potentials as
\begin{subequations}
    \begin{equation}
    \vb{E}=-\grad V-\pdv{\vb{A}}{t}
    \end{equation}
    and
    \begin{equation}\label{eq:B_eq_curlA}
    \vb{B}=\curl\vb{A},
    \end{equation}
\end{subequations}
which are invariant under the gauge transformation~\cite{Jackson_2000_electrodyn,Jackson_2001_Historical_Gauge}
\begin{subequations}\label{eq:s_gauge_transf}
    \begin{equation}
    V\rightarrow V^\prime=V-\pdv{\Lambda}{t}
    \end{equation}
    and
    \begin{equation}
    \vb{A}\rightarrow\vb{A}^\prime=\vb{A}+\grad \Lambda,
    \end{equation}
\end{subequations}
where $\Lambda$ is an arbitrary scalar function. The potential spatial step $V(z)$ is equivalent to the transformation
\begin{equation}
    V'=V_1+\Delta V\theta(z-z_0)
    \quad\textrm{and}\quad
    \vb{A}^\prime=0,
\end{equation}
Consistency with the gauge~\eqref{eq:s_gauge_transf} would demand that 
\begin{equation}
    \pdv{\Lambda}{t}=-\Delta{}V\theta(z-z_0)
    \quad\textrm{and}\quad
    \grad \Lambda=0.
\end{equation}
which imply
\begin{equation}
    \Lambda=-\Delta{}Vt\theta(z-z_0)
    \quad\textrm{and}\quad
    \Lambda\neq\Lambda(z).
\end{equation}
respectively. The incompatibility between these two conditions on $\Lambda$ indicates that the transformation indeed breaks the symmetry of the gauge~\eqref{eq:s_gauge_transf}, which explains why the corresponding potential leads to electron back-scattering~\cite{Griffiths_2018_Quantum} (Sec.~\ref{sec:gauge_sm_Vz}).

\subsection{Vector Potential Spatial Step \texorpdfstring{$A(z)$}{A(z)}}\label{sec:gauge_sm_Az}
The vector potential spatial step $\vb{A}(z)=A(z)\vb{\hat{z}}$ is equivalent to the transformation
\begin{equation}
    V'=0
    \quad\textrm{and}\quad
    \vb{A}^\prime=A_1+\Delta{}A\theta(z-z_0),
\end{equation}
which is a particular case of the gauge transformation~$\eqref{eq:s_gauge_transf}$ with 
\begin{equation}
    \pdv{\Lambda}{t}=0
    \quad\textrm{and}\quad
    \grad \Lambda=\Delta{}A\theta(z-z_0),
\end{equation}
corresponding to
\begin{equation}
    \Lambda=\Delta{}A\theta(z-z_0)z=\Lambda(z)
\end{equation}
Therefore, the step $A(z)$ does not involve any change in the fields, and hence also in the impedance, which explains why we found that it produces no (reflected wave) back-scattering (Sec.~\ref{subsec:solution_Dirac_spatial_vector}).

\clearpage
\section{Phase and Group Velocities}\label{sec:phase_group_vel}

The phase and group velocities may be computed from the dispersion relation~\eqref{eq:disp_rel}, i.e.,
\begin{equation}\label{eq:disp_rel_app}
    (E-qA_0)^{2}=(p-qA_3)^{2}+m^{2}.
\end{equation}

The phase velocity is defined as
\begin{equation}\label{eq:vp_def}
    v_\mathrm{p}=\frac{E}{p}.
\end{equation}
In general, it may be found by solving Eq.~\eqref{eq:disp_rel_app} for $p$ and substituting the result into Eq.~\eqref{eq:vp_def}, which yields
\begin{equation}\label{eq:vp_A0_A3}
    v_\mathrm{p}=\frac{E}{\sqrt{(E-qA_0)^{2}-m^{2}}+qA_3}.
\end{equation}
For the cases of the scalar potential spatial step [$V(z)$] (Sec.~\ref{subsec:solution_Dirac_spatial_scalar}) and the vector potential temporal step [$A(t)$] (Sec.~\ref{subsec:solution_Dirac_temporal_vector}), the scattered phase velocities may be directly obtained from Eq.~\eqref{eq:vp_def} as
\begin{equation}\label{eq:vpEpt}
    v_\mathrm{p,t}=\frac{E}{p_\mathrm{t}}
    \quad\textrm{and}\quad
    v_\mathrm{p.r}=\frac{E}{p_\mathrm{r}},
\end{equation}
with $p_\mathrm{t}$ and $p_\mathrm{r}$ given by Eq.~\eqref{eq:s_scalar_spatial_momenta_after} and Eq.~\eqref{eq:s_scalar_spatial_momenta_before}, and
\begin{equation}\label{eq:vpEfp}
    v_\mathrm{p,f}=\frac{E_\mathrm{f}}{p}
    \quad\textrm{and}\quad
    v_\mathrm{p,b}=\frac{E_\mathrm{b}}{p},
\end{equation}
with $E_\mathrm{f}$ and $E_\mathrm{b}$ given by Eq.~\eqref{eq:s_vector_temporal_energies_after} and Eq.~\eqref{eq:s_vector_temporal_energies_after_b}, respectively.

The group velocity is defined as
\begin{equation}\label{eq:s_group_vel}
    v_\mathrm{g}=\pdv{E}{p}.
\end{equation}
Its general expression may be found by taking the derivative of Eq.~\eqref{eq:disp_rel_app} versus $p$ and isolating $\partial E/\partial p$, which results in
\begin{equation}\label{eq:s_group_vel_open}
    v_\mathrm{g}=\frac{p-qA_3}{E-qA_0}.
\end{equation}

For the cases of the scalar potential spatial step [$V(z)$] (Sec.~\ref{subsec:solution_Dirac_spatial_scalar}) and the vector potential temporal step [$A(t)$] (Sec.~\ref{subsec:solution_Dirac_temporal_vector}), the scattered group velocities may be obtained from Eq.~\eqref{eq:s_group_vel_open} as
\begin{equation}\label{eq:vgptEqV}
    v_\mathrm{g,t}=\frac{p_\mathrm{t}}{E-qV}
    \quad\textrm{and}\quad
    v_\mathrm{g,r}=\frac{p_\mathrm{r}}{E-qV},
\end{equation}
with $p_\mathrm{t}$ and $p_\mathrm{r}$ given by Eq.~\eqref{eq:s_scalar_spatial_momenta_after} and Eq.~\eqref{eq:s_scalar_spatial_momenta_before}, and
\begin{equation}\label{eq:vgpqAEf}
    v_\mathrm{g,f}=\frac{p-qA}{E_\mathrm{f}}
    \quad\textrm{and}\quad
    v_\mathrm{g,b}=\frac{p-qA}{E_\mathrm{b}},
\end{equation}
with $E_\mathrm{f}$ and $E_\mathrm{b}$ given by Eq.~\eqref{eq:s_vector_temporal_energies_after} and Eq.~\eqref{eq:s_vector_temporal_energies_after_b}, respectively.

In the problem of the scalar potential spatial step [$V(z)$] (Sec.~\ref{subsec:solution_Dirac_spatial_scalar}), assuming $qV_1<E_\mathrm{i}-m$ and $qV_1<qV_2$, we find, using Eqs.~\eqref{eq:vpEpt} and Eq.~\eqref{eq:vgptEqV}
\begin{enumerate}
    \item $qV_2<E-m$ $\rightarrow$ $v_{\mathrm{g},2}<v_{\mathrm{g},1}$ and $v_{\mathrm{p},1}<v_{\mathrm{p},2}$,
    \item $E-m<qV_2<E+m$ $\rightarrow$ $v_{\mathrm{g},2}=0$ and $v_{\mathrm{p},2}=\infty$ (Klein region),
    \item $E+m<qV_2<2E-V_1$ $\rightarrow$ $v_{\mathrm{g} ,2}<v_{\mathrm{g},1}$ and $v_{\mathrm{p},1}<v_{\mathrm{p},2}$,
    \item $2E-V_1<V_2$ $\rightarrow$ $v_{\mathrm{g},1}<v_{\mathrm{g},2}$ and $v_{\mathrm{p},2}<v_{\mathrm{p},1}$,
\end{enumerate}
while in the problem of the vector potential temporal step [$A(t)$] (Sec.~\ref{subsec:solution_Dirac_temporal_vector}), assuming $qA_1<p_\mathrm{i}$ and $qA_1<qA_2$, we find, using Eqs.~\eqref{eq:vpEfp} and Eq.~\eqref{eq:vgpqAEf},
\begin{enumerate}
    \item $qA_2<p_ \mathrm{i}$ $\rightarrow$ $v_{\mathrm{g},2}<v_{\mathrm{g},1}$ and $v_{\mathrm{p},2}<v_{\mathrm{p},1}$,
    \item $p_\mathrm{i}<qA_2<2p_\mathrm{i}-A_1$ $\rightarrow$ $v_{\mathrm{g},2}<v_{\mathrm{g},1}$ and $v_{\mathrm{p},2}<v_{\mathrm{p},1}$,
    \item $2p_\mathrm{i}-A_1<qA_2$ $\rightarrow$ $v_{\mathrm{g},1}<v_{\mathrm{g},2}$ and $v_{\mathrm{p},1}<v_{\mathrm{p},2}$,
\end{enumerate}
where $1$ and $2$ refer to the two regions.

\clearpage
\section{Spatial and Temporal Step Electromagnetic Problems}\label{sec:EM_pure_time_space}
We provide here the main results pertaining to the 1+1D spatial and temporal step electromagnetic problems for the sake of comparison. Related details are available in~\cite{Caloz_2020_spacetimeI,Caloz_2020_spacetimeII}.

\subsection{Spatial Step Problem}
The wave equation may be written as
\begin{equation}
    \left[\frac{1}{n^2} \pdv[2]{z} - \pdv[2]{t} \right] \left\{\begin{array}{l}
    E(z,t) \\
    H(z,t)
    \end{array}\right\} = 0.
\end{equation}
The $E$ and $H$ fields before ($z<z_0$, region~1) and after ($z>z_0$, region~2) the step are
\begin{subequations}\label{eq:s_EM_spatial_plane_wave_sol}
    \begin{equation}\label{eq:s_E_field_spatial_plane_wave_sol_before}
    E_1 = \mathrm{e}^{-i \omega_\mathrm{i} t} \mathrm{e}^{i k_\mathrm{i} z}  
    +r \mathrm{e}^{-i \omega_\mathrm{r} t} \mathrm{e}^{i k_\mathrm{r} z},
    \end{equation}
    \begin{equation}\label{eq:s_H_field_spatial_plane_wave_sol_before}
    H_1 = \left(\mathrm{e}^{-i \omega_\mathrm{i} t} \mathrm{e}^{i k_\mathrm{i} z}  
    -r \mathrm{e}^{-i \omega_\mathrm{r} t} \mathrm{e}^{i k_\mathrm{r} z}\right)/\eta_1,
    \end{equation}   
    \begin{equation}\label{eq:s_E_field_spatial_plane_wave_sol_after}
    E_2 = t \mathrm{e}^{-i \omega_\mathrm{t} t} \mathrm{e}^{i k_\mathrm{t} z},
    \end{equation}
    \begin{equation}\label{eq:s_H_field_spatial_plane_wave_sol_after}
    H_2 = t \mathrm{e}^{-i \omega_\mathrm{t} t} \mathrm{e}^{i k_\mathrm{t} z} / \eta_2,
    \end{equation}
\end{subequations}
where 
\begin{equation}
        \eta_{1,2}=\sqrt{\frac{\mu_{1,2}}{\epsilon_{1,2}}}
\end{equation}
is the intrinsic medium impedance, and
\begin{subequations}\label{eq:s_EM_space_frequencies_wavenumbers}
    \begin{equation}\label{eq:s_EM_space_frequencies}
        \omega_\mathrm{i}=\omega_\mathrm{r}=\omega_\mathrm{t},
    \end{equation}
    \begin{equation}\label{eq:s_EM_space_wavenumbers}
        k_\mathrm{i}=-k_\mathrm{r}=n_1\frac{\omega_\mathrm{i}}{c}, \quad \text{and} \quad k_\mathrm{t}=n_2\frac{\omega_\mathrm{i}}{c}.
    \end{equation}
\end{subequations}
Inserting Eqs.~\eqref{eq:s_EM_spatial_plane_wave_sol} to~\eqref{eq:s_EM_space_frequencies_wavenumbers} into the boundary condition relation
\begin{equation}\label{EM_space_boundary_conditions}
    \left.E_{1}\right|_{z=z_0}=\left.E_{2}\right|_{z=z_0} \quad \text{and} \quad \left.H_{1}\right|_{z=z_0}=\left.H_{2}\right|_{z=z_0}
\end{equation}
yields then the scattering amplitude coefficients
\begin{equation}\label{space_scattering_amplitudes}
    r=\frac{1-N}{1+N} \quad \text{and} \quad t=\frac{2}{1+N},
\end{equation}
while the reflectance and transmittance are found, using the Poynting vector definitions
\begin{equation}
    R=\left|\frac{\vb{E_\mathrm{r}}\times \vb{H_\mathrm{r}}}{\vb{E_\mathrm{i}}\times \vb{H_\mathrm{i}}}\right|
    \quad \text{and} \quad
    T=\left|\frac{\vb{E_\mathrm{t}}\times \vb{H_\mathrm{t}}}{\vb{E_\mathrm{i}}\times \vb{H_\mathrm{i}}}\right|,
\end{equation}
and substituting Eqs.~\eqref{eq:s_EM_spatial_plane_wave_sol} in these relations, as
\begin{equation}\label{eq:s_reflectance_transmittance}
    R=|r|^2 =\left|\frac{1-N}{1+N}\right|^2
    \quad \text{and} \quad 
    T=|t|^2N = \left|\frac{2}{1+N}\right|^2N,
\end{equation}
where $N=\dfrac{n_2}{n_1}=\dfrac{\eta_1}{\eta_2}$, assuming nonmagnetic materials.

The space-time and dispersion diagrams corresponding to these results are provided in Fig.~\ref{fig:EM_TEM_interf}(a).

\subsection{Temporal Step Problem}
The wave equation may be written as
\begin{equation}
    \left[\pdv[2]{z} - n^2  \pdv[2]{t} \right] \left\{\begin{array}{l}
    D(z,t) \\
    B(z,t)
    \end{array}\right\} = 0.
\end{equation}
The $D$ and $B$ fields before ($t<t_0$, region~1) and after ($t>t_0$, region~2) the step are
\begin{subequations}\label{eq:s_EM_temporal_plane_wave_sol}
    \begin{equation}\label{eq:s_D_field_temporal_plane_wave_sol_before}
    D_1 = \mathrm{e}^{-i \omega_\mathrm{i} t} \mathrm{e}^{i k_\mathrm{i} z},  
    \end{equation}
    \begin{equation}\label{eq:s_B_field_temporal_plane_wave_sol_before}
    B_1 = \left(\mathrm{e}^{-i \omega_\mathrm{i} t} \mathrm{e}^{i k_\mathrm{i} z}\right)/\eta_1,  
    \end{equation}    
    \begin{equation}\label{eq:s_D_field_temporal_plane_wave_sol_after}
    D_2 = f \mathrm{e}^{-i \omega_\mathrm{f} t} \mathrm{e}^{i k_\mathrm{f} z}+b \mathrm{e}^{-i \omega_\mathrm{b} t} \mathrm{e}^{i k_\mathrm{b} z},
    \end{equation}
    \begin{equation}\label{eq:s_B_field_temporal_plane_wave_sol_after}
    B_2 = \left(f \mathrm{e}^{-i \omega_\mathrm{f} t} \mathrm{e}^{i k_\mathrm{f} z}-b \mathrm{e}^{-i \omega_\mathrm{b} t} \mathrm{e}^{i k_\mathrm{b} z}\right)/\eta_2,
    \end{equation}
\end{subequations}
and
\begin{subequations}\label{eq:s_EM_time_frequencies_wavenumbers}
    \begin{equation}\label{eq:s_EM_time_wavenumbers}
        k_\mathrm{i} = k_\mathrm{f} = k_\mathrm{b},
    \end{equation}
    \begin{equation}\label{eq:s_EM_time_frequencies}
        \omega_\mathrm{i}=\frac{c}{n_1}k_\mathrm{i} \quad \text{and} \quad\omega_\mathrm{f}=-\omega_\mathrm{b}=\frac{c}{n_2}k_\mathrm{i}.
    \end{equation}
\end{subequations}
Inserting Eqs.~\eqref{eq:s_EM_temporal_plane_wave_sol} to~\eqref{eq:s_EM_time_frequencies_wavenumbers} into the boundary condition relation
\begin{equation}\label{EM_time_boundary_conditions}
    \left.D_{1}\right|_{t=t_0}=\left.D_{2}\right|_{t=t_0} \quad \text{and} \quad \left.B_{1}\right|_{t=t_0}=\left.B_{2}\right|_{t=t_0},
\end{equation}
yields then the scattering amplitude coefficients
\begin{equation}\label{time_scattering_amplitudes}
    f=\frac{1+N}{2N} \quad \text{and} \quad b=\frac{N-1}{2N},
\end{equation}
while the reflectance and transmittance are found, using the Poynting vector definitions
\begin{subequations}
    \begin{equation}
    F =\left|\frac{\vb{E_\mathrm{f}}\times \vb{H_\mathrm{f}}}{\vb{E_\mathrm{i}}\times \vb{H_\mathrm{i}}}\right| 
      =\left|\frac{\vb{D_\mathrm{f}}/\epsilon_2\times \vb{B_\mathrm{f}}/\mu_2}{\vb{D_\mathrm{i}}/\epsilon_1\times \vb{B_\mathrm{i}}/\mu_1}\right| 
     = \left|\frac{\vb{D_\mathrm{f}}\times \vb{B_\mathrm{f}}}{\vb{D_\mathrm{i}}\times \vb{B_\mathrm{i}}}\right|\left(\frac{n_1}{n_2}\right)^2 
    \end{equation}
    \begin{equation}
    B =\left|\frac{\vb{E_\mathrm{b}}\times \vb{H_\mathrm{b}}}{\vb{E_\mathrm{i}}\times \vb{H_\mathrm{i}}}\right| 
    =\left|\frac{\vb{D_\mathrm{b}}/\epsilon_2\times \vb{B_\mathrm{b}}/\mu_2}{\vb{D_\mathrm{i}}/\epsilon_1\times \vb{B_\mathrm{i}}/\mu_1}\right| 
    = \left|\frac{\vb{D_\mathrm{b}}\times \vb{B_\mathrm{b}}}{\vb{D_\mathrm{i}}\times \vb{B_\mathrm{i}}}\right|\left(\frac{n_1}{n_2}\right)^2,
    \end{equation}
\end{subequations}
and substituting Eqs.~\eqref{eq:s_EM_temporal_plane_wave_sol} in these relations, as
\begin{equation}\label{eq:s_refrectance_transmittance_temporal}
    F = |f|^2\frac{1}{N}= \left|\frac{1+N}{2N}\right|^2\frac{1}{N}
    \quad \text{and} \quad
    B = |b|^2\frac{1}{N}= \left|\frac{N-1}{2N}\right|^2 \frac{1}{N},
\end{equation}    
where $N=\dfrac{n_2}{n_1}=\dfrac{\eta_1}{\eta_2}$, assuming nonmagnetic materials.

\begin{figure}[ht!]
    \centering
    \includegraphics[width=0.7\textwidth]{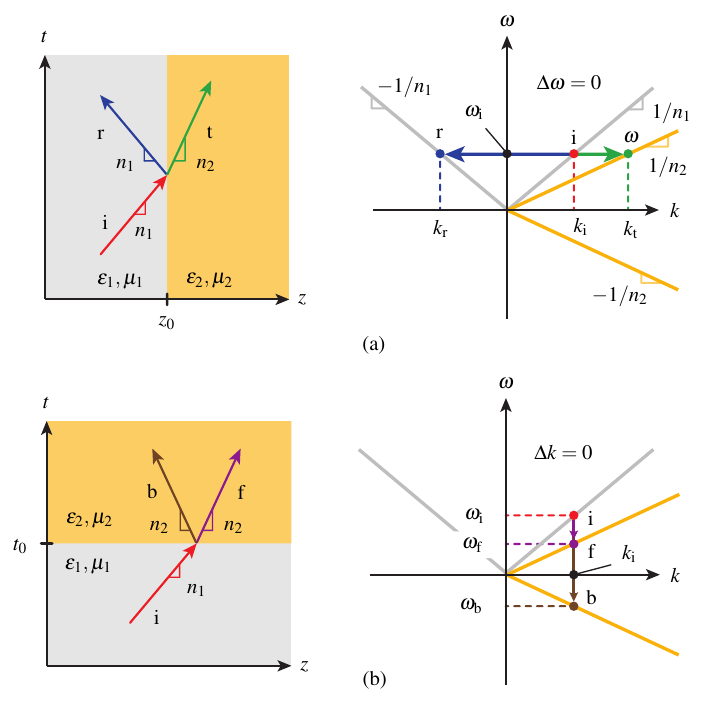}
    \caption{Electromagnetic wave scattering at (a)~a spatial and (b)~a temporal refractive index step discontinuity, represented in space-time diagrams (left panels) and dispersion diagrams with transitions (right panels).}
    \label{fig:EM_TEM_interf}
\end{figure}

The space-time and dispersion diagrams corresponding to these results are provided in Fig.~\ref{fig:EM_TEM_interf}(b).

\clearpage

\section{Smooth Temporal Step}\label{sec:smoothstep}

\subsection{Dirac Equation for a Time-Varying Vector Potential}

\pati{Dirac Equation} 
We start with the Dirac equation in its covariant form~\cite{Greiner_2000_rel_quant}, viz.,
\begin{subequations}
\begin{equation}\label{eq:Dirac}
    \left[\gamma^\mu(i\partial_\mu-qA_\mu)-m\right]\psi
    =0,
\end{equation}
which involves the four-gradient
\begin{equation}
    \partial_{\mu} \equiv\left(\partial_{0},\partial_{i}\right),
    \quad\text{with}\quad i=1,2,3,
\end{equation}
and the gamma matrices\footnote{Note that we have chosen here the Weyl representation for these matrices~\cite{Greiner_2000_rel_quant}. Other choices would have been the Dirac and Majorana representations~\cite{Greiner_2000_rel_quant}. The reason for that choice is that it leads -- as we found in painstaking derivations -- to convenient hypergeometric expressions and ultimately to closed-form solutions to the problem.}
\begin{equation}\label{eq:gamma0_smooth}
	\gamma^{0}=
        \begin{pmatrix}
            0 & I \\
            I & 0
        \end{pmatrix},
        \quad\text{with}\quad
        I=
        \begin{pmatrix}
            1 & 0 \\
            0 & 1
        \end{pmatrix},
\end{equation}
and
\begin{equation}\label{eq:gammai_smooth}
    \gamma^{i}=
    \begin{pmatrix}
        0 & -\sigma^i \\
        \sigma^i & 0
    \end{pmatrix},
\end{equation}
with the Pauli matrices
\begin{equation}\label{eq:sigma_smooth}
    \sigma^1=
    \begin{pmatrix}
        0 & 1 \\
        1 & 0 
    \end{pmatrix},
    \quad
    \sigma^2=
    \begin{pmatrix}
        0 & -i \\
        i & 0 
    \end{pmatrix}
    \quad\text{and}\quad
    \sigma^3=
    \begin{pmatrix}
        1 & 0 \\
        0 & -1 
    \end{pmatrix}.
\end{equation}
\end{subequations}

\pati{Assumptions and Wavefunction Ansatz}
In the sequel of this section, we assume a 1+1D ($t$ and $z$) spacetime system and a $z$-directed magnetic vector potential, viz., $A_0=A_1=A_2=0$ and $A_3=A_z(t)$. These assumption reduce Eq.~\eqref{eq:Dirac} to
\begin{equation}\label{eq:Dirac_expl}
    \left[\gamma^0i\partial_0+\gamma^3(i\partial_3+qA_z(t))-m\right]\psi
    =0,
\end{equation}
which, using the spinor ansatz
\begin{equation}\label{eq:momentum_wave_ansatz}
    \psi=
        \begin{pmatrix}
            \varphi \\
            \vartheta
        \end{pmatrix}
        e^{i p z},
\end{equation}
takes the more explicit form
\begin{equation}\label{eq:Dirac_after_momentum_wave_ansatz}
    \left[\gamma^0 i\partial_0-\gamma^3(p-qA_z(t))-m\right]\begin{pmatrix}
    \varphi \\
    \vartheta
    \end{pmatrix}
    =0.
\end{equation}
Inserting the gamma matrices~\eqref{eq:gamma0_smooth} and~\eqref{eq:gammai_smooth} with the sigma matrices~\eqref{eq:sigma_smooth} into Eq.~\eqref{eq:Dirac_after_momentum_wave_ansatz} yields then
\begin{equation}\label{eq:Dirac_after_putting_sigmas}
    \left[\begin{pmatrix}
            0 & 1 \\
            1 & 0
        \end{pmatrix} i\partial_0 -\begin{pmatrix}
    0 & -1 \\
    1 & 0
    \end{pmatrix}(p-qA_z(t))-m\right]\begin{pmatrix}
    \varphi \\
    \vartheta
    \end{pmatrix}
    =0,
\end{equation}
which splits into the two equations
\begin{subequations} \label{eq:split_smooth_Dirac}
    \begin{equation} \label{eq:split_smooth_Dirac1}
        i\dv{\vartheta}{t}+(p-qA_z(t))\vartheta-m\varphi=0
    \end{equation}
    \text{and}
    \begin{equation} \label{eq:split_smooth_Dirac2}
        i\dv{\varphi}{t}-(p-qA_z(t))\varphi-m\vartheta=0.
    \end{equation}
\end{subequations}
Isolating $\vartheta$ in Eq.~\eqref{eq:split_smooth_Dirac2},
\begin{equation}\label{eq:theta_of_phi}
    \vartheta=\frac{1}{m}\left[i\dv{\varphi}{t}-(p-qA_z(t))\varphi\right],
\end{equation}
and substituting this expression as well as its time derivative,
\begin{equation}
    \dv{\vartheta}{t}=\frac{1}{m}\left[i\dv[2]{\varphi}{t}+q\dv{A_z(t)}{t}\varphi-(p-qA_z(t))\dv{\varphi}{t}\right],
\end{equation}
into Eq.~\eqref{eq:split_smooth_Dirac1}, we get

\begin{equation} \label{eq:phi_time_dif_eq}
    \dv[2]{\varphi}{t}+\left[(p-qA_z(t))^2+m^2-iq\dv{A_z(t)}{t}\right]\varphi=0.
\end{equation}

\pati{Smooth Time-Dependent Vector Potential}
Now, we select for the potential $A_z(t)$ the smooth vector potential corresponding to the hyperbolic tangent function
\begin{equation}\label{eq:smooth_pot}
    A_z(t) = A_1 + \frac{A_2-A_1}{2} \left[1 + \tanh\left(\frac{t-t_0}{\tau}\right)\right] = \frac{A_1+A_2\mathrm{e}^{2\frac{t-t_0}{\tau}}}{1+\mathrm{e}^{2\frac{t-t_0}{\tau}}},
\end{equation}
which corresponds to the electric field
\begin{equation}\label{eq:E_smooth_pot}
    E_z(t) 
    = -\frac{\partial A_z(t)}{\partial t} 
    = -\frac{(A_2-A_1)}{2\tau} \sech^2\left(\frac{t-t_0}{\tau}\right) = -\frac{2(A_2-A_1)}{\tau}\frac{\mathrm{e}^{2\frac{t-t_0}{\tau}}}{\left(1+\mathrm{e}^{2\frac{t-t_0}{\tau}}\right)^2},
\end{equation}
where $\tau$ is a time constant.

\begin{figure}[h!]
    \centering 
    \includegraphics[width=1\textwidth]{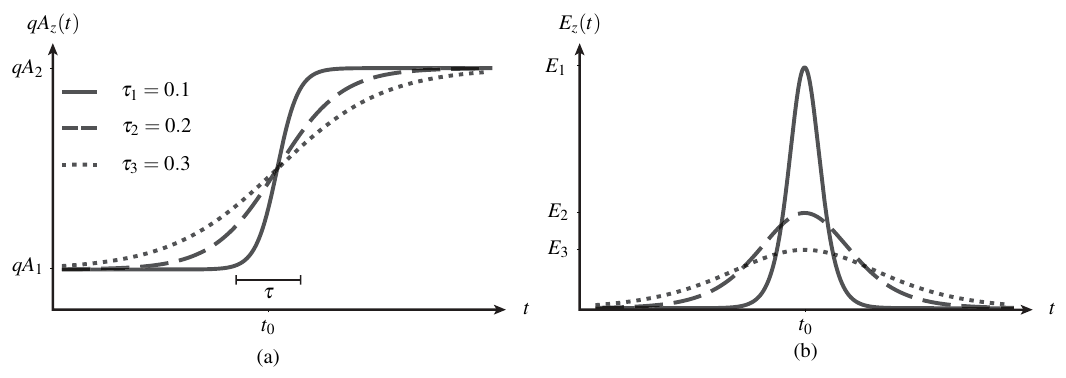}
    \caption{Functions associated with the smooth potential investigated in Sec.~\ref{sec:smoothstep} for different values of the time parameter $\tau$. (a)~Potential function itself, $A_z(t)$ [Eq.~\eqref{eq:smooth_pot}]. (b)~Corresponding electric field, $E_z(t)$ [Eq.~\eqref{eq:E_smooth_pot}].}
    \label{fig:A_pot_smooth}
\end{figure}

We shall next solve Eq.~\eqref{eq:phi_time_dif_eq} for the earlier medium and later medium separately so as to be able to subsequently apply boundary conditions between them. For each case, we shall define a new variable that will simplify the vector potential in Eq.~\eqref{eq:smooth_pot} to the point that Eq.~\eqref{eq:phi_time_dif_eq} will progressively transform into a hypergeometric equation, which admits analytical solutions.  

\subsection{Solution for the Earlier Medium (\texorpdfstring{$t<t_0$}{t<t0})}

\pati{New Variable from $t$ to $\zeta_e$}
For the earlier (e) medium, we use the variable
\begin{equation}\label{eq:zeta_e}
    \zeta_\mathrm{e} = -\mathrm{e}^{2\frac{t-t_0}{\tau}},
\end{equation}
which simplifies Eq.~\eqref{eq:smooth_pot} to
\begin{subequations}\label{eq:A_dt_d2t_e}
\begin{equation}\label{eq:A_of_zeta_e}
    A_z(\zeta_\mathrm{e}) = \frac{A_1-A_2 \zeta_\mathrm{e}}{1-\zeta_\mathrm{e}},
\end{equation}
and whose first and second time derivatives lead to
\begin{equation}\label{dt_dzeta_e}
    \dv{t} = \frac{2}{\tau}\zeta_\mathrm{e} \dv{\zeta_\mathrm{e}},
\end{equation}
and
\begin{equation}\label{d2t_dzeta2_e}
    \dv[2]{t} = \frac{4}{\tau^2} \zeta_\mathrm{e} \dv{\zeta_\mathrm{e}} +  \frac{4}{\tau^2} \zeta_\mathrm{e}^2 \dv[2]{\zeta_\mathrm{e}},
\end{equation}
\end{subequations}
respectively. 

Inserting Eqs.~\eqref{eq:A_dt_d2t_e} into Eq.~\eqref{eq:phi_time_dif_eq} and multiplying the result by $\frac{\tau^2}{4}\frac{1-\zeta_\mathrm{e}}{\zeta_\mathrm{e}}$, we get, after some algebraic manipulations,
\begin{subequations}\label{eq:phi_zeta_e_dif_eq}
    \begin{equation}\label{eq:phi_zeta_e_dif_eq_0}
        \zeta_\mathrm{e} (1-\zeta_\mathrm{e}) \dv[2]{\varphi_\mathrm{e}}{\zeta_\mathrm{e}} + (1-\zeta_\mathrm{e}) \dv{\varphi_\mathrm{e}}{\zeta_\mathrm{e}}
        +\frac{1}{\zeta_\mathrm{e}(1-\zeta_\mathrm{e})}\left [\alpha_2 \zeta_\mathrm{e}^2 
        + \alpha_0 \zeta_\mathrm{e} 
        + \alpha_1 \right]\varphi_\mathrm{e}=0,
    \end{equation}
    where
    \begin{equation}\label{eq:alpha_0}
        \alpha_0 = -\frac{\tau^2}{4}\left[((p-qA_2)^2+m^2) + ((p-qA_1)^2+m^2)-(qA_2-qA_1)^2 - 2i \frac{qA_2-qA_1}{\tau}\right],
    \end{equation}
    \begin{equation}\label{eq:alpha_1}
        \alpha_1=\frac{\tau^2}{4}\left[(p-qA_1)^2+m^2\right]
        = \frac{\tau^2}{4}E_1^2,
    \end{equation}
    and
    \begin{equation}\label{eq:alpha_2}
        \alpha_2=\frac{\tau^2}{4}\left[(p-qA_2)^2+m^2\right]
        = \frac{\tau^2}{4}E_2^2,
    \end{equation}
\end{subequations}
and whose singularities at $\zeta_\mathrm{e}=0$ and $1-\zeta_\mathrm{e}=0$ can be removed by inserting the expression
\begin{equation}\label{eq:phi_of_zeta_e}
    \varphi_\mathrm{e}(\zeta_\mathrm{e})=\zeta_\mathrm{e}^\mu(1-\zeta_\mathrm{e})^\nu f(\zeta_\mathrm{e}),
\end{equation}
its derivative,
\begin{subequations}
\begin{equation}
    \dv{\varphi_\mathrm{e}}{\zeta_\mathrm{e}} = \zeta_\mathrm{e}^\mu(1-\zeta_\mathrm{e})^\nu \left [\left(\mu \frac{1}{\zeta_\mathrm{e}} - \nu \frac{1}{1-\zeta_\mathrm{e}}\right) f + \dv{f}{\zeta_\mathrm{e}} \right],
\end{equation}
and its second derivative,
\begin{equation}
    \dv[2]{\varphi_\mathrm{e}}{\zeta_\mathrm{e}} = \zeta_\mathrm{e}^\mu(1-\zeta_\mathrm{e})^\nu \left [\left(\mu(\mu-1)\frac{1}{\zeta_\mathrm{e}^2} + \nu(\nu-1)\frac{1}{(1-\zeta_\mathrm{e})^2} -2\mu\nu \frac{1}{\zeta_\mathrm{e}(1-\zeta_\mathrm{e})}\right) f + \left(2\mu \frac{1}{\zeta_\mathrm{e}} - 2\nu \frac{1}{1-\zeta_\mathrm{e}} \right) \dv{f}{\zeta_\mathrm{e}} + \dv[2]{f}{\zeta_\mathrm{e}} \right].
\end{equation}
\end{subequations}
Dividing the resulting equation by $\zeta_\mathrm{e}^\mu(1-\zeta_\mathrm{e})^\nu$ finally yields the hypergeometric-type equation
\begin{subequations}\label{eq:f_zeta_e_dif_eq}
    \begin{equation}
        \zeta_\mathrm{e} (1-\zeta_\mathrm{e}) \dv[2]{f}{\zeta_\mathrm{e}} + \left[(2\mu+1) - (2\mu + 2\nu + 1)\zeta_\mathrm{e} \right] \dv{f}{\zeta_\mathrm{e}} -\left [(\mu + \nu + \lambda)(\mu + \nu - \lambda) \right] f = 0,
    \end{equation}
    where
    \begin{equation}\label{eq:mu_E1}
        \mu = i \sqrt{\alpha_1} = i \frac{\tau}{2} E_1,
    \end{equation}
    \begin{equation}\label{eq:nu_A2_A1}
        \nu = \frac{1 - \sqrt{1-4(\alpha_0+\alpha_1+\alpha_2)}}{2} = i \frac{\tau}{2} (qA_2-qA_1),
    \end{equation}
    and
    \begin{equation}\label{eq:lambda_E2}
        \lambda = i \sqrt{\alpha_2} = i \frac{\tau}{2} E_2.
    \end{equation}
\end{subequations}

\pati{Solution of Hypergeometric $f(\zeta_\mathrm{e})$ Differential Equation}

The solution to Eqs.~\eqref{eq:f_zeta_e_dif_eq} can be expressed in terms of Gauss hypergeometric functions $_2F_1(\cdot)$~\cite{gradshteyn2014table} as
\begin{equation}
    f(\zeta_\mathrm{e}) = C_1 {_2F_1}(\mu+\nu-\lambda, \mu+\nu+\lambda;1+2\mu;\zeta_\mathrm{e}) + C_2 \zeta_\mathrm{e}^{-2\mu} {_2F_1}(-\mu+\nu-\lambda, -\mu+\nu+\lambda;1-2\mu;\zeta_\mathrm{e}),
\end{equation}
and substituting this equation into Eq.~\eqref{eq:phi_of_zeta_e} allows to determine the wavefunction $\varphi_\mathrm{e}(\zeta_\mathrm{e})$ in terms of Gauss hypergeometric functions as
\begin{equation}\label{eq:phi_e_c1_c2}
\begin{split}
    \varphi_\mathrm{e}(\zeta_\mathrm{e}) 
    & = C_1 \zeta_\mathrm{e}^{\mu}(1-\zeta_\mathrm{e})^{\nu} {_2F_1}(\mu+\nu-\lambda, \mu+\nu+\lambda;1+2\mu;\zeta_\mathrm{e}) \\ 
    & + C_2 \zeta_\mathrm{e}^{-\mu}(1-\zeta_\mathrm{e})^{\nu} {_2F_1}(-\mu+\nu-\lambda, -\mu+\nu+\lambda;1-2\mu;\zeta_\mathrm{e}).
\end{split}
\end{equation}
Since there is only an incident wave in the earlier medium, considering Eq.~\eqref{eq:mu_E1}, the coefficient $C_1$ must be zero, which reduces Eq.~\eqref{eq:phi_e_c1_c2} to
\begin{equation}\label{eq:phi_e_c2}
    \varphi_\mathrm{e}(\zeta_\mathrm{e}) = C_2 \zeta_\mathrm{e}^{-\mu}(1-\zeta_\mathrm{e})^{\nu} {_2F_1}(-\mu+\nu-\lambda, -\mu+\nu+\lambda;1-2\mu;\zeta_\mathrm{e}).
\end{equation}

\pati{Calculation of $\vartheta_\mathrm{e}$}
To complete the spinor wavefunction, we also need to calculate $\vartheta_\mathrm{e}$. This may be accomplished by inserting Eq.~\eqref{eq:phi_e_c2} into Eq.~\eqref{eq:theta_of_phi} and using Eqs.~\eqref{eq:A_of_zeta_e} and \eqref{dt_dzeta_e}. This includes the intermediate step
    \begin{equation}
    \begin{split}
        \dv{\varphi_\mathrm{e}}{\zeta_\mathrm{e}} = & -C_2 \mu \zeta_\mathrm{e}^{-\mu-1}(1-\zeta_\mathrm{e})^\nu {_2F_1}(-\mu+\nu-\lambda, -\mu+\nu+\lambda;1-2\mu;\zeta_\mathrm{e}) \\
        & - C_2 \zeta_\mathrm{e}^{-\mu} \nu (1-\zeta_\mathrm{e})^{\nu-1} {_2F_1}(-\mu+\nu-\lambda, -\mu+\nu+\lambda;1-2\mu;\zeta_\mathrm{e}) \\
        & + C_2 \zeta_\mathrm{e}^{-\mu} (1-\zeta_\mathrm{e})^{\nu} \frac{(-\mu+\nu-\lambda) (-\mu+\nu+\lambda)}{1-2\mu} {_2F_1}(-\mu+\nu-\lambda+1, -\mu+\nu+\lambda+1;1-2\mu+1;\zeta_\mathrm{e}),
    \end{split}
    \end{equation}
    where using the derivative property of hypergeometric functions~\cite{gradshteyn2014table},
    \begin{equation}\label{eq:deriv_hypgeo}
        \dv{\zeta} [{_2F_1}  (a,b;c;\zeta)]=\frac{ab}{c}{_2F_1}(a+1,b+1;c+1,\zeta),
    \end{equation}
leads to
\begin{equation}\label{eq:theta_e_C2}
\begin{split}
    \vartheta_\mathrm{e} =\frac{C_2}{m} & \left[ -i \frac{2}{\tau} \mu \zeta_\mathrm{e}^{-\mu}(1-\zeta_\mathrm{e})^\nu {_2F_1}(-\mu+\nu-\lambda, -\mu+\nu+\lambda;1-2\mu;\zeta_\mathrm{e}) \right. \\
    & \left. - i \frac{2}{\tau}  \zeta_\mathrm{e}^{-\mu+1} \nu (1-\zeta_\mathrm{e})^{\nu-1} {_2F_1}(-\mu+\nu-\lambda, -\mu+\nu+\lambda;1-2\mu;\zeta_\mathrm{e}) \right.\\
    & \left. + i \frac{2}{\tau}  \zeta_\mathrm{e}^{-\mu+1} (1-\zeta_\mathrm{e})^{\nu} \frac{(-\mu+\nu-\lambda) (-\mu+\nu+\lambda)}{1-2\mu} {_2F_1}(-\mu+\nu-\lambda+1, -\mu+\nu+\lambda+1;1-2\mu+1;\zeta_\mathrm{e}) \right. \\
    & \left. -\left(p-\frac{qA_1-qA_2 \zeta_\mathrm{e}}{1-\zeta_\mathrm{e}}\right)  \zeta_\mathrm{e}^{-\mu}(1-\zeta_\mathrm{e})^{\nu} {_2F_1}(-\mu+\nu-\lambda, -\mu+\nu+\lambda;1-2\mu;\zeta_\mathrm{e}) \right].
\end{split}
\end{equation}

\subsection{Solution for the Later Medium (\texorpdfstring{$t>t_0$}{t>t0})}

\pati{New Variable from $t$ to $\zeta_\mathrm{l}$}
For the later (l) medium, we use the variable
\begin{equation}\label{eq:zeta_l}
    \zeta_\mathrm{l} = -\mathrm{e}^{-2\frac{t-t_0}{\tau}},
\end{equation}
which simplifies Eq.~\eqref{eq:smooth_pot} to
\begin{subequations}\label{eq:A_dt_d2t_l}
\begin{equation}\label{eq:A_of_zeta_l}
    A_z(\zeta_\mathrm{l}) = \frac{A_2-A_1 \zeta_\mathrm{l}}{1-\zeta_\mathrm{l}},
\end{equation}
and whose first and second time derivatives lead to
\begin{equation}\label{dt_dzeta_l}
    \dv{t} = -\frac{2}{\tau}\zeta_\mathrm{l} \dv{\zeta_\mathrm{l}}
\end{equation}
and
\begin{equation}\label{d2t_dzeta2_l}
    \dv[2]{t} = \frac{4}{\tau^2} \zeta_\mathrm{l} \dv{\zeta_\mathrm{l}} +  \frac{4}{\tau^2} \zeta_\mathrm{l}^2 \dv[2]{\zeta_\mathrm{l}}
\end{equation}
\end{subequations}
respectively.

Inserting Eqs.~\eqref{eq:A_dt_d2t_l} into Eq.~\eqref{eq:phi_time_dif_eq} and multiplying the result by $\frac{\tau^2}{4}\frac{1-\zeta_\mathrm{l}}{\zeta_\mathrm{l}}$, we get, after some algebraic manipulations,
\begin{subequations}\label{eq:phi_zeta_l_dif_eq}
    \begin{equation}\label{eq:phi_zeta_l_dif_eq_0}
        \zeta_\mathrm{l} (1-\zeta_\mathrm{l}) \dv[2]{\varphi_\mathrm{l}}{\zeta_\mathrm{l}} + (1-\zeta_\mathrm{l}) \dv{\varphi_\mathrm{l}}{\zeta_\mathrm{l}}
        +\frac{1}{\zeta_\mathrm{l}(1-\zeta_\mathrm{l})}\left [\beta_1 \zeta_\mathrm{l}^2 
        + \beta_0 \zeta_\mathrm{l} 
        + \beta_2 \right]\varphi_\mathrm{l}=0,
    \end{equation}
    where
    \begin{equation}\label{eq:beta_0}
        \beta_0 = -\frac{\tau^2}{4}\left[((p-qA_2)^2+m^2) + ((p-qA_1)^2+m^2)-(qA_2-qA_1)^2 - 2i \frac{qA_2-qA_1}{\tau}\right],
    \end{equation}
    \begin{equation}\label{eq:beta_1}
        \beta_1=\frac{\tau^2}{4}\left[(p-qA_1)^2+m^2\right]
        = \frac{\tau^2}{4}E_1^2,
    \end{equation}
    and
    \begin{equation}\label{eq:beta_2}
        \beta_2=\frac{\tau^2}{4}\left[(p-qA_2)^2+m^2\right]
        = \frac{\tau^2}{4}E_2^2,
    \end{equation}
\end{subequations}
and whose singularities at $\zeta_\mathrm{l}=0$ and $1-\zeta_\mathrm{l}=0$ can be removed by inserting the expression
\begin{equation}\label{eq:phi_of_zeta_l}
    \varphi_\mathrm{l}(\zeta_\mathrm{l})=\zeta_\mathrm{l}^\sigma(1-\zeta_\mathrm{l})^\rho f(\zeta_\mathrm{l}),
\end{equation}
its derivative,
\begin{subequations}
\begin{equation}
    \dv{\varphi_\mathrm{l}}{\zeta_\mathrm{l}} = \zeta_\mathrm{l}^\sigma(1-\zeta_\mathrm{l})^\rho \left [\left(\sigma \frac{1}{\zeta_\mathrm{l}} - \rho \frac{1}{1-\zeta_\mathrm{l}}\right) f + \dv{f}{\zeta_\mathrm{l}} \right],
\end{equation}
and its second derivative,
\begin{equation}
    \dv[2]{\varphi_\mathrm{l}}{\zeta_\mathrm{l}} = \zeta_\mathrm{l}^\sigma(1-\zeta_\mathrm{l})^\rho \left [\left(\sigma(\sigma-1)\frac{1}{\zeta_\mathrm{l}^2} + \rho(\rho-1)\frac{1}{(1-\zeta_\mathrm{l})^2} -2\sigma\rho \frac{1}{\zeta_\mathrm{l}(1-\zeta_\mathrm{l})}\right) f + \left(2\sigma \frac{1}{\zeta_\mathrm{l}} - 2\rho \frac{1}{1-\zeta_\mathrm{l}} \right) \dv{f}{\zeta_\mathrm{l}} + \dv[2]{f}{\zeta_\mathrm{l}} \right].
\end{equation}
\end{subequations}
Dividing the resulting equation by $\zeta_\mathrm{l}^\sigma(1-\zeta_\mathrm{l})^\rho$ finally yields the hypergeometric-type equation
\begin{subequations}\label{eq:f_zeta_l_dif_eq}
    \begin{equation}
        \zeta_\mathrm{l} (1-\zeta_\mathrm{l}) \dv[2]{f}{\zeta_\mathrm{l}} + \left[(2\sigma+1) - (2\sigma + 2\rho + 1)\zeta_\mathrm{l} \right] \dv{f}{\zeta_\mathrm{l}} -\left [(\sigma + \rho + \eta)(\sigma + \rho - \eta) \right] f = 0,
    \end{equation}
    where
    \begin{equation}\label{eq:sigma_E2}
        \sigma = i \sqrt{\beta_2} = i \frac{\tau}{2} E_2,
    \end{equation}
    \begin{equation}\label{eq:rho_A2,A1}
        \rho = \frac{1 - \sqrt{1-4(\beta_0+\beta_1+\beta_2)}}{2}= i \frac{\tau}{2} (qA_2-qA_1),
    \end{equation}
    and
    \begin{equation}\label{eq:eta_E1}
        \eta = i \sqrt{\beta_1} = i \frac{\tau}{2} E_1.
    \end{equation}
\end{subequations}

\pati{Solution of Hypergeometric $f(\zeta_\mathrm{l})$ Differential Equation}

The solution to Eqs.~\eqref{eq:f_zeta_l_dif_eq} can be expressed in terms of Gauss hypergeometric functions $_2F_1(\cdot)$~\cite{gradshteyn2014table} as
\begin{equation}
    f(\zeta_\mathrm{l}) = C_3 {_2F_1}(\sigma+\rho-\eta, \sigma+\rho+\eta;1+2\sigma;\zeta_\mathrm{l}) + C_4 \zeta_\mathrm{l}^{-2\sigma} {_2F_1}(-\sigma+\rho-\eta, -\sigma+\rho+\eta;1-2\sigma;\zeta_\mathrm{l}).
\end{equation}
and substituting this equation into Eq.~\eqref{eq:phi_of_zeta_l} allows to determine the wavefunction $\varphi_\mathrm{l}(\zeta_\mathrm{l})$ in terms of Gauss hypergeometric functions as
\begin{equation}\label{eq:phi_l_c3_c4}
\begin{split}
    \varphi_\mathrm{l}(\zeta_\mathrm{l}) 
    & = C_3 \zeta_\mathrm{l}^{\sigma}(1-\zeta_\mathrm{l})^{\rho} {_2F_1}(\sigma+\rho-\eta, \sigma+\rho+\eta;1+2\sigma;\zeta_\mathrm{l}) \\ 
    & + C_4 \zeta_\mathrm{l}^{-\sigma}(1-\zeta_\mathrm{l})^{\rho} {_2F_1}(-\sigma+\rho-\eta, -\sigma+\rho+\eta;1-2\sigma;\zeta_\mathrm{l}).
\end{split}
\end{equation}

\pati{Calculation of $\vartheta_\mathrm{l}$}
To complete the spinor wavefunction, we also need to calculate $\vartheta_\mathrm{l}$. This may be accomplished by inserting Eq.~\eqref{eq:phi_l_c3_c4} into Eq.~\eqref{eq:theta_of_phi} and using Eqs.~\eqref{eq:A_of_zeta_l} and \eqref{dt_dzeta_l}. This includes the intermediate step
    \begin{equation}
    \begin{split}
        \dv{\varphi_\mathrm{l}}{\zeta_\mathrm{l}} & = C_3  \sigma \zeta_\mathrm{l}^{\sigma-1}(1-\zeta_\mathrm{l})^\rho {_2F_1}(\sigma+\rho-\eta, \sigma+\rho+\eta;1+2\sigma;\zeta_\mathrm{l}) \\
        & -C_3  \zeta_\mathrm{l}^{\sigma} \rho (1-\zeta_\mathrm{l})^{\rho-1} {_2F_1}(\sigma+\rho-\eta, \sigma+\rho+\eta;1+2\sigma;\zeta_\mathrm{l}) \\
        & +C_3  \zeta_\mathrm{l}^{\sigma} (1-\zeta_\mathrm{l})^{\rho} \frac{(\sigma+\rho-\eta) (\sigma+\rho+\eta)}{1-2\sigma} {_2F_1}(\sigma+\rho-\eta +1, \sigma+\rho+\eta +1;1+2\sigma +1;\zeta_\mathrm{l}) \\
        & -C_4  \sigma \zeta_\mathrm{l}^{-\sigma-1}(1-\zeta_\mathrm{l})^\rho {_2F_1}(-\sigma+\rho-\eta, -\sigma+\rho+\eta;1-2\sigma;\zeta_\mathrm{l}) \\
        & -C_4  \zeta_\mathrm{l}^{-\sigma} \rho (1-\zeta_\mathrm{l})^{\rho-1} {_2F_1}(-\sigma+\rho-\eta, -\sigma+\rho+\eta;1-2\sigma;\zeta_\mathrm{l}) \\
        & +C_4  \zeta_\mathrm{l}^{-\sigma} (1-\zeta_\mathrm{l})^{\rho} \frac{(-\sigma+\rho-\eta) (-\sigma+\rho+\eta)}{1-2\sigma} {_2F_1}(-\sigma+\rho-\eta +1, -\sigma+\rho+\eta +1;1-2\sigma +1;\zeta_\mathrm{l}),
    \end{split}
    \end{equation}
    where using the derivative property of hypergeometric functions given in Eq.~\eqref{eq:deriv_hypgeo} leads to
\begin{equation}\label{eq:theta_l_c3_c4}
\begin{split}
    \vartheta_\mathrm{l} =\frac{C_3}{m} & \left[-i \frac{2}{\tau}   \sigma \zeta_\mathrm{l}^{\sigma}(1-\zeta_\mathrm{l})^\rho {_2F_1}(\sigma+\rho-\eta, \sigma+\rho+\eta;1+2\sigma;\zeta_\mathrm{l}) \right. \\
    & \left. +i \frac{2}{\tau}  \zeta_\mathrm{l}^{\sigma+1} \rho (1-\zeta_\mathrm{l})^{\rho-1} {_2F_1}(\sigma+\rho-\eta, \sigma+\rho+\eta;1+2\sigma;\zeta_\mathrm{l}) \right. \\
    & \left. -i \frac{2}{\tau} \zeta_\mathrm{l}^{\sigma+1} (1-\zeta_\mathrm{l})^{\rho} \frac{(\sigma+\rho-\eta) (\sigma+\rho+\eta)}{1-2\sigma} {_2F_1}(\sigma+\rho-\eta +1, \sigma+\rho+\eta +1;1+2\sigma +1;\zeta_\mathrm{l}) \right. \\
    & \left. -\left(p-\frac{qA_2-qA_1 \zeta_\mathrm{l}}{1-\zeta_\mathrm{l}}\right)\zeta_\mathrm{l}^{\sigma}(1-\zeta_\mathrm{l})^{\rho} {_2F_1}(\sigma+\rho-\eta, \sigma+\rho+\eta;1+2\sigma;\zeta_\mathrm{l}) \right]  \\
     +\frac{C_4}{m} & \left[i \frac{2}{\tau} \sigma \zeta_\mathrm{l}^{-\sigma}(1-\zeta_\mathrm{l})^\rho {_2F_1}(-\sigma+\rho-\eta, -\sigma+\rho+\eta;1-2\sigma;\zeta_\mathrm{l}) \right. \\
    & \left. +i \frac{2}{\tau} \zeta_\mathrm{l}^{-\sigma+1} \rho (1-\zeta_\mathrm{l})^{\rho-1} {_2F_1}(-\sigma+\rho-\eta, -\sigma+\rho+\eta;1-2\sigma;\zeta_\mathrm{l}) \right. \\
    & \left. -i \frac{2}{\tau} \zeta_\mathrm{l}^{-\sigma+1} (1-\zeta_\mathrm{l})^{\rho} \frac{(-\sigma+\rho-\eta) (-\sigma+\rho+\eta)}{1-2\sigma} {_2F_1}(-\sigma+\rho-\eta +1, -\sigma+\rho+\eta +1;1-2\sigma +1;\zeta_\mathrm{l}) \right. \\
    & \left. -\left(p-\frac{qA_2-qA_1 \zeta_\mathrm{l}}{1-\zeta_\mathrm{l}}\right)\zeta_\mathrm{l}^{-\sigma}(1-\zeta_\mathrm{l})^{\rho} {_2F_1}(-\sigma+\rho-\eta, -\sigma+\rho+\eta;1-2\sigma;\zeta_\mathrm{l}) \right].
\end{split}
\end{equation}

\subsection{Asymptotic Forms}
Being fundamentally different from its infinitely sharp counterpart in the vicinity of its transition, the smooth potential [Fig.~\ref{fig:A_pot_smooth}(a)] can obviously not be compared with the infinitely sharp potential in that region. Instead, the comparison must be carried out sufficiently far (in time) from the transition, and we should therefore resort to an asymptotic evaluation of the above exact solutions for the smooth potential step.

\subsubsection{Asymptotic Forms for the Earlier Medium (\texorpdfstring{$t<t_0$}{t<t0})}
\pati{Asymptotic Form of $\varphi_\mathrm{e}$}
The asymptotic regime for the earlier medium corresponds to the limit $t\rightarrow-\infty$, where $\zeta_\mathrm{e}\rightarrow{0}$ and $(1-\zeta_\mathrm{e})\rightarrow 1$, and $F(a,b;c;\zeta_\mathrm{e})\rightarrow 1$~\cite{gradshteyn2014table}. Inserting these limits into Eq.~\eqref{eq:phi_e_c1_c2} yields the asymptotic form of $\varphi_\mathrm{e}$ in terms of $\zeta_\mathrm{e}$,
\begin{equation}
    \varphi_\mathrm{e}(\zeta_\mathrm{e}\rightarrow 0) = C_1 \zeta_\mathrm{e}^{\mu} + C_2 \zeta_\mathrm{e}^{-\mu},
\end{equation}
and subsequently in terms of $t$
\begin{equation}\label{eq:phi_e_asym}
    \varphi_\mathrm{e}(t\rightarrow -\infty) = C_1 (-\mathrm{e}^{2\frac{t-t_0}{\tau}})^{i \sqrt{\alpha_1}} + C_2 (-\mathrm{e}^{2\frac{t-t_0}{\tau}})^{-i \sqrt{\alpha_1}},
\end{equation}
\pati{Asymptotic Form of $\vartheta_\mathrm{e}$}
while inserting Eq.~\eqref{eq:phi_e_asym} and Eq.~\eqref{eq:smooth_pot} into Eq.~\eqref{eq:theta_of_phi} yields
\begin{equation}\label{eq:theta_e_asym}
    \vartheta_\mathrm{e}(t\rightarrow -\infty)=\frac{C_1}{m}\left[-\frac{2}{\tau} \sqrt{\alpha_1} - (p-qA_1) \right] (-\mathrm{e}^{2\frac{t-t_0}{\tau}})^{i \sqrt{\alpha_1}} + \frac{C_2}{m}\left[\frac{2}{\tau} \sqrt{\alpha_1} - (p-qA_1) \right] (-\mathrm{e}^{2\frac{t-t_0}{\tau}})^{-i \sqrt{\alpha_1}}.
\end{equation}
Finally, since there is only the incident wave in the earlier medium, the coefficient $C_1$ must be zero, which reduces Eq.~\eqref{eq:phi_e_asym} and \eqref{eq:theta_e_asym} to
\begin{subequations}
\begin{equation}\label{eq:phi_e_asym_C2}
    \varphi_\mathrm{e}(t\rightarrow -\infty) = C_2 (-\mathrm{e}^{2\frac{t-t_0}{\tau}})^{-i \sqrt{\alpha_1}}
\end{equation}
and
\begin{equation}\label{eq:theta_e_asym_C2}
    \vartheta_\mathrm{e}(t\rightarrow -\infty)=\frac{C_2}{m}\left[\frac{2}{\tau} \sqrt{\alpha_1} - (p-qA_1) \right] (-\mathrm{e}^{2\frac{t-t_0}{\tau}})^{-i \sqrt{\alpha_1}}.
\end{equation}
\end{subequations}

\subsubsection{Asymptotic Forms for the Later Medium (\texorpdfstring{$t>t_0$}{t>t0})}
\pati{Asymptotic Form of $\varphi_\mathrm{l}$}
The asymptotic regime for the later medium corresponds to the limit $t\rightarrow \infty$, where $\zeta_\mathrm{l}\rightarrow{0}$ and $(1-\zeta_\mathrm{e})\rightarrow 1$, and $F(a,b;c;\zeta_\mathrm{l})\rightarrow 1$~\cite{gradshteyn2014table}. Inserting these limits into Eq.~\eqref{eq:phi_l_c3_c4} yields the asymptotic form of $\varphi_\mathrm{l}$ in terms of $\zeta_\mathrm{l}$,
\begin{equation}
    \varphi_\mathrm{l}(\zeta_\mathrm{l}\rightarrow 0) = C_3 \zeta_\mathrm{l}^{\sigma} + C_4 \zeta_\mathrm{l}^{-\sigma},
\end{equation}
and subsequently in terms of $t$
\begin{equation}\label{eq:phi_l_asym}
    \varphi_\mathrm{l}(t\rightarrow +\infty) = C_3 (-\mathrm{e}^{-2\frac{t-t_0}{\tau}})^{i \sqrt{\beta_2}} + C_4 (-\mathrm{e}^{-2\frac{t-t_0}{\tau}})^{-i \sqrt{\beta_2}},
\end{equation}
\pati{Asymptotic Form of $\vartheta_\mathrm{l}$}
while inserting Eq.~\eqref{eq:phi_l_asym} and Eq.~\eqref{eq:smooth_pot} into Eq.~\eqref{eq:theta_of_phi} yields
\begin{equation}\label{eq:theta_l_asym}
    \vartheta_\mathrm{l}(t\rightarrow +\infty)=\frac{C_3}{m}\left[\frac{2}{\tau} \sqrt{\beta_2} - (p-qA_2) \right] (-\mathrm{e}^{-2\frac{t-t_0}{\tau}})^{i \sqrt{\beta_2}} + \frac{C_4}{m}\left[-\frac{2}{\tau} \sqrt{\beta_2} - (p-qA_2) \right] (-\mathrm{e}^{-2\frac{t-t_0}{\tau}})^{-i \sqrt{\beta_2}}.
\end{equation}

\subsection{Weyl to Dirac Spinor Transformation}

For consistency with the mathematical forms of the spinors used throughout this paper, we transform our spinors from the Weyl representation to the Dirac representation. This ensures that our calculations remain consistent and comparable with the results established in the paper. This transformation~\cite{das2020lectures} may be written as
\begin{equation}\label{DWtransf}
    \psi(\mathrm{Dirac})=U\psi(\mathrm{Weyl}),
        \quad\text{where}\quad
        U=\frac{1}{\sqrt{2}}\begin{pmatrix}
            1 & 1\\
            1 & -1
        \end{pmatrix}.
\end{equation}
If we denote the Dirac and Weyl spinors as
\begin{equation}
    \psi(\mathrm{Dirac})=\begin{pmatrix}
        \varphi^\mathrm{D}\\
        \vartheta^\mathrm{D}
        \end{pmatrix}
    \quad\text{and}\quad
    \psi(\mathrm{Weyl})=\begin{pmatrix}
        \varphi\\
        \vartheta
    \end{pmatrix},
\end{equation}
we find using Eq.~\eqref{DWtransf} that the spinor in the Dirac representation in terms of the elements of the spinor in the Weyl representation read
\begin{equation}\label{eq:Weyl_to_Dirac_transform}
    \begin{pmatrix}
    \varphi^\mathrm{D}\\
    \vartheta^\mathrm{D}
    \end{pmatrix}= \frac{1}{\sqrt{2}}\begin{pmatrix}
    \varphi+\vartheta\\
    \varphi-\vartheta
    \end{pmatrix}.
\end{equation}

\pati{Weyl to Dirac for Asymptotic Form of $\varphi_\mathrm{e}$}
For the earlier medium, inserting Eq.~\eqref{eq:phi_e_asym_C2} and Eq.~\eqref{eq:theta_e_asym_C2} into the first row of Eq.~\eqref{eq:Weyl_to_Dirac_transform}, we find 
\begin{equation}\label{eq:phi_e_asym_Dirac_inter}
    \varphi_\mathrm{e}^\mathrm{D}(t\rightarrow -\infty)=\frac{1}{\sqrt{2}}\left(C_2 (-\mathrm{e}^{2\frac{t-t_0}{\tau}})^{-i \sqrt{\alpha_1}}+\frac{C_2}{m}\left[\frac{2}{\tau} \sqrt{\alpha_1} - (p-qA_1) \right] (-\mathrm{e}^{2\frac{t-t_0}{\tau}})^{-i \sqrt{\alpha_1}}\right),
\end{equation}
which, upon substituting Eq.~\eqref{eq:alpha_1} and using $(-1)^a=\mathrm{e}^{i\pi a}$, transforms to the the asymptotic form
\begin{equation}
    \varphi_\mathrm{e}^\mathrm{D}(t\rightarrow -\infty)=\frac{1}{\sqrt{2}}\frac{C_2}{m}\mathrm{e}^{\pi \frac{\tau}{2}E_1}\left(m + E_1 - (p-qA_1) \right)\mathrm{e}^{-i E_1(t-t_0)}.
\end{equation}

\pati{Weyl to Dirac for an Asymptotic Form of $\vartheta_\mathrm{e}$}
Similarly, inserting Eq.~\eqref{eq:phi_e_asym_C2} and Eq.~\eqref{eq:theta_e_asym_C2} into the second row of Eq.~\eqref{eq:Weyl_to_Dirac_transform}, then substituting Eq.~\eqref{eq:alpha_1} into the resulting expression and using $(-1)^a=\mathrm{e}^{i\pi a}$, we find
\begin{equation}
    \vartheta_\mathrm{e}^\mathrm{D}(t\rightarrow -\infty)=\frac{1}{\sqrt{2}}\frac{C_2}{m}\mathrm{e}^{\pi \frac{\tau}{2}E_1}\left(m - E_1 + (p-qA_1) \right)\mathrm{e}^{-i E_1(t-t_0)}.
\end{equation}
This equation may be algebraically manipulated so as to take a form including an expression that resembles the wave function for the infinitely sharp step [Eq.~\eqref{eq:s_vector_temporal_plane_wave_sol_before}], viz.,
\begin{equation}
    \vartheta_\mathrm{e}^\mathrm{D}(t\rightarrow -\infty)=\frac{1}{\sqrt{2}}\frac{C_2}{m}\mathrm{e}^{\pi \frac{\tau}{2}E_1}\left(m + E_1 - (p-qA_1) \right)\left(\frac{E_1-m}{p-qA_1}\right)\mathrm{e}^{-i E_1(t-t_0)}.
\end{equation}

Thus, the Dirac spinor wavefunction for the earlier medium is
\begin{equation}\label{eq:psi_e_asym_Dirac}
    \psi_\mathrm{e}^\mathrm{D}(t\rightarrow -\infty,z)=
        \begin{pmatrix}
            \varphi_\mathrm{e}^\mathrm{D}(t\rightarrow -\infty) \\
            \vartheta_\mathrm{e}^\mathrm{D}(t\rightarrow -\infty)
        \end{pmatrix}
        e^{i p z}=\frac{1}{\sqrt{2}}\frac{C_2}{m}\mathrm{e}^{\pi \frac{\tau}{2}E_1}\left(m + E_1 - (p-qA_1) \right)
        \begin{pmatrix}
            1 \\
            \frac{E_1-m}{p-qA_1}
        \end{pmatrix}
        \mathrm{e}^{-i E_1(t-t_0)} e^{i p z}.
\end{equation}

\pati{Weyl to Dirac for Asymptotic Form of $\varphi_\mathrm{l}$}
For the later medium, inserting Eq.~\eqref{eq:phi_l_asym} and Eq.~\eqref{eq:theta_l_asym} into the first row of Eq.~\eqref{eq:Weyl_to_Dirac_transform}, we find 
\begin{equation}\label{eq:phi_l_asym_Dirac_inter}
\begin{split}
    \varphi_\mathrm{l}^\mathrm{D}(t\rightarrow +\infty)=\frac{1}{\sqrt{2}} & \left(C_3 (-\mathrm{e}^{-2\frac{t-t_0}{\tau}})^{i \sqrt{\beta_2}} + C_4 (-\mathrm{e}^{-2\frac{t-t_0}{\tau}})^{-i \sqrt{\beta_2}} \right. \\
    & \left. + \frac{C_3}{m}\left[\frac{2}{\tau} \sqrt{\beta_2} - (p-qA_2) \right] (-\mathrm{e}^{-2\frac{t-t_0}{\tau}})^{i \sqrt{\beta_2}} + \frac{C_4}{m}\left[-\frac{2}{\tau} \sqrt{\beta_2} - (p-qA_2) \right] (-\mathrm{e}^{-2\frac{t-t_0}{\tau}})^{-i \sqrt{\beta_2}}\right),
\end{split}
\end{equation}
which, upon substituting Eq.~\eqref{eq:beta_2} and using $(-1)^a=\mathrm{e}^{i\pi a}$, transforms to the the asymptotic form
\begin{equation}
\begin{split}
    \varphi_\mathrm{l}^\mathrm{D}(t\rightarrow +\infty) = & \frac{1}{\sqrt{2}} \frac{C_3}{m} \mathrm{e}^{-\pi \frac{\tau}{2}E_2} \left(m + E_2 - (p-qA_2) \right) \mathrm{e}^{-i E_2(t-t_0)} \\
    &  + \frac{1}{\sqrt{2}} \frac{C_4}{m} \mathrm{e}^{\pi \frac{\tau}{2}E_2} \left(m - E_2 - (p-qA_2) \right) \mathrm{e}^{i E_2(t-t_0)}.
\end{split}
\end{equation}

\pati{Weyl to Dirac for Asymptotic Form of $\vartheta_\mathrm{l}$}
Similarly, inserting Eq.~\eqref{eq:phi_l_asym} and Eq.~\eqref{eq:theta_l_asym} into the second row of Eq.~\eqref{eq:Weyl_to_Dirac_transform}, then substituting Eq.~\eqref{eq:alpha_1} into the resulting expression and using $(-1)^a=\mathrm{e}^{i\pi a}$, we find
\begin{equation}
\begin{split}
    \vartheta_\mathrm{l}^\mathrm{D}(t\rightarrow +\infty) = & \frac{1}{\sqrt{2}} \frac{C_3}{m} \mathrm{e}^{-\pi \frac{\tau}{2}E_2} \left(m - E_2 + (p-qA_2) \right) \mathrm{e}^{-i E_2(t-t_0)} \\
    &  + \frac{1}{\sqrt{2}} \frac{C_4}{m} \mathrm{e}^{\pi \frac{\tau}{2}E_2} \left(m + E_2 + (p-qA_2) \right) \mathrm{e}^{i E_2(t-t_0)}.
\end{split}
\end{equation}
This equation may be algebraically manipulated so as to take a form including an expression that resembles the wave function
for the infinitely sharp step [Eq.~\eqref{eq:s_vector_temporal_plane_wave_sol_after}], viz.,
\begin{equation}
\begin{split}
    \vartheta_\mathrm{l}^\mathrm{D}(t\rightarrow +\infty) = & \frac{1}{\sqrt{2}} \frac{C_3}{m} \mathrm{e}^{-\pi \frac{\tau}{2}E_2} \left(m + E_2 - (p-qA_2) \right)\left(\frac{E_2-m}{p-qA_2}\right)\mathrm{e}^{-i E_2(t-t_0)} \\
    &  + \frac{1}{\sqrt{2}} \frac{C_4}{m} \mathrm{e}^{\pi \frac{\tau}{2}E_2} \left(m - E_2 - (p-qA_2) \right)\left(\frac{-E_2-m}{p-qA_2}\right)\mathrm{e}^{i E_2(t-t_0)},
\end{split}
\end{equation}

Thus, the Dirac spinor wavefunction for the later medium is
\begin{equation}\label{eq:psi_l_asym_Dirac}
    \begin{split}
        \psi_\mathrm{l}^\mathrm{D}(t\rightarrow +\infty,z)=
        \begin{pmatrix}
            \varphi_\mathrm{l}^\mathrm{D}(t\rightarrow +\infty) \\
            \vartheta_\mathrm{l}^\mathrm{D}(t\rightarrow +\infty)
        \end{pmatrix}
        e^{i p z}
        & =\frac{1}{\sqrt{2}} \frac{C_3}{m} \mathrm{e}^{-\pi \frac{\tau}{2}E_2} \left(m + E_2 - (p-qA_2) \right)
        \begin{pmatrix}
            1 \\
            \frac{E_2-m}{p-qA_2}
        \end{pmatrix}
        \mathrm{e}^{-i E_2(t-t_0)} e^{i p z} \\
        & + \frac{1}{\sqrt{2}} \frac{C_4}{m} \mathrm{e}^{\pi \frac{\tau}{2}E_2} \left(m - E_2 - (p-qA_2) \right)
        \begin{pmatrix}
            1 \\
            \frac{-E_2-m}{p-qA_2}
        \end{pmatrix}
        \mathrm{e}^{i E_2(t-t_0)} e^{i p z}.
        \end{split}
\end{equation}

\pati{Weyl to Dirac for $\varphi_\mathrm{e}$}
Inserting Eq.~\eqref{eq:phi_e_c2} and Eq.~\eqref{eq:theta_e_C2} into the first row of Eq.~\eqref{eq:Weyl_to_Dirac_transform}, we find the \emph{general form} of $\varphi_\mathrm{e}$ in the Dirac representation as
\begin{equation}\label{eq:phi_e_C2_Dirac}
\begin{split}
    \varphi_\mathrm{e}^\mathrm{D} =\frac{1}{\sqrt{2}}\frac{C_2}{m} & \left[m \zeta_\mathrm{e}^{-\mu}(1-\zeta_\mathrm{e})^{\nu} {_2F_1}(-\mu+\nu-\lambda, -\mu+\nu+\lambda;1-2\mu;\zeta_\mathrm{e}) \right. \\ 
    & \left. -i \frac{2}{\tau} \mu \zeta_\mathrm{e}^{-\mu}(1-\zeta_\mathrm{e})^\nu {_2F_1}(-\mu+\nu-\lambda, -\mu+\nu+\lambda;1-2\mu;\zeta_\mathrm{e}) \right. \\
    & \left. - i \frac{2}{\tau}  \zeta_\mathrm{e}^{-\mu+1} \nu (1-\zeta_\mathrm{e})^{\nu-1} {_2F_1}(-\mu+\nu-\lambda, -\mu+\nu+\lambda;1-2\mu;\zeta_\mathrm{e}) \right.\\
    & \left. + i \frac{2}{\tau}  \zeta_\mathrm{e}^{-\mu+1} (1-\zeta_\mathrm{e})^{\nu} \frac{(-\mu+\nu-\lambda) (-\mu+\nu+\lambda)}{1-2\mu} {_2F_1}(-\mu+\nu-\lambda+1, -\mu+\nu+\lambda+1;1-2\mu+1;\zeta_\mathrm{e}) \right. \\
    & \left. -\left(p-\frac{qA_1-qA_2 \zeta_\mathrm{e}}{1-\zeta_\mathrm{e}}\right)  \zeta_\mathrm{e}^{-\mu}(1-\zeta_\mathrm{e})^{\nu} {_2F_1}(-\mu+\nu-\lambda, -\mu+\nu+\lambda;1-2\mu;\zeta_\mathrm{e}) \right].
\end{split}
\end{equation}

\pati{Weyl to Dirac for $\vartheta_\mathrm{e}$}
Similarly, inserting Eq.~\eqref{eq:phi_e_c2} and Eq.~\eqref{eq:theta_e_C2} into the second row of Eq.~\eqref{eq:Weyl_to_Dirac_transform},  we find the general form of $\vartheta_\mathrm{e}$ in the Dirac representation as
\begin{equation}\label{eq:theta_e_C2_Dirac}
\begin{split}
    \vartheta_\mathrm{e}^\mathrm{D} =\frac{1}{\sqrt{2}}\frac{C_2}{m} & \left[m \zeta_\mathrm{e}^{-\mu}(1-\zeta_\mathrm{e})^{\nu} {_2F_1}(-\mu+\nu-\lambda, -\mu+\nu+\lambda;1-2\mu;\zeta_\mathrm{e}) \right. \\ 
    & \left. +i \frac{2}{\tau} \mu \zeta_\mathrm{e}^{-\mu}(1-\zeta_\mathrm{e})^\nu {_2F_1}(-\mu+\nu-\lambda, -\mu+\nu+\lambda;1-2\mu;\zeta_\mathrm{e}) \right. \\
    & \left. + i \frac{2}{\tau}  \zeta_\mathrm{e}^{-\mu+1} \nu (1-\zeta_\mathrm{e})^{\nu-1} {_2F_1}(-\mu+\nu-\lambda, -\mu+\nu+\lambda;1-2\mu;\zeta_\mathrm{e}) \right.\\
    & \left. - i \frac{2}{\tau}  \zeta_\mathrm{e}^{-\mu+1} (1-\zeta_\mathrm{e})^{\nu} \frac{(-\mu+\nu-\lambda) (-\mu+\nu+\lambda)}{1-2\mu} {_2F_1}(-\mu+\nu-\lambda+1, -\mu+\nu+\lambda+1;1-2\mu+1;\zeta_\mathrm{e}) \right. \\
    & \left. + \left(p-\frac{qA_1-qA_2 \zeta_\mathrm{e}}{1-\zeta_\mathrm{e}}\right)  \zeta_\mathrm{e}^{-\mu}(1-\zeta_\mathrm{e})^{\nu} {_2F_1}(-\mu+\nu-\lambda, -\mu+\nu+\lambda;1-2\mu;\zeta_\mathrm{e}) \right].
\end{split}
\end{equation}

\pati{Weyl to Dirac for $\varphi_\mathrm{l}$}
Still similarly, inserting Eq.~\eqref{eq:phi_l_c3_c4} and Eq.~\eqref{eq:theta_l_c3_c4} into the first row of Eq.~\eqref{eq:Weyl_to_Dirac_transform}, we find the general form of $\varphi_\mathrm{l}$ in the Dirac representation as
\begin{equation}\label{eq:phi_l_C3_C4_Dirac}
\begin{split}
    \varphi_\mathrm{l}^\mathrm{D} =\frac{1}{\sqrt{2}}\frac{C_3}{m} & \left[m \zeta_\mathrm{l}^{\sigma}(1-\zeta_\mathrm{l})^{\rho} {_2F_1}(\sigma+\rho-\eta, \sigma+\rho+\eta;1+2\sigma;\zeta_\mathrm{l}) \right. \\
    & \left. -i \frac{2}{\tau}   \sigma \zeta_\mathrm{l}^{\sigma}(1-\zeta_\mathrm{l})^\rho {_2F_1}(\sigma+\rho-\eta, \sigma+\rho+\eta;1+2\sigma;\zeta_\mathrm{l}) \right. \\
    & \left. +i \frac{2}{\tau}  \zeta_\mathrm{l}^{\sigma+1} \rho (1-\zeta_\mathrm{l})^{\rho-1} {_2F_1}(\sigma+\rho-\eta, \sigma+\rho+\eta;1+2\sigma;\zeta_\mathrm{l}) \right. \\
    & \left. -i \frac{2}{\tau} \zeta_\mathrm{l}^{\sigma+1} (1-\zeta_\mathrm{l})^{\rho} \frac{(\sigma+\rho-\eta) (\sigma+\rho+\eta)}{1-2\sigma} {_2F_1}(\sigma+\rho-\eta +1, \sigma+\rho+\eta +1;1+2\sigma +1;\zeta_\mathrm{l}) \right. \\
    & \left. -\left(p-\frac{qA_2-qA_1 \zeta_\mathrm{l}}{1-\zeta_\mathrm{l}}\right)\zeta_\mathrm{l}^{\sigma}(1-\zeta_\mathrm{l})^{\rho} {_2F_1}(\sigma+\rho-\eta, \sigma+\rho+\eta;1+2\sigma;\zeta_\mathrm{l}) \right]  \\
     +\frac{1}{\sqrt{2}}\frac{C_4}{m} & \left[m \zeta_\mathrm{l}^{-\sigma}(1-\zeta_\mathrm{l})^{\rho} {_2F_1}(-\sigma+\rho-\eta, -\sigma+\rho+\eta;1-2\sigma;\zeta_\mathrm{l}) \right. \\
     & \left. + i \frac{2}{\tau} \sigma \zeta_\mathrm{l}^{-\sigma}(1-\zeta_\mathrm{l})^\rho {_2F_1}(-\sigma+\rho-\eta, -\sigma+\rho+\eta;1-2\sigma;\zeta_\mathrm{l}) \right. \\
    & \left. +i \frac{2}{\tau} \zeta_\mathrm{l}^{-\sigma+1} \rho (1-\zeta_\mathrm{l})^{\rho-1} {_2F_1}(-\sigma+\rho-\eta, -\sigma+\rho+\eta;1-2\sigma;\zeta_\mathrm{l}) \right. \\
    & \left. -i \frac{2}{\tau} \zeta_\mathrm{l}^{-\sigma+1} (1-\zeta_\mathrm{l})^{\rho} \frac{(-\sigma+\rho-\eta) (-\sigma+\rho+\eta)}{1-2\sigma} {_2F_1}(-\sigma+\rho-\eta +1, -\sigma+\rho+\eta +1;1-2\sigma +1;\zeta_\mathrm{l}) \right. \\
    & \left. -\left(p-\frac{qA_2-qA_1 \zeta_\mathrm{l}}{1-\zeta_\mathrm{l}}\right)\zeta_\mathrm{l}^{-\sigma}(1-\zeta_\mathrm{l})^{\rho} {_2F_1}(-\sigma+\rho-\eta, -\sigma+\rho+\eta;1-2\sigma;\zeta_\mathrm{l}) \right],
\end{split}
\end{equation}

\pati{Weyl to Dirac for $\vartheta_\mathrm{l}$}
and finally inserting Eq.~\eqref{eq:phi_l_c3_c4} and Eq.~\eqref{eq:theta_l_c3_c4} into the second row of Eq.~\eqref{eq:Weyl_to_Dirac_transform}, we find the general form of $\vartheta_\mathrm{l}$ in the Dirac representation as
\begin{equation}\label{eq:theta_l_C3_C4_Dirac}
\begin{split}
    \vartheta_\mathrm{l}^\mathrm{D} =\frac{1}{\sqrt{2}}\frac{C_3}{m} & \left[m \zeta_\mathrm{l}^{\sigma}(1-\zeta_\mathrm{l})^{\rho} {_2F_1}(\sigma+\rho-\eta, \sigma+\rho+\eta;1+2\sigma;\zeta_\mathrm{l}) \right. \\
    & \left. +i \frac{2}{\tau} \sigma \zeta_\mathrm{l}^{\sigma}(1-\zeta_\mathrm{l})^\rho {_2F_1}(\sigma+\rho-\eta, \sigma+\rho+\eta;1+2\sigma;\zeta_\mathrm{l}) \right. \\
    & \left. -i \frac{2}{\tau}  \zeta_\mathrm{l}^{\sigma+1} \rho (1-\zeta_\mathrm{l})^{\rho-1} {_2F_1}(\sigma+\rho-\eta, \sigma+\rho+\eta;1+2\sigma;\zeta_\mathrm{l}) \right. \\
    & \left. +i \frac{2}{\tau} \zeta_\mathrm{l}^{\sigma+1} (1-\zeta_\mathrm{l})^{\rho} \frac{(\sigma+\rho-\eta) (\sigma+\rho+\eta)}{1-2\sigma} {_2F_1}(\sigma+\rho-\eta +1, \sigma+\rho+\eta +1;1+2\sigma +1;\zeta_\mathrm{l}) \right. \\
    & \left. +\left(p-\frac{qA_2-qA_1 \zeta_\mathrm{l}}{1-\zeta_\mathrm{l}}\right)\zeta_\mathrm{l}^{\sigma}(1-\zeta_\mathrm{l})^{\rho} {_2F_1}(\sigma+\rho-\eta, \sigma+\rho+\eta;1+2\sigma;\zeta_\mathrm{l}) \right]  \\
     +\frac{1}{\sqrt{2}} \frac{C_4}{m} & \left[m \zeta_\mathrm{l}^{-\sigma}(1-\zeta_\mathrm{l})^{\rho} {_2F_1}(-\sigma+\rho-\eta, -\sigma+\rho+\eta;1-2\sigma;\zeta_\mathrm{l}) \right. \\
     & \left. - i \frac{2}{\tau} \sigma \zeta_\mathrm{l}^{-\sigma}(1-\zeta_\mathrm{l})^\rho {_2F_1}(-\sigma+\rho-\eta, -\sigma+\rho+\eta;1-2\sigma;\zeta_\mathrm{l}) \right. \\
    & \left. - i \frac{2}{\tau} \zeta_\mathrm{l}^{-\sigma+1} \rho (1-\zeta_\mathrm{l})^{\rho-1} {_2F_1}(-\sigma+\rho-\eta, -\sigma+\rho+\eta;1-2\sigma;\zeta_\mathrm{l}) \right. \\
    & \left. + i \frac{2}{\tau} \zeta_\mathrm{l}^{-\sigma+1} (1-\zeta_\mathrm{l})^{\rho} \frac{(-\sigma+\rho-\eta) (-\sigma+\rho+\eta)}{1-2\sigma} {_2F_1}(-\sigma+\rho-\eta +1, -\sigma+\rho+\eta +1;1-2\sigma +1;\zeta_\mathrm{l}) \right. \\
    & \left. + \left(p-\frac{qA_2-qA_1 \zeta_\mathrm{l}}{1-\zeta_\mathrm{l}}\right)\zeta_\mathrm{l}^{-\sigma}(1-\zeta_\mathrm{l})^{\rho} {_2F_1}(-\sigma+\rho-\eta, -\sigma+\rho+\eta;1-2\sigma;\zeta_\mathrm{l}) \right].
\end{split}
\end{equation}

\subsection{Boundary Conditions}
At this point, we need to determine the coefficient ratios  $C_3/C_2$ and $C_4/C_2$ to be able to calculate probability coefficients. For this purpose, we apply the boundary conditions to the wavefunctions in the general forms. Let us first write the Dirac wavefunctions in (Eqs.~\eqref{eq:phi_e_C2_Dirac}, \eqref{eq:phi_l_C3_C4_Dirac}, \eqref{eq:theta_e_C2_Dirac}, and \eqref{eq:theta_l_C3_C4_Dirac}) at $t=t_0$, which will imply the use of the following relations:
\begin{itemize}
    \item at $t=t_0$, $\zeta_\mathrm{e}=\zeta_\mathrm{l}=-1$ since $\zeta_\mathrm{e} = -\mathrm{e}^{2\frac{t-t_0}{\tau}}$ and $\zeta_\mathrm{l} = -\mathrm{e}^{-2\frac{t-t_0}{\tau}}$,
    \item $(-1)^a=\mathrm{e}^{i\pi a}$ and $\mathrm{e}^{i\pi}=-1$,
    \item $\rho=\nu$, $\eta=\mu$, and $\sigma=\lambda$, according to Eqs.~\eqref{eq:mu_E1}, \eqref{eq:nu_A2_A1}, \eqref{eq:lambda_E2}, \eqref{eq:sigma_E2}, \eqref{eq:rho_A2,A1} and~\eqref{eq:eta_E1}.
\end{itemize}

\pati{Dirac Wavefunctions at $t=t_0$}
At $t=t_0$, the wavefunctions in Eqs.~\eqref{eq:phi_e_C2_Dirac}, \eqref{eq:theta_e_C2_Dirac}, \eqref{eq:phi_l_C3_C4_Dirac}, and \eqref{eq:theta_l_C3_C4_Dirac} may be written as
\begin{subequations}
\begin{equation}\label{eq:phi_e_C2_Dirac_t_0}
    \varphi_\mathrm{e}^\mathrm{D}(t=t_0) =\frac{2^{\nu-1} }{\sqrt{2}}\frac{C_2}{m} \mathrm{e}^{\pi\frac{\tau}{2}E_1} \left[ \left(2m + \mathcal{D}_1 \right) \mathcal{F}_1 - \mathcal{D}_2 \mathcal{F}_2 \right],
\end{equation}
\begin{equation}\label{eq:phi_l_C3_c4_Dirac_t_0}
    \varphi_\mathrm{l}^\mathrm{D}(t=t_0) =\frac{2^{\nu-1}}{\sqrt{2}}\frac{C_3}{m} \mathrm{e}^{-\pi\frac{\tau}{2}E_2}  \left[\left(2m + \mathcal{D}_3 \right)  \mathcal{F}_3 + \mathcal{D}_4 \mathcal{F}_4 \right] 
     +\frac{2^{\nu-1}}{\sqrt{2}}\frac{C_4}{m} \mathrm{e}^{\pi\frac{\tau}{2}E_2} \left[\left(2m - \mathcal{D}_5  \right) \mathcal{F}_5 + \mathcal{D}_6 \mathcal{F}_6  \right],
\end{equation}
\begin{equation}\label{eq:theta_e_C2_Dirac_t_0}
    \vartheta_\mathrm{e}^\mathrm{D}(t=t_0) =\frac{2^{\nu-1}}{\sqrt{2}}\frac{C_2}{m}  \mathrm{e}^{\pi\frac{\tau}{2}E_1}\left[\left(2m - \mathcal{D}_1\right)  \mathcal{F}_1 + \mathcal{D}_2 \mathcal{F}_2  \right],
\end{equation}
\begin{equation}\label{eq:theta_l_C3_C4_Dirac_t_0}
    \vartheta_\mathrm{l}^\mathrm{D}(t=t_0) =\frac{2^{\nu-1}}{\sqrt{2}}\frac{C_3}{m} \mathrm{e}^{-\pi\frac{\tau}{2}E_2} \left[\left(2m - \mathcal{D}_3\right) \mathcal{F}_3 - \mathcal{D}_4 \mathcal{F}_4 \right]
     +\frac{2^{\nu-1}}{\sqrt{2}} \frac{C_4}{m} \mathrm{e}^{\pi\frac{\tau}{2}E_2} \left[\left(2m + \mathcal{D}_5 \right) \mathcal{F}_5 - \mathcal{D}_6 \mathcal{F}_6  \right],
\end{equation}
\end{subequations}
where
\begin{subequations}
    \begin{equation}
        \mathcal{F}_1={_2F_1}(-\mu+\nu-\lambda, -\mu+\nu+\lambda;1-2\mu;-1),
    \end{equation}
    \begin{equation}
        \mathcal{F}_2={_2F_1}(-\mu+\nu-\lambda+1, -\mu+\nu+\lambda+1;1-2\mu+1;-1),
    \end{equation}
    \begin{equation}
        \mathcal{F}_3={_2F_1}(\lambda+\nu-\mu, \lambda+\nu+\mu;1+2\lambda;-1),
    \end{equation}
    \begin{equation}
        \mathcal{F}_4={_2F_1}(\lambda+\nu-\mu +1, \lambda+\nu+\mu +1;1+2\lambda +1;-1),
    \end{equation}
    \begin{equation}
        \mathcal{F}_5={_2F_1}(-\lambda+\nu-\mu, -\lambda+\nu+\mu;1-2\lambda;-1),
    \end{equation}
    \begin{equation}
        \mathcal{F}_6={_2F_1}(-\lambda+\nu-\mu +1, -\lambda+\nu+\mu +1;1-2\lambda +1;-1)
    \end{equation}
\end{subequations}
and
\begin{subequations}
    \begin{equation}
    \mathcal{D}_1 = 2 E_1 - (qA_2-qA_1) -\left(p-qA_2\right) -\left(p-qA_1\right),
\end{equation}
\begin{equation}
    \mathcal{D}_2 = i \frac{4}{\tau} \frac{(-\mu+\nu-\lambda) (-\mu+\nu+\lambda)}{1-2\mu},
\end{equation}
\begin{equation}
    \mathcal{D}_3 = 2 E_2 + (qA_2-qA_1) - \left(p-qA_2\right) - \left(p-qA_1\right),
\end{equation}
\begin{equation}
    \mathcal{D}_4 = i \frac{4}{\tau} \frac{(\lambda+\nu-\mu) (\lambda+\nu+\mu)}{1-2\lambda},
\end{equation}
\begin{equation}
    \mathcal{D}_5 = 2 E_2 - (qA_2-qA_1) + \left(p-qA_2\right) + \left(p-qA_1\right),
\end{equation}
\begin{equation}
    \mathcal{D}_6 = i \frac{4}{\tau} \frac{(-\lambda+\nu-\mu) (-\lambda+\nu+\mu)}{1-2\lambda}.
\end{equation}
\end{subequations}

\pati{Coefficients by Matching Wavefunctions at $t=t_0$}
We can now apply the boundary conditions to the wave functions~\eqref{eq:phi_e_C2_Dirac_t_0} to \eqref{eq:phi_l_C3_c4_Dirac_t_0}, and \eqref{eq:theta_e_C2_Dirac_t_0} to \eqref{eq:theta_l_C3_C4_Dirac_t_0} as
\begin{subequations}
    \begin{equation}
    \varphi_\mathrm{e}^\mathrm{D}(t=t_0) = \varphi_\mathrm{l}^\mathrm{D}(t=t_0)
    \end{equation}
    and
    \begin{equation}
    \vartheta_\mathrm{e}^\mathrm{D}(t=t_0) = \vartheta_\mathrm{l}^\mathrm{D}(t=t_0),
    \end{equation}
\end{subequations}
which yields
\begin{subequations}
    \begin{equation}
    \begin{split}
        \mathrm{e}^{\pi\frac{\tau}{2}E_1} \left[ \left(2m + \mathcal{D}_1 \right) \mathcal{F}_1 - \mathcal{D}_2 \mathcal{F}_2 \right]
        & = \frac{C_3}{C_2} \mathrm{e}^{-\pi\frac{\tau}{2}E_2}  \left[\left(2m + \mathcal{D}_3 \right)  \mathcal{F}_3 + \mathcal{D}_4 \mathcal{F}_4 \right] \\
        & + \frac{C_4}{C_2} \mathrm{e}^{\pi\frac{\tau}{2}E_2} \left[\left(2m - \mathcal{D}_5  \right) \mathcal{F}_5 + \mathcal{D}_6 \mathcal{F}_6  \right]
    \end{split} 
    \end{equation}
    and
    \begin{equation}
    \begin{split}
         \mathrm{e}^{\pi\frac{\tau}{2}E_1} \left[\left(2m - \mathcal{D}_1\right) \mathcal{F}_1 + \mathcal{D}_2 \mathcal{F}_2  \right]
         & = \frac{C_3}{C_2} \mathrm{e}^{-\pi\frac{\tau}{2}E_2} \left[\left(2m - \mathcal{D}_3\right) \mathcal{F}_3 - \mathcal{D}_4 \mathcal{F}_4 \right]  \\
         & + \frac{C_4}{C_2} \mathrm{e}^{\pi\frac{\tau}{2}E_2} \left[\left(2m + \mathcal{D}_5 \right) \mathcal{F}_5 - \mathcal{D}_6 \mathcal{F}_6 \right],
    \end{split}
    \end{equation}
\end{subequations}
which resolve to
\begin{subequations}
    \begin{equation}
        \frac{C_3}{C_2} = \mathrm{e}^{\pi\frac{\tau}{2}(E_1+E_2)}\frac{  \left[\left(2m - \mathcal{D}_1\right) \mathcal{F}_1 + \mathcal{D}_2 \mathcal{F}_2  \right] \left[\left(2m - \mathcal{D}_5  \right) \mathcal{F}_5 + \mathcal{D}_6 \mathcal{F}_6  \right] - \left[ \left(2m + \mathcal{D}_1 \right) \mathcal{F}_1 - \mathcal{D}_2 \mathcal{F}_2 \right]  \left[\left(2m + \mathcal{D}_5 \right) \mathcal{F}_5 - \mathcal{D}_6 \mathcal{F}_6  \right]}{ \left[\left(2m - \mathcal{D}_3\right) \mathcal{F}_3 - \mathcal{D}_4 \mathcal{F}_4 \right] \left[\left(2m - \mathcal{D}_5  \right) \mathcal{F}_5 + \mathcal{D}_6 \mathcal{F}_6  \right]-\left[\left(2m + \mathcal{D}_3 \right)  \mathcal{F}_3 + \mathcal{D}_4 \mathcal{F}_4 \right] \left[\left(2m + \mathcal{D}_5 \right) \mathcal{F}_5 - \mathcal{D}_6 \mathcal{F}_6  \right]}
    \end{equation}
    and
    \begin{equation}
         \frac{C_4}{C_2} = \mathrm{e}^{\pi\frac{\tau}{2}(E_1-E_2)}\frac{ \left[ \left(2m + \mathcal{D}_1 \right) \mathcal{F}_1 - \mathcal{D}_2 \mathcal{F}_2 \right]  \left[\left(2m - \mathcal{D}_3\right) \mathcal{F}_3 - \mathcal{D}_4 \mathcal{F}_4 \right] -  \left[\left(2m - \mathcal{D}_1\right) \mathcal{F}_1 + \mathcal{D}_2 \mathcal{F}_2  \right]\left[\left(2m + \mathcal{D}_3 \right)  \mathcal{F}_3 + \mathcal{D}_4 \mathcal{F}_4 \right]}{ \left[\left(2m - \mathcal{D}_3\right) \mathcal{F}_3 - \mathcal{D}_4 \mathcal{F}_4 \right]\left[\left(2m - \mathcal{D}_5  \right) \mathcal{F}_5 + \mathcal{D}_6 \mathcal{F}_6  \right]- \left[\left(2m + \mathcal{D}_3 \right)  \mathcal{F}_3 + \mathcal{D}_4 \mathcal{F}_4 \right] \left[\left(2m + \mathcal{D}_5 \right) \mathcal{F}_5 - \mathcal{D}_6 \mathcal{F}_6 \right]}.
    \end{equation}
\end{subequations}

\subsection{Asymptotic Spinor Wavefunctions and Scattering Probabilities}
We shall finally calculate here the sought after scattering probabilities. For this purpose, we first recall the asymptotic Dirac spinors in Eqs.~\eqref{eq:psi_e_asym_Dirac} and \eqref{eq:psi_l_asym_Dirac}:
\begin{subequations}\label{eq:psi_e_l_asym_Dirac}
\begin{equation}
    \psi_\mathrm{e}^\mathrm{D}(t\rightarrow -\infty,z) = G_\mathrm{i} \begin{pmatrix}
            1 \\
            \frac{E_1-m}{p-qA_1}
        \end{pmatrix}
        \mathrm{e}^{-i E_1(t-t_0)} e^{i p z}
\end{equation}
and
\begin{equation}
    \begin{split}
        \psi_\mathrm{l}^\mathrm{D}(t\rightarrow +\infty,z) = G_\mathrm{f} \begin{pmatrix}
            1 \\
            \frac{E_2-m}{p-qA_2}
        \end{pmatrix}
        \mathrm{e}^{-i E_2(t-t_0)} e^{i p z} 
        + G_\mathrm{b} \begin{pmatrix}
            1 \\
            \frac{-E_2-m}{p-qA_2}
        \end{pmatrix}
        \mathrm{e}^{i E_2(t-t_0)} e^{i p z},
        \end{split}
\end{equation}
where
    \begin{equation}\label{eq:Gi}
        G_\mathrm{i} = \frac{1}{\sqrt{2}}\frac{C_2}{m}\mathrm{e}^{\pi \frac{\tau}{2}E_1}\left(m + E_1 - (p-qA_1) \right),
    \end{equation}
    \begin{equation}\label{eq:Gf}
        G_\mathrm{f} = \frac{1}{\sqrt{2}} \frac{C_3}{m} \mathrm{e}^{-\pi \frac{\tau}{2}E_2} \left(m + E_2 - (p-qA_2) \right),
    \end{equation}
    and
    \begin{equation}\label{eq:Gb}
        G_\mathrm{b} = \frac{1}{\sqrt{2}} \frac{C_4}{m} \mathrm{e}^{\pi \frac{\tau}{2}E_2} \left(m - E_2 - (p-qA_2) \right).
    \end{equation}
\end{subequations}

The later-forward and later-backward scattering coefficients may then directly be obtained from Eqs.~\eqref{eq:Gi}, \eqref{eq:Gf} and~\eqref{eq:Gb} according to their definitions, as
\begin{subequations}
    \begin{equation}
        f = \frac{G_\mathrm{f}}{G_\mathrm{i}} = \frac{\frac{1}{\sqrt{2}} \frac{C_3}{m} \mathrm{e}^{-\pi \frac{\tau}{2}E_2} \left(m + E_2 - (p-qA_2) \right)}{\frac{1}{\sqrt{2}}\frac{C_2}{m}\mathrm{e}^{\pi \frac{\tau}{2}E_1}\left(m + E_1 - (p-qA_1) \right)} = \frac{C_3}{C_2} \mathrm{e}^{-\pi\frac{\tau}{2}(E_1+E_2)} \frac{\left(m + E_2 - (p-qA_2) \right)}{\left(m + E_1 - (p-qA_1) \right)}
    \end{equation}
    and
    \begin{equation}
        b = \frac{G_\mathrm{b}}{G_\mathrm{i}} = \frac{\frac{1}{\sqrt{2}} \frac{C_4}{m} \mathrm{e}^{\pi \frac{\tau}{2}E_2} \left(m - E_2 - (p-qA_2) \right)}{\frac{1}{\sqrt{2}}\frac{C_2}{m}\mathrm{e}^{\pi \frac{\tau}{2}E_1}\left(m + E_1 - (p-qA_1) \right)} = \frac{C_4}{C_2} \mathrm{e}^{-\pi\frac{\tau}{2}(E_1-E_2)} \frac{\left(m - E_2 - (p-qA_2) \right)}{\left(m + E_1 - (p-qA_1) \right)},
    \end{equation}
\end{subequations}
which leads, using the approach in Sec.~\ref{subsec:solution_Dirac_temporal_vector}, to the probabilities
\begin{subequations}
    \begin{equation}
        F = \frac{f^2}{b^2+f^2}
    \end{equation}
    and
    \begin{equation}
        B = \frac{b^2}{b^2+f^2},
    \end{equation}
\end{subequations}
where
\begin{subequations}
    \begin{equation}
        \frac{C_3}{C_2} = \mathrm{e}^{\pi\frac{\tau}{2}(E_1+E_2)}\frac{  \left[\left(2m - \mathcal{D}_1\right) \mathcal{F}_1 + \mathcal{D}_2 \mathcal{F}_2  \right] \left[\left(2m - \mathcal{D}_5  \right) \mathcal{F}_5 + \mathcal{D}_6 \mathcal{F}_6  \right] - \left[ \left(2m + \mathcal{D}_1 \right) \mathcal{F}_1 - \mathcal{D}_2 \mathcal{F}_2 \right]  \left[\left(2m + \mathcal{D}_5 \right) \mathcal{F}_5 - \mathcal{D}_6 \mathcal{F}_6  \right]}{ \left[\left(2m - \mathcal{D}_3\right) \mathcal{F}_3 - \mathcal{D}_4 \mathcal{F}_4 \right] \left[\left(2m - \mathcal{D}_5  \right) \mathcal{F}_5 + \mathcal{D}_6 \mathcal{F}_6  \right]-\left[\left(2m + \mathcal{D}_3 \right)  \mathcal{F}_3 + \mathcal{D}_4 \mathcal{F}_4 \right] \left[\left(2m + \mathcal{D}_5 \right) \mathcal{F}_5 - \mathcal{D}_6 \mathcal{F}_6  \right]}
    \end{equation}
    and
    \begin{equation}
         \frac{C_4}{C_2} = \mathrm{e}^{\pi\frac{\tau}{2}(E_1-E_2)}\frac{ \left[ \left(2m + \mathcal{D}_1 \right) \mathcal{F}_1 - \mathcal{D}_2 \mathcal{F}_2 \right]  \left[\left(2m - \mathcal{D}_3\right) \mathcal{F}_3 - \mathcal{D}_4 \mathcal{F}_4 \right] -  \left[\left(2m - \mathcal{D}_1\right) \mathcal{F}_1 + \mathcal{D}_2 \mathcal{F}_2  \right]\left[\left(2m + \mathcal{D}_3 \right)  \mathcal{F}_3 + \mathcal{D}_4 \mathcal{F}_4 \right]}{ \left[\left(2m - \mathcal{D}_3\right) \mathcal{F}_3 - \mathcal{D}_4 \mathcal{F}_4 \right]\left[\left(2m - \mathcal{D}_5  \right) \mathcal{F}_5 + \mathcal{D}_6 \mathcal{F}_6  \right]- \left[\left(2m + \mathcal{D}_3 \right)  \mathcal{F}_3 + \mathcal{D}_4 \mathcal{F}_4 \right] \left[\left(2m + \mathcal{D}_5 \right) \mathcal{F}_5 - \mathcal{D}_6 \mathcal{F}_6 \right]},
    \end{equation}
\end{subequations}
where
\begin{subequations}
    \begin{equation}
        \mathcal{F}_1={_2F_1}(-\mu+\nu-\lambda, -\mu+\nu+\lambda;1-2\mu;-1),
    \end{equation}
    \begin{equation}
        \mathcal{F}_2={_2F_1}(-\mu+\nu-\lambda+1, -\mu+\nu+\lambda+1;1-2\mu+1;-1),
    \end{equation}
    \begin{equation}
        \mathcal{F}_3={_2F_1}(\lambda+\nu-\mu, \lambda+\nu+\mu;1+2\lambda;-1),
    \end{equation}
    \begin{equation}
        \mathcal{F}_4={_2F_1}(\lambda+\nu-\mu +1, \lambda+\nu+\mu +1;1+2\lambda +1;-1),
    \end{equation}
    \begin{equation}
        \mathcal{F}_5={_2F_1}(-\lambda+\nu-\mu, -\lambda+\nu+\mu;1-2\lambda;-1),
    \end{equation}
    \begin{equation}
        \mathcal{F}_6={_2F_1}(-\lambda+\nu-\mu +1, -\lambda+\nu+\mu +1;1-2\lambda +1;-1)
    \end{equation}
\end{subequations}
and
\begin{subequations}
    \begin{equation}
    \mathcal{D}_1 = 2 E_1 - (qA_2-qA_1) -\left(p-qA_2\right) -\left(p-qA_1\right),
\end{equation}
\begin{equation}
    \mathcal{D}_2 = i \frac{4}{\tau} \frac{(-\mu+\nu-\lambda) (-\mu+\nu+\lambda)}{1-2\mu},
\end{equation}
\begin{equation}
    \mathcal{D}_3 = 2 E_2 + (qA_2-qA_1) - \left(p-qA_2\right) - \left(p-qA_1\right),
\end{equation}
\begin{equation}
    \mathcal{D}_4 = i \frac{4}{\tau} \frac{(\lambda+\nu-\mu) (\lambda+\nu+\mu)}{1-2\lambda},
\end{equation}
\begin{equation}
    \mathcal{D}_5 = 2 E_2 - (qA_2-qA_1) + \left(p-qA_2\right) + \left(p-qA_1\right),
\end{equation}
\begin{equation}
    \mathcal{D}_6 = i \frac{4}{\tau} \frac{(-\lambda+\nu-\mu) (-\lambda+\nu+\mu)}{1-2\lambda},
\end{equation}
\end{subequations}
with
\begin{subequations}
    \begin{equation}
        \mu = i \sqrt{\alpha_1} = i \frac{\tau}{2} E_1,
    \end{equation}
    \begin{equation}
        \nu = \frac{1 - \sqrt{1-4(\alpha_0+\alpha_1+\alpha_2)}}{2} = i \frac{\tau}{2} (qA_2-qA_1),
    \end{equation}
    \begin{equation}
        \lambda = i \sqrt{\alpha_2} = i \frac{\tau}{2} E_2.
    \end{equation}
\end{subequations}

In this analysis, $E_1$ and $E_2$ represent the incident electron energy $E_\mathrm{i}$ and the later-forward energy $E_\mathrm{f}$, respectively, and the later-backward energy is defined as $E_\mathrm{b}=-E_\mathrm{f}$.

\subsection{Dimensional Analysis for Time Constant \texorpdfstring{$\tau$}{tau}}

We have used natural units ($c=\hbar=1$) for convenience throughout the calculations. However, to compare the theoretical results with experimental results, it is necessary to denormalize some of the obtained formulas. Specifically, we will need here to denormalize time in order to compare the transition times in the smooth potential with the de Broglie period of the electron.

This denormalization can be performed as follows. In \emph{standard} units,  which we shall distinguish from our natural units by tilde accents, the temporal part of the wavefunction takes the form~\cite{Griffiths_2018_Quantum,zettili2009quantum,shankar2012principles,miller2008quantum,landau_2013_quantum,sakurai_napolitano_2020}
\begin{equation}\label{eq:su_psi_a}
    \psi\propto\mathrm{e}^{-i\tilde{E}\tilde{t}/\hbar},
\end{equation}
with $\tilde{E}$ and $\tilde{t}$ measured in joules and seconds, respectively. For simplicity, we express the energy, $\tilde{E}$, in  terms of multiple of the electron rest mass energy, i.e., $\tilde{E}=\kappa\left(mc^2\right)$, which transforms Eq.~\eqref{eq:su_psi_a} into
\begin{equation}\label{eq:su_psi_b}
    \psi\propto\mathrm{e}^{-i\kappa mc^2\tilde{t}/\hbar}.
\end{equation}
Moreover, it makes sense to also express the \emph{natural}-unit energy, $E$, in terms of that multiple ($\kappa$) of the electron mass, i.e., $E=\kappa m$, which alters Eq.~\eqref{eq:su_psi_b} to
\begin{equation}\label{eq:su_psi_d}
    \psi\propto\mathrm{e}^{-iEc^2\tilde{t}/\hbar}.
\end{equation}
Comparing this expression with its \emph{natural}-unit counterpart,
\begin{equation}\label{eq:su_psi_c}
    \psi\propto\mathrm{e}^{-iEt},
\end{equation}
finally yields
\begin{equation}
    t=\frac{c^2\tilde{t}}{\hbar},
\end{equation}
whose dimension is $1/m$ since the dimension of $E$ is $m$.

Thus, the natural-unit dimensionless parameter $\tau$ in Eq.~\eqref{eq:smooth_pot} denormalizes as
\begin{equation}
    \tilde{\tau} = \tau\frac{\hbar}{mc^2},
\end{equation}
measured in seconds, which we call $\eta$ in the paper.

\newpage

\section{Non-relativistic Regime}\label{sec:nonrelregime}

\subsection{Non-relativistic Limit of the Dirac Equation}\label{subsec:nonrelDirac}
In order to determine the \emph{non-relativistic limit},
\begin{equation}\label{eq:nr_lim_vc}
    v\ll{c},
\end{equation}
of the Dirac equation, we shall first recast that equation in its (non-covariant) Schr\"{o}dinger form because the covariant form, introduced in Sec.~\ref{sec:Dirac_equation} and applied in the report to address the core of the problem, does not provide explicit access to the dynamic energy, which needs to be compared with the rest energy to determine that limit.

Let us start by recalling the covariant form of the Dirac equation, given by Eq.~\eqref{eq:Dirac_potential}:
\begin{equation}
    \left[\gamma^\mu(i\partial_\mu-qA_\mu)-m\right]\psi
    =0.
\end{equation}
Separating in this equation the temporal and spatial operators according to Einstein summation convention yields
\begin{equation}
    \left[\gamma^{0}(i\partial_{0}-q A_{0})
    +\gamma^{i}(i\partial_{i}-q A_{i})-m\right]\psi
    =0,
\end{equation}
or, factoring out $\gamma^{0}$,
\begin{equation}
    \gamma^{0}\left[i \partial_{0}+\left(\gamma^{0}\right)^{-1} \gamma^{i} i \partial_{i}-q A_{0}-\left(\gamma^{0}\right)^{-1} \gamma^{i} qA_{i}-\left(\gamma^{0}\right)^{-1} m\right] \psi=0.
\end{equation}
Dropping $\gamma^{0}$ and subsequently using Eq.~\eqref{eq:gamma_i} with the fact that, according to Eq.~\eqref{eq:gamma_0}, $\left(\gamma^{0}\right)^{-1}=\gamma^{0}$, simplifies that equation to
\begin{equation}
    \left[i \partial_{0}+\alpha^{i} i \partial_{i}-q A_{0}- \alpha^{i} qA_{i}-\gamma^{0} m\right] \psi=0,
\end{equation}
or, isolating the temporal derivative term,
\begin{equation}
    i \partial_{0} \psi=\left[\alpha^{i}\left(-i \partial_{i}+q A_{i}\right)+\gamma^{0} m + qA_0 \right] \psi.
\end{equation}

We may now write the last equation in vector form using the definitions
\begin{subequations}
    \begin{equation}
    \left(\partial_{0}, \partial_{i}\right) \equiv\left(\pdv{t}, \grad\right),
    \end{equation}
    \begin{equation}
    \alpha^{i} \equiv \vb*{\alpha},
    \end{equation}
    and
    \begin{equation}
    \left(A_{0}, A_{i}\right) \equiv (V,-\vb{A}).
    \end{equation}
\end{subequations}
This yields
\begin{equation}
    i \pdv{t}\psi=\left[\vb*{\alpha} \cdot(-i \grad-q \vb{A})+\gamma^{0} m+q V\right] \psi,
\end{equation}
which, using $\vb{\hat{p}}=-i \grad$, becomes
\begin{subequations}\label{eq:Dirac_eq}
\begin{equation}
    i \pdv{t}\psi=\mathcal{H}\psi,
\end{equation}
with the Hamiltonian
\begin{equation}\label{eq:Dirac_Hamiltonian2}
   \mathcal{H}=\vb*{\alpha} \cdot(\vb{\hat{p}}-q \vb{A})+\gamma^0m+q V,
\end{equation}
where
\begin{equation} \label{eq:alpha_beta}
     \quad 
    \vb*{\alpha}= \begin{pmatrix}
        0 &\vb*{\sigma} \\
        \vb*{\sigma} & 0
    \end{pmatrix},
\end{equation}
\text{with} 
\begin{equation}\label{eq:sigma_vect_of_matr}
    \quad
    \vb*{\sigma}=(\sigma^1, \sigma^2,\sigma^3)
\end{equation}
\end{subequations}
being a vector whose components are the Pauli matrices given in Eq.~\eqref{eq:Pauli_matrix}; specifically, it is a $1\times{3}$ vector of $2\times{2}$ matrices, so that $\vb*{\alpha}$ is a $4\times{4}$ matrix. Note that, in the above relations, the `$\cdot$' symbol represents the scalar product defined as $\vb{u}\cdot\vb{v}\triangleq\vb{u}\vb{v}^\mathrm{T}$, where $\mathrm{T}$ represents the transpose operation. 

The Hamiltonian in Eq.~\eqref{eq:Dirac_Hamiltonian2} has eigenvalues corresponding to the total relativistic energy of the particle,
\begin{equation} \label{eq:totalenergy1}
    E=E_\mathrm{k}+m+E_\mathrm{p},
\end{equation}
where $E_\mathrm{k}$, $m$ and
\begin{equation} \label{eq:totalenergy2}
    E_\mathrm{p}=qV
\end{equation}
are the relativistic kinetic energy, rest mass energy, and potential energy, respectively. The wavefunction $\psi$ solution to Eq.~\eqref{eq:Dirac_eq} for positive energies may then be expressed as
\begin{equation} \label{eq:spinor_exp_m}
    \psi=\bar{\psi} \mathrm{e}^{-imt}
        =\begin{pmatrix} 
        \varphi \\
        \vartheta 
        \end{pmatrix}\mathrm{e}^{-imt},
\end{equation}
with separate the time evolution due to the rest mass, $\mathrm{e}^{-imt}$, and kinetic and spatial energy dependencies, embedded in the modified wavefunction, $\bar{\psi}$. That separation will next allow us to get rid of the mass-related temporal dependence $\mathrm{e}^{-imt}$. Indeed, inserting Eqs.~\eqref{eq:spinor_exp_m},~\eqref{eq:alpha_beta} and~\eqref{eq:gamma_0} into Eq.~\eqref{eq:Dirac_eq} yields
\begin{equation}
    i \pdv{t} 
    \left\{\begin{pmatrix} 
        \varphi \\
        \vartheta
    \end{pmatrix}
    \mathrm{e}^{-i m t}\right\}=\begin{pmatrix}
    \vb*{\sigma} \cdot(\vb{\hat{p}}-q \vb{A}) \vartheta \\
    \vb*{\sigma} \cdot(\vb{\hat{p}}-q \vb{A}) \varphi
    \end{pmatrix} \mathrm{e}^{-i m t}+m\begin{pmatrix}
    \varphi \\
    -\vartheta
    \end{pmatrix} \mathrm{e}^{-i m t}+qV\begin{pmatrix}
    \varphi \\
    \vartheta
    \end{pmatrix} \mathrm{e}^{-i m t},
\end{equation}
which, upon applying the product rule to the left-hand side derivative and multiplying the resulting equation by $e^{imt}$ (and hence breaking covariance), simplifies to
\begin{equation}\label{eq:Dirac_syst}
    i \pdv{t} 
    \begin{pmatrix} 
        \varphi \\
        \vartheta
    \end{pmatrix}
    =\begin{pmatrix}
    \vb*{\sigma} \cdot(\vb{\hat{p}}-q \vb{A}) \vartheta \\
    \vb*{\sigma} \cdot(\vb{\hat{p}}-q \vb{A}) \varphi
    \end{pmatrix} -2m\begin{pmatrix}
    0 \\
    \vartheta
    \end{pmatrix} +qV\begin{pmatrix}
    \varphi \\
    \vartheta
    \end{pmatrix}.
\end{equation}
The suppression of the exponential time evolution associated with the rest mass energy in this relation corresponds to a redefinition of the zero of energy that will not change the observable physics in the considered non-relativistic limit. In the non-relativistic regime, only energy differences, viz., $\Delta{E}=E_2-E_1=\hbar\omega$, with $\omega$ being the absorption or emission frequency, are observable. Changing the reference energy, such as by the rest-mass energy, $E_\mathrm{m}$, does not change the observable, since $\Delta{E}=(E_2+E_\mathrm{m})-(E_1+E_\mathrm{m})=E_2-E_1=\hbar\omega$ represents the same (observable) absorption or emission frequency, whereas $E_\mathrm{m}t$ represents only a non-observable phase shift, $\Delta\phi=m$. This is different from the relativistic regime, where the rest-mass energy contributes to Lorentz invariance and can therefore not be simply subtracted. Another perspective is that no mass-to-energy conversion [$E_\mathrm{k}=mc^2(\gamma-1)$ ($c=1$ in natural units)] occurs in the non-relativistic limit, so the mass-related energy does not contribute to observable quantities in that limit. The observable physics will thus depend only on the kinetic and potential energies, associated with the modified wavefunction $\bar{\psi}$, and hence with $\varphi$ and $\vartheta$. Equation~\eqref{eq:Dirac_syst} is in fact a system of two coupled equations, which splits into
\begin{subequations}\label{eq:split_Dirac}
    \begin{equation}\label{eq:first_dirac}
        i \pdv{t} \varphi=\vb*{\sigma} \cdot(\vb{\hat{p}}-q \vb{A})\vartheta +E_\mathrm{p}\varphi
    \end{equation}
    and
    \begin{equation}\label{eq:second_dirac}
        i \pdv{t} \vartheta=\vb*{\sigma} \cdot(\vb{\hat{p}}-q \vb{A})\varphi - 2m\vartheta + E_\mathrm{p}\vartheta
    \end{equation}
\end{subequations}

Let us now consider the non-relativistic limit. For this purpose, let us write the relativistic energy, in \emph{standard units}, viz.,
\begin{subequations}
    \begin{equation}\label{eq:en_rel_nu}
        E=\sqrt{(mc^2)^2+(pc)^2}+E_\mathrm{p},
    \end{equation}
    where
    \begin{equation}
        p=\frac{1}{\sqrt {1-\frac{v^2}{c^2}}} mv.
    \end{equation}
\end{subequations}
In the non-relativistic limit [Eq.~\eqref{eq:nr_lim_vc}], $v\ll{c}$, the latter relation becomes 
\begin{subequations}
    \begin{equation}\label{eq:pmv_nr}
        p\simeq{mv},
    \end{equation}
    and inserting this new relation into Eq.~\eqref{eq:en_rel_nu} yields
\begin{equation}\label{eq:Enrlim_normun}
    E=\sqrt{(mc^2)^2+(mvc)^2}+E_\mathrm{p}=mc^2\sqrt{1+\frac{v^2}{c^2}}+E_\mathrm{p}.
\end{equation}
Since $v\ll{c}$, the square root in the last expression may be approximated by its second-order Taylor expansion, which leads to
    \begin{equation}
    \begin{split}
        E
        &\simeq mc^2\left(1+\frac{1}{2}\frac{v^2}{c^2}\right)+E_\mathrm{p} \\
        & = mc^2+\frac{1}{2}mv^2 +E_\mathrm{p} \\
        & = mc^2+\frac{p^2}{2m}+E_\mathrm{p} \\
        & = mc^2+E_\mathrm{k}+E_\mathrm{p},
        \end{split}
    \end{equation}
\end{subequations}
or, in natural units ($c=1$),
\begin{equation} \label{eq:tot_ener_nonrel}
    E = m+E_\mathrm{k}+E_\mathrm{p},
\end{equation}
where (in that non-relativistic limit approximation) the rest energy term ($m$) is much larger than the (now simply $p^2/(2m)$) kinetic energy ($E_\mathrm{k}$) and potential energy ($E_\mathrm{p}$) contributions (see numerical example in Sec.~\ref{sec:num_ex}).

With the above redefinition of the zero-energy level, the eigen-equation corresponding to the function $\vartheta$ in Eq.~\eqref{eq:split_Dirac} may written
\begin{subequations}
    \begin{equation}\label{eq:theta_eigeneq}
        i\pdv{t}\vartheta
         =W_\vartheta\vartheta,
    \end{equation}
    where, according to Eq.~\eqref{eq:tot_ener_nonrel},
    \begin{equation}
        W_\vartheta
         =E-m
         =E_\mathrm{k}+E_\mathrm{p},
    \end{equation}
    is the total mass-shifted energy, corresponding to the sum of the kinetic and potential energies.
\end{subequations}
In the non-relativistic limit, since the kinetic and potential energies are negligible compared to the mass term, i.e.,
\begin{equation}\label{eq:Wmsm}
    W_\vartheta\ll{m},  
\end{equation}
so that, according to Eq.~\eqref{eq:theta_eigeneq},
\begin{equation}\label{eq:nr_approx_thetam}
    i\pdv{t}\vartheta\ll{m},
\end{equation}
we find, with $E_\mathrm{p}=qV\ll{m}$, that Eq.~\eqref{eq:second_dirac} reduces to
\begin{equation} \label{eq:nonrel_theta_phi}
    \vartheta=\frac{\vb*{\sigma} \cdot(\vb{\hat{p}}-e \vb{A})}{2 m} \varphi,
\end{equation}
whose insertion into Eq.~\eqref{eq:first_dirac} yields
\begin{equation} \label{eq:nonrel_preSch}
    i \pdv{t} \varphi=\frac{1}{2 m}[\vb*{\sigma} \cdot(\vb{\hat{p}}-q \vb{A})] [\vb*{\sigma} \cdot(\vb{\hat{p}}-q \vb{A})]\varphi+q V \varphi.
\end{equation}

The first term on the right-hand side of this equation may be written as $(\vb*{\sigma}\cdot\vb{a})(\vb*{\sigma}\cdot\vb{b})\varphi/(2m)$, where $\vb*{\sigma}$ is the $1\times{3}$ vector of $2\times{2}$ matrices given by Eq.~\eqref{eq:sigma_vect_of_matr} and $\vb{a}=\vb{b}=(\vb{\hat{p}}-q\vb{A})$ are $1\times{3}$ vectors. This expression may be decomposed into components with
\begin{equation}\label{eq:aibjsisj}
    (\vb*{\sigma}\cdot\vb{a})(\vb*{\sigma}\cdot\vb{b})
    =(\sigma^{i}a^{i})(\sigma^{j}b^{j})
    =a^{i} b^{j}\sigma^{i}\sigma^{j},
\end{equation}
which is a $2\times{2}$ matrix, since $a^i$ and $b^j$ are scalar vector components and $\sigma^i$ and $\sigma^j$ are $2\times{2}$ matrix vector components [the Pauli matrices, given by Eq.~\eqref{eq:Pauli_matrix}]. The term $\sigma^{i}\sigma^{j}$ in the last equality can be written in an alternative, more convenient fashion, upon using the commutation and anticommuntation relations between the Pauli matrices~\cite{ryder1996quantum}, viz.,
\begin{subequations}
    \begin{equation}
        \left[\sigma^{i}, \sigma^{j}\right]=\sigma^{i} \sigma^{j}-\sigma^{j} \sigma^{i}=2 i \varepsilon^{i j k} \sigma^{k},
    \end{equation}
    where $\varepsilon^{i j k}$ is the Levi-Civita symbol, and
    \begin{equation}
        \left\{\sigma^{i}, \sigma^{j}\right\}=\sigma^{i} \sigma^{j}+\sigma^{j} \sigma^{i}=2 \delta^{i j},
    \end{equation}
\end{subequations}
where $\delta^{i j}$ is the Kronecker delta, whose summation leads to
\begin{equation} \label{eq:sigma_i_j}
    \sigma^{i} \sigma^{j}= i \varepsilon^{i j k} \sigma^{k}+\delta^{i j}.
\end{equation}
Inserting Eq.~\eqref{eq:sigma_i_j} into Eq.~\eqref{eq:aibjsisj} yields
\begin{equation}
\begin{split}
    (\vb*{\sigma}\cdot\vb{a})(\vb*{\sigma}\cdot\vb{b})
    &=a^{i} b^{j}\left(i \varepsilon^{i j k} \sigma^{k}+\delta^{i j}\right) \\
    &=i \varepsilon^{i j k} a^{i} b^{j} \sigma^{k}+a^{i} b^{j} \delta^{i j} \\
    &=i \underbrace{\varepsilon^{k i j} a^{i} b^{j}}_{(\vb{a} \times \vb{b})^{k}} \sigma^{k}+a^{i} b^{i}\\
    &=i \sigma^{k}(\vb{a} \times \vb{b})^{k}+a^{i} b^{i},
\end{split}
\end{equation}
which can alternatively be written, using again Einstein summation convention, in the vectorial form
\begin{equation}\label{eq:sigma_identity}
    (\vb*{\sigma} \cdot \vb{a})(\vb*{\sigma} \cdot \vb{b})=
    i \vb*{\sigma} \cdot(\vb{a} \times \vb{b})+\vb{a} \cdot \vb{b}.
\end{equation}
Applying this result to the first term of the right-hand side of Eq.~\eqref{eq:nonrel_preSch} yields
\begin{equation} \label{eq:sigma_p_qA}
    \begin{split}
    [\vb*{\sigma} \cdot(\vb{\hat{p}}-q \vb{A})] [\vb*{\sigma} \cdot(\vb{\hat{p}}-q \vb{A})]
    & = i \vb*{\sigma} \cdot[(\vb{\hat{p}}-q \vb{A}) \times(\vb{\hat{p}}-q \vb{A})]+(\vb{\hat{p}}-q \vb{A}) \cdot(\vb{\hat{p}}-q \vb{A}) \\
    & = i \vb*{\sigma} \cdot[\underbrace{(\vb{\hat{p}} \times \vb{\hat{p}})}_{=0}+q^{2}(\underbrace{\vb{A} \times \vb{A} )}_{=0}-q(\vb{\hat{p}} \times \vb{A} )-q(\vb{A} \times \vb{\hat{p}})]+(\vb{\hat{p}}-q \vb{A})^{2} \\
    & = i \vb*{\sigma} \cdot[-q(\vb{\hat{p}} \times \vb{A})-q(\vb{A} \times \vb{\hat{p}})]+(\vb{\hat{p}}-q \vb{A})^{2} \\
    & = -q \vb*{\sigma} \cdot[({\grad} \times \vb{A})+(\vb{A} \times \grad)]+(\vb{\hat{p}}-q {\vb{A}})^{2},
    \end{split}
\end{equation}
where $\vb{\hat{p}}=-i\grad$ has been used in the last equality. In the final expression, the term $[({\grad} \times \vb{A})+(\vb{A} \times \grad)]$ is an operator and must therefore be evaluated conjointly with the wavefunction upon which it acts. Calling that wavefunction $f$, we find then
\begin{equation}
    \begin{split}
    [({\grad} \times \vb{A}f)+(\vb{A} \times \grad f)]^i 
    &=\varepsilon^{i j k}\partial^j A^k f + \varepsilon^{i j k}A^j \partial^k f \\
    &=\varepsilon^{i j k}(\partial^j A^k) f + \varepsilon^{i j k}(\partial^j f) A^k + \varepsilon^{i j k}A^j(\partial^k f) \\
    &=\varepsilon^{i j k}(\partial^j A^k) f + \varepsilon^{i j k}(\partial^j f) A^k + \varepsilon^{i k j}A^k (\partial^j f) \\
    &=\varepsilon^{i j k}(\partial^j A^k) f + \cancel{\varepsilon^{i j k}(\partial^j f) A^k} -\cancel{\varepsilon^{i j k} (\partial^j f) A^k} \\
    &=\underbrace{\varepsilon^{i j k}(\partial^j A^k)}_{(\grad \times \vb{A})^i} f,
    \end{split}
\end{equation}
which reveals that
\begin{equation}
    [({\grad} \times \vb{A} f)+(\vb{A} \times \grad f)] = (\grad \times \vb{A})f.
\end{equation}
Setting in this relation $\grad\times\vb{A}=\vb{B}$, where $\vb{B}$ is the magnetic field, according to Eq.~\eqref{eq:B_eq_curlA}, and inserting the result into Eq.~\eqref{eq:sigma_p_qA}, we get
\begin{equation} \label{eq:2first_term_onRHS}
    [\vb*{\sigma} \cdot(\vb{\hat{p}}-q \vb{A})] [\vb*{\sigma} \cdot(\vb{\hat{p}}-q \vb{A})] 
    =-q \vb*{\sigma} \cdot\vb{B} 
    +(\vb{\hat{p}}-q {\vb{A}})^{2}.
\end{equation}
Finally, substituting this identity into Eq~\eqref{eq:nonrel_preSch} yields 
\begin{equation} \label{eq:Sch_Pauli}
    i \pdv{t} \varphi=\left[\frac{(\vb{\hat{p}}-q {\vb{A}})^{2}}{2 m}-\frac{q}{2m} \vb*{\sigma} \cdot\vb{B}+q V \right] \varphi,
\end{equation}
which is the \emph{Schr{\"o}dinger-Pauli} equation, where $q=-|q|=-e$ in the case of the electron.

In the absence of magnetic field ($\vb{B}=0$), the term $\vb*{\sigma} \cdot\vb{B} $ disappears, and Eq.~\eqref{eq:Sch_Pauli} reduces to the ordinary Schr{\"o}dinger equation,
\begin{equation}\label{eq:Schroedinger_eq}
    i \pdv{t} \varphi=\left[\frac{(\vb{\hat{p}}-q {\vb{A}})^{2}}{2 m}+q V \right] \varphi.
\end{equation}

The non-relativistic limit $v\ll{c}$ [Eq.~\eqref{eq:nr_lim_vc}] is naturally valid in a large \emph{range} of velocities ($\sim{v<c/10}$), which we hereafter refer to as the \emph{non-relativistic regime}.

\subsection{Numerical Example for the Relativistic Regime}\label{sec:num_ex}

Assuming standard units and the SI (Système International) system:

\begin{itemize}
    \item electron rest mass: $m\simeq9.109 \times10^{-31}$~kg
    \item electron charge: $e\simeq1.602 \times 10^{-19}$~C
    \item speed of light in vacuum: $c\simeq 2.998 \times 10^8$~m/s
    \item rest mass energy: $E_\mathrm{m}=m c^2
    \simeq(9.109\times 10^{-31})(2.998\times 10^8)^2 \simeq 8.187 \times 10^{-14}$~J
    \item non-relativistic kinetic energy for $v=c/100$: \\
    $E_\mathrm{k}=\frac{1}{2}mv^2
    =\frac{1}{2}\times(9.109\times 10^{-31}) \times (2.998 \times 10^6)^2  \simeq 4.094 \times 10^{-18}$~J
    \item potential energy for $V=7$~V ($10$ times the built-in potential of a silicon p-n junction): \\
    $E_\mathrm{p}=eV = (1.602 \times 10^{-19})\times 7 \simeq 1.121 \times 10^{-18}$~J
    \item energy ratio: $\dfrac{E_\mathrm{k}+E_\mathrm{p}}{E_\mathrm{m}}=\dfrac{4.094 \times 10^{-18}+1.121 \times 10^{-18}}{8.187 \times 10^{-14}} \simeq 6.370\times 10^{-5} \simeq 0.00637~\%$
    \item non-relativistic total energy with rest mass energy: \\ $E_\mathrm{nr}=E_\mathrm{m}+E_\mathrm{k}+E_\mathrm{p}=8.187\times 10^{-14}+4.094 \times 10^{-18}+1.121 \times 10^{-18} \simeq 8.1875\times 10^{-14}$
    \item relativistic total energy: 
    $E_\mathrm{r}=\sqrt{(mc^2)^2+(pc)^2}+qV$, where $p=\gamma mv$ with $\gamma = \frac{1}{\sqrt{1 - \left(\frac{v}{c}\right)^2}} \simeq 1.00005$ \\
    $\rightarrow  p \simeq 2.731 \times 10^{-24}$~kg~m/s \\  
    $\begin{aligned}
    \Rightarrow E_\mathrm{r} 
    & =  \sqrt{(9.109 \times 10^{-31})^2 (2.998 \times 10^8)^4 + (2.731 \times 10^{-24})^2 (2.998 \times 10^8)^2} + 1.121 \times 10^{-18} \\ 
    & \simeq 8.1877 \times 10^{-14}~J
    \end{aligned}$
    \item error: $\dfrac{\left|E_\mathrm{r}-E_\mathrm{nr}\right|}{E_\mathrm{r}}
    =\dfrac{\left|8.1877 \times 10^{-14} - 8.1875 \times 10^{-14}\right|}{8.1877 \times 10^{-14}} \simeq 2.4427 \times 10^{-5} \triangleq 0.0024427~\%$
\end{itemize}

\subsection{Spatial Scattering Coefficients in the Non-relativistic Regime}

For the sake of completeness, let us first consider the relativistic scattering coefficients for the scalar potential \emph{spatial} step, which were given by Eq.~\eqref{eq:s_scalar_spatial_coef_gamma} as
\begin{subequations}
\begin{equation}\label{eq:r_and_t}
    r=\frac{1-\Gamma_\mathrm{s}}{1+\Gamma_\mathrm{s}} 
    \quad \text{and}\quad 
    t=\frac{2}{1+\Gamma_\mathrm{s}},
\end{equation}
where
\begin{equation}\label{eq:gammas_1}
\Gamma_\mathrm{s}=\dfrac{\left(E-q V_{2}-m\right) \sqrt{\left(E-q V_{1}\right)^{2}-m^{2}}}{\left(E-q V_{1}-m\right) \sqrt{\left(E-q V_{2}\right)^{2}-m^{2}}}.
\end{equation}
\end{subequations}

Squaring Eq.~\eqref{eq:gammas_1} and algebraically manipulating the resulting expression yields
\begin{equation} \label{eq:gammasqr1}
\begin{split}
    \Gamma_\mathrm{s}^{2} 
    & =\frac{\left(E-q V_{2}-m\right)^{2}\left[\left(E-q V_{1}\right)^{2}-m^{2}\right]}{\left(E-q V_{1}-m\right)^{2}\left[\left(E-q V_{2}\right)^{2}-m^{2}\right]} \\
    & =\frac{\left(E-q V_{2}-m\right)^{2}\left[\left(E-q V_{1}-m\right)\left(E-q V_{1}+m\right)\right]}{\left(E-q V_{1}-m\right)^{2}\left[\left(E-q V_{2}-m\right)\left(E-q V_{2}+m\right)\right]} \\
    & =\frac{\left(E-q V_{2}-m\right)\left(E-q V_{1}+m\right)}{\left(E-q V_{1}-m\right)\left(E-q V_{2}+m\right)} \\
    & =\frac{E^{2}-q V_{1} E-q V_{2} E+qV_{1} qV_{2}-m q V_{2}+m q V_{1}-m^{2}}{E^{2}-q V_{2} E-q V_{1} E+qV_{1} qV_{2}-m q V_{1}+m q V_{2}-m^{2}} \\
    & =\frac{E^{2}-q V_{1} E-q V_{2} E+qV_{1} qV_{2}-m q V_{2}+m q V_{1}-m^{2}}{E^{2}-q V_{2} E-q V_{1} E+qV_{1} qV_{2}-m q V_{1}+m q V_{2}-m^{2}}-1+1 \\
    & =1-\frac{2 m (qV_{2}-qV_{1})}{E^{2}-q V_{2} E-q V_{1} E+ qV_{1} qV_{2}-m q V_{1}+m q V_{2}-m^{2}} \\
    & =1-\frac{2 m (qV_{2}-qV_{1})}{(E-qV_{1})^2-m^2-(qV_1)^2+qV_1E-qV_2E+qV_1 qV_2+m(qV_2-qV_1)} \\
    & =1-\frac{2 m (qV_{2}-qV_{1})}{(E-V_{1})^2-m^2-(qV_2-qV_1)(E-qV_1-m)} \\
    & =1-\frac{2 m (qV_2-qV_1)}{\left[(E-qV_{1})^2-m^2\right]\left[1-(qV_2-qV_1)\dfrac{E-qV_1-m}{(E-qV_{1})^2-m^2}\right]} \\
    & =1-\frac{(qV_2-qV_1)}{\dfrac{(E-qV_{1})^2-m^2}{2m}\left[1-(qV_2-qV_1)\dfrac{E-qV_1-m}{(E-qV_{1})^2-m^2}\right]} \\
    & =1-\frac{(qV_2-qV_1)}{\dfrac{p_\mathrm{i}^2}{2m}\left[1-(qV_2-qV_1)\dfrac{1}{E-qV_1+m}\right]}.
\end{split}
\end{equation}
In the non-relativistic regime [Eq.~\eqref{eq:Wmsm}],
\begin{equation}\label{eq:gamma1_nr_approx}
    E_\mathrm{(r)}-\cancel{qV_1}+m >>qV_2-\cancel{qV_1},
\end{equation}
and the kinetic energy $p_\mathrm{i}^2/2m$ is simply
\begin{equation}\label{eq:nonrelkin}
        \frac{p_\mathrm{i}^2}{2m} = E_\mathrm{(nr)}-qV_1,
\end{equation}
where $E_\mathrm{(nr)}$ is non-relativistic total energy.

Inserting Eqs.~\eqref{eq:gamma1_nr_approx} and~\eqref{eq:nonrelkin} into Eq.~\eqref{eq:gammasqr1} yields then
\begin{equation}
    \Gamma_\mathrm{s}^{2} \simeq 1-\dfrac{qV_2-qV_1}{E-qV_1}
\end{equation}
and
\begin{equation}\label{eq:gams_mEqv}
    \Gamma_\mathrm{s} \simeq \sqrt{1-\dfrac{qV_2-qV_1}{E-qV_1}} = \sqrt{\dfrac{E-qV_2}{E-qV_1}} = \sqrt{\dfrac{2m(E-qV_2)}{2m(E-qV_1)}}.
\end{equation}
Since the non-relativistic total energy is $E_\mathrm{(nr)}=p^2/2m+qV=\hbar^2 k^2/2m+qV$, where $\hbar=1$ in natural units,
\begin{equation}
    \sqrt{2m(E-qV_{1,2})}=k_{1,2},
\end{equation}
where $k_{1,2}$ are wave numbers, so that Eq.~\eqref{eq:gams_mEqv} reduces to
\begin{equation} \label{eq:gamma1_Sch}
    \Gamma_\mathrm{s} 
    \simeq \dfrac{k_2}{k_1}.
\end{equation}
Inserting Eq.~\eqref{eq:gamma1_Sch} into Eq.~\eqref{eq:r_and_t} leads to the non-relativistic reduction of the Dirac reflection and transmission coefficients to  the Schr{\"o}dinger scattering coefficients
\begin{equation}\label{eq:Sch_Vz_coef_nonrel}
    r=\dfrac{1-\dfrac{k_2}{k_1}}{1+\dfrac{k_2}{k_1}}=\dfrac{k_1-k_2}{k_1+k_2} \quad \text { and } \quad t=\dfrac{2}{1+\dfrac{k_2}{k_1}}=\dfrac{2k_1}{k_1+k_2}.
\end{equation}

Note that the corresponding solution to the Schr{\"o}dinger equation [Eq.~\eqref{eq:Schroedinger_eq}] is here [for the potential given in~\eqref{eq:s_scalar_pot_spatial}] readily found as
\begin{subequations}
    \begin{equation}
        \psi_{1} = \mathrm{e}^{-i E_\mathrm{i} t} \mathrm{e}^{i p_\mathrm{i} z}  
        +r \mathrm{e}^{-i E_\mathrm{r} t} \mathrm{e}^{i p_\mathrm{r} z}
        \end{equation}
        and
        \begin{equation}
        \psi_{2} = t \mathrm{e}^{-i E_\mathrm{t} t} \mathrm{e}^{i p_\mathrm{t} z},
    \end{equation}
where
    \begin{equation}
    E_\mathrm{i}=E_\mathrm{r}=E_\mathrm{t}=E, \quad
        p_\mathrm{i}=\sqrt{2m(E-qV_1)}, \quad
        p_\mathrm{r}=-p_\mathrm{i} \quad
        \textrm{and} \quad
        p_\mathrm{t}=\sqrt{2m(E-qV_2)}.
    \end{equation}
\end{subequations}
which implies the same scattering coefficients $r$ and $t$ as in Eq.~\eqref{eq:Sch_Vz_coef_nonrel}.

\subsection{Temporal Scattering Coefficients in the Non-relativistic Regime}

The relativistic scattering coefficients for the vector potential temporal step were given by Eq.~\eqref{eq:s_vector_temporal_coef_gamma} as
\begin{subequations}
\begin{equation}\label{eq:f_and_b}
    f=\frac{1+\Gamma_\mathrm{t}}{2 \Gamma_\mathrm{t}} \quad \text{and} \quad b=\frac{\Gamma_\mathrm{t}-1}{2 \Gamma_\mathrm{t}},
\end{equation}
with
\begin{equation}\label{eq:gammas_2}
    \Gamma_\mathrm{t}=\dfrac{\dfrac{E_{f}}{p-q A_{2}}}{\dfrac{E_\mathrm{i}-m}{p-q A_{1}}+\dfrac{m}{p-q A_{2}}}=\dfrac{\dfrac{\sqrt{\left(p-q A_{2}\right)^{2}+m^{2}}}{p-q A_{2}}}{\dfrac{\sqrt{\left(p-q A_{1}\right)^{2}+m^{2}}-m}{p-q A_{1}}+\dfrac{m}{p-q A_{2}}},
\end{equation}
\end{subequations}
where Eqs.~\eqref{eq:s_vector_temporal_energies_before} and~\eqref{eq:s_vector_temporal_energies_after} has been used in the last relation.

Squaring Eq.~\eqref{eq:gammas_2} and algebraically manipulating the resulting expression yields
\begin{equation}
\begin{split}
    \Gamma_\mathrm{t}^{2} 
    & =\dfrac{\left(p-q A_{2}\right)^{2}+m^{2}}{\left[\dfrac{p-q A_{2}}{p-q A_{1}}\left(\sqrt{\left(p-q A_{1}\right)^{2}+m^{2}}-m\right)+m\right]^{2}} \\
    & =\dfrac{\left(p-q A_{2}\right)^{2}+m^{2}}{\left(\dfrac{p-q A_{2}}{p-q A_{1}}\right)^2\left(\sqrt{\left(p-q A_{1}\right)^{2}+m^{2}}-m\right)^2+2\dfrac{p-q A_{2}}{p-q A_{1}}\left(\sqrt{\left(p-q A_{1}\right)^{2}+m^{2}}-m\right)m+m^2} \\
    & =\dfrac{\left(\dfrac{p-q A_{2}}{m}\right)^{2}+1}{\left(\dfrac{p-q A_{2}}{p-q A_{1}}\right)^2\left(\sqrt{\left(\dfrac{p-q A_{1}}{m}\right)^{2}+1}-1\right)^2+2\dfrac{p-q A_{2}}{p-q A_{1}}\left(\sqrt{\left(\dfrac{p-q A_{1}}{m}\right)^{2}+1}-1\right)+1}.
\end{split}
\end{equation}
In the non-relativistic regime [Eq.~\eqref{eq:Wmsm}],
\begin{equation}
    m(c^2)\gg p(c)-qA_{1,2}=(c)\sqrt{2mE_\mathrm{k}},
    \quad\text{(standard units)}
\end{equation}
and the square roots in the last expression may therefore be approximated by their second-order Taylor expansion, which leads to
\begin{equation}
\begin{split}
    \Gamma_\mathrm{t}^{2} 
    & \simeq \dfrac{\left(\dfrac{p-q A_{2}}{m}\right)^{2}+1}{\left(\dfrac{p-q A_{2}}{p-q A_{1}}\right)^2 \left(\dfrac{1}{2}\left(\dfrac{p-q A_{1}}{m}\right)^{2}+1-1\right)^2+2\dfrac{p-q A_{2}}{p-q A_{1}}\left(\dfrac{1}{2}\left(\dfrac{p-q A_{1}}{m}\right)^{2}+1-1\right)+1} \\
    & =\dfrac{\left(\dfrac{p-q A_{2}}{m}\right)^{2}+1}{\left(\dfrac{p-q A_{2}}{p-q A_{1}}\right)^2 \left(\dfrac{1}{2}\left(\dfrac{p-q A_{1}}{m}\right)^{2}\right)^2+2\dfrac{p-q A_{2}}{p-q A_{1}}\left(\dfrac{1}{2}\left(\dfrac{p-q A_{1}}{m}\right)^{2}\right)+1} \\
    & =\dfrac{\left(\dfrac{p-q A_{2}}{m}\right)^{2}+1}{\dfrac{1}{4}\left(\dfrac{p-q A_{1}}{m}\right)^2\left(\dfrac{p-q A_{2}}{m}\right)^2+\left(\dfrac{p-q A_{1}}{m}\right)\left(\dfrac{p-q A_{2}}{m}\right)+1} \\
    & =\dfrac{\left(\dfrac{p-q A_{2}}{m}\right)^{2}+1}{\left(\dfrac{1}{2}\left(\dfrac{p-q A_{1}}{m}\right)\left(\dfrac{p-q A_{2}}{m}\right)+1\right)^2}
\end{split}
\end{equation}
and
\begin{equation}
\begin{split}
    \Gamma_\mathrm{t}
    & =\dfrac{\sqrt{\left(\dfrac{p-q A_{2}}{m}\right)^{2}+1}}{\dfrac{1}{2}\left(\dfrac{p-q A_{1}}{m}\right)\left(\dfrac{p-q A_{2}}{m}\right)+1} \\
    & \simeq \dfrac{\dfrac{1}{2}\left(\dfrac{p-q A_{2}}{m}\right)^{2}+1}{\dfrac{1}{2}\left(\dfrac{p-q A_{1}}{m}\right)\left(\dfrac{p-q A_{2}}{m}\right)+1}.
\end{split}
\end{equation}

Since $m(c^2)\gg p(c)-qA_{1,2}$, the terms $[(p-qA_{1,2})/m]^2$ are negligible. Therefore,
\begin{equation}\label{eq:gamma2_Sch}
    \Gamma_\mathrm{t} \simeq 1.
\end{equation}

Inserting Eq.~\eqref{eq:gamma2_Sch} into Eq.~\eqref{eq:f_and_b} leads to the non-relativistic reduction of the Dirac reflection and transmission coefficients 
\begin{equation} \label{eq:Sch_At_coef_nonrel}
    f=1 \quad \text { and } \quad b=0,
\end{equation}
which are the same as those obtained from the Schr{\"o}dinger equation.

Note that the corresponding solution to the Schr{\"o}dinger equation [Eq.~\eqref{eq:Schroedinger_eq}] is here [for the potential given in~\eqref{eq:s_vector_pot_temporal}] readily found as
\begin{subequations}
    \begin{equation}
        \psi_{1} = \mathrm{e}^{-i E_\mathrm{i} t} \mathrm{e}^{i p_\mathrm{i} z}
    \end{equation}
    and
    \begin{equation}
        \psi_{2} = f \mathrm{e}^{-i E_\mathrm{f} t} \mathrm{e}^{i p_\mathrm{f} z},
    \end{equation}
where
    \begin{equation}
        p_\mathrm{i}=p_\mathrm{f}=p, \quad
        E_\mathrm{i}=\frac{(p-qA_1)^2}{2m} \quad
        \textrm{and} \quad
        E_\mathrm{f}=\frac{(p-qA_2)^2}{2m}.
    \end{equation}
\end{subequations}
which implies the later-forward coefficient $f=1$ and later-backward coefficient $b=0$, consistently with Eq.~\eqref{eq:Sch_At_coef_nonrel}.

\end{document}